\documentclass[longauth]{aa}

\usepackage{graphicx}
\usepackage{txfonts}

\usepackage{subcaption}         
\usepackage{placeins}           
                             
\usepackage[]{hyperref}
\usepackage{xcolor}
\usepackage{multirow}

\usepackage{amsmath,mathtools}
\usepackage{threeparttable}

\newcommand{\Prot}{P_{\mathrm{rot},\star}}
\newcommand{\ttt}{\texttt{t}}
\newcommand{\ttx}{\texttt{x}}
\newcommand{\tty}{\texttt{y}}
\newcommand{\ttdx}{\texttt{pc1}}
\newcommand{\ttdy}{\texttt{pc2}}
\newcommand{\ttxy}{(\texttt{xy})}
\newcommand{\GProll}{GP(\texttt{roll})}
\newcommand{\ttbg}{\texttt{bg}}
\newcommand{\ttco}{\texttt{conta}}
\newcommand{\ttsm}{\texttt{smear}}

\newcommand{\trades}[0]{\texttt{TRADES}}
\newcommand{\pyde}[0]{\texttt{PyDE}}
\newcommand{\emcee}[0]{\texttt{emcee}}

\newcommand{\unif}[2]{\ensuremath{\mathcal{U} (#1,#2)}}

\newcommand{\ms}{$\mathrm{m\,s^{-1}}$}
\newcommand{\kms}{$\mathrm{km\,s^{-1}}$}

\newcommand{\Tzerob}[1][days]  {$59649.1244\pm0.0011$~#1} 
\newcommand{\Pb}[1][days]  {$3.390364\pm0.000011$~#1} 
\newcommand{\kb}[1][${\rm m\,s^{-1}}$]  {$4.54_{-0.78}^{+0.77}$~#1} 
\newcommand{\mpb}[1][$M_{\oplus}$]  {$9.6\pm1.7$~#1} 
\newcommand{\Tzeroc}[1][days]  {$59650.9224\pm0.0010$~#1} 
\newcommand{\Pc}[1][days]  {$7.885393\pm0.000039$~#1} 
\newcommand{\kc}[1][${\rm m\,s^{-1}}$]  {$2.50_{-0.82}^{+0.84}$~#1} 
\newcommand{\mpc}[1][$M_{\oplus}$]  {$7.0_{-2.3}^{+2.4}$~#1}

\newcommand{\Tzerod}[1][days] {$59655.2044\pm0.0020$~#1} 
\newcommand{\Pd}[1][days]   {$13.73151\pm0.00014$~#1} 
\newcommand{\kd}[1][${\rm m\,s^{-1}}$]   {$2.13_{-0.81}^{+0.80}$~#1} 
\newcommand{\mpd}[1][$M_{\oplus}$]   {$7.2\pm2.7$~#1} 
\newcommand{\Tzeroe}[1][days]   {$60687.1358\pm0.0019$~#1} 
\newcommand{\Pe}[1][days]   {$21.49030\pm0.00012$~#1} 
\newcommand{\ke}[1][${\rm m\,s^{-1}}$]  {$1.74_{-0.85}^{+0.88}$~#1} 
\newcommand{\mpe}[1][$M_{\oplus}$] {$6.8_{-3.3}^{+3.5}$~#1} 
\newcommand{\Tzerof}[1][days] {$60384.4\pm3.2$~#1} 
\newcommand{\Pf}[1][days]   {$44.41_{-0.82}^{+0.85}$~#1} 
\newcommand{\kf}[1][${\rm m\,s^{-1}}$]   {$2.46_{-0.83}^{+0.78}$~#1} 
\newcommand{\mpf}[1][$M_{\oplus}$]   {$12.3_{-4.2}^{+4.0}$~#1} 
\newcommand{\OffsetRV}[1][${\rm m\,s^{-1}}$]   {$1.1 \pm 1.7$~#1} 
\newcommand{\OffsetFWHM}[1][${\rm m\,s^{-1}}$]   {$7125.2 \pm 8.3$~#1} 
\newcommand{\jHARPSNN}[1][${\rm m\,s^{-1}}$]   {$4.06_{-0.49}^{+0.52}$~#1} 
\newcommand{\jFWHM}[1][${\rm m\,s^{-1}}$]   {$4.00_{-0.64}^{+0.74}$~#1} 
\newcommand{\AmpBRV}[1][]   {$5.6 _{-1.1}^{+1.3}$~#1} 
\newcommand{\AmpCRV}[1][]   {$19.3_{-3.2}^{+4.2}$~#1} 
\newcommand{\AmpBFWHM}[1][]   {$28.8_{-3.9}^{+5.3}$~#1} 
\newcommand{\jlambdae}[1][]   {$30.2 _{-3.9 }^{+4.1}$~#1} 
\newcommand{\jlambdap}[1][]   {$0.459_{-0.056}^{+0.064}$~#1} 
\newcommand{\jPGP}[1][]   {$19.47\pm0.25$~#1} 

\begin{document}

   \title{The multi-planet system TOI-5624: Four transiting sub-Neptunes with an outer companion revealed by transit-timing variations\thanks{This study uses CHEOPS data observed as part of the Guaranteed Time Observation (GTO) programmes CH\_PR110045 (PI: Serrano) and CH\_PR140083 (PI: Gandolfi). Based on observations made with the Italian Telescopio Nazionale Galileo (TNG) operated on the island of La Palma by the Fundación Galileo Galilei of the INAF (Istituto Nazionale di Astrofisica) at the Spanish Observatorio del Roque de los Muchachos of the Instituto de Astrofisica de Canaria (programme ID: A47TAC\_45; PI: Serrano).}}
  
\author{
A.~Bonfanti\inst{\ref{inst:1}}\and
D.~Gandolfi\inst{\ref{inst:2}}\and
P.~Leonardi\inst{\ref{inst:3},\ref{inst:4}}\and
H.~P.~Osborn\inst{\ref{inst:5},\ref{inst:6}}\and
L.~M.~Serrano\inst{\ref{inst:2}}\and
G.~Hébrard\inst{\ref{inst:7},\ref{inst:8}}\and
N.~Billot\inst{\ref{inst:9}}\and
A.~Bekkelien\inst{\ref{inst:9}}\and
G.~Olofsson\inst{\ref{inst:10}}\and
C.~Broeg\inst{\ref{inst:11},\ref{inst:5}}\and
D.~Nardiello\inst{\ref{inst:12},\ref{inst:13},\ref{inst:14}}\and
S.~G.~Sousa\inst{\ref{inst:15},\ref{inst:16}}\and
T.~G.~Wilson\inst{\ref{inst:17}}\and
A.~C.~M.~Correia\inst{\ref{inst:18}}\and
C.~Pezzotti\inst{\ref{inst:19}}\and
A.~Brandeker\inst{\ref{inst:10}}\and
L.~Fossati\inst{\ref{inst:1}}\and
M.~Gillon\inst{\ref{inst:20}}\and
M.~Stalport\inst{\ref{inst:19},\ref{inst:20}}\and
B.~Akinsanmi\inst{\ref{inst:9}}\and
Y.~Alibert\inst{\ref{inst:5},\ref{inst:11}}\and
R.~Alonso\inst{\ref{inst:21},\ref{inst:22}}\and
J.~Asquier\inst{\ref{inst:23}}\and
T.~Bárczy\inst{\ref{inst:24}}\and
D.~Barrado\inst{\ref{inst:25}}\and
S.~C.~C.~Barros\inst{\ref{inst:15},\ref{inst:16}}\and
W.~Baumjohann\inst{\ref{inst:1}}\and
W.~Benz\inst{\ref{inst:11},\ref{inst:5}}\and
L.~Borsato\inst{\ref{inst:13}}\and
A.~Castro-González\inst{\ref{inst:9}}\and
A.~Collier~Cameron\inst{\ref{inst:26}}\and
Sz.~Csizmadia\inst{\ref{inst:27}}\and
P.~E.~Cubillos\inst{\ref{inst:1},\ref{inst:28}}\and
M.~B.~Davies\inst{\ref{inst:29}}\and
M.~Deleuil\inst{\ref{inst:30}}\and
X.~Delfosse\inst{\ref{inst:31}}\and
A.~Deline\inst{\ref{inst:9}}\and
O.~D.~S.~Demangeon\inst{\ref{inst:15},\ref{inst:16}}\and
B.-O.~Demory\inst{\ref{inst:5},\ref{inst:32},\ref{inst:11}}\and
A.~Derekas\inst{\ref{inst:33}}\and
F.~Destriez\inst{\ref{inst:34},\ref{inst:7}}\and
B.~Edwards\inst{\ref{inst:35}}\and
D.~Ehrenreich\inst{\ref{inst:9},\ref{inst:36}}\and
A.~Erikson\inst{\ref{inst:27}}\and
A.~Fortier\inst{\ref{inst:11},\ref{inst:5}}\and
M.~Fridlund\inst{\ref{inst:37},\ref{inst:38}}\and
K.~Gazeas\inst{\ref{inst:39}}\and
M.~Güdel\inst{\ref{inst:40}}\and
M.~N.~Günther\inst{\ref{inst:23}}\and
N.~Hara\inst{\ref{inst:41}}\and
N.~Heidari\inst{\ref{inst:7}}\and
A.~Heitzmann\inst{\ref{inst:9}}\and
Ch.~Helling\inst{\ref{inst:1},\ref{inst:42}}\and
K.~G.~Isaak\inst{\ref{inst:23}}\and
T.~Keller\inst{\ref{inst:11},\ref{inst:5}}\and
L.~L.~Kiss\inst{\ref{inst:43},\ref{inst:44}}\and
D.~Kitzmann\inst{\ref{inst:11},\ref{inst:5}}\and
J.~Korth\inst{\ref{inst:9}}\and
G.~Lacedelli\inst{\ref{inst:21}}\and
K.~W.~F.~Lam\inst{\ref{inst:27}}\and
J.~Laskar\inst{\ref{inst:45}}\and
A.~Lecavelier~des~Etangs\inst{\ref{inst:46}}\and
A.~Leleu\inst{\ref{inst:9},\ref{inst:11}}\and
M.~Lendl\inst{\ref{inst:9}}\and
D.~Magrin\inst{\ref{inst:13}}\and
P.~F.~L.~Maxted\inst{\ref{inst:47}}\and
M.~Mecina\inst{\ref{inst:48}}\and
B.~Merín\inst{\ref{inst:49}}\and
C.~Mordasini\inst{\ref{inst:11},\ref{inst:5}}\and
V.~Nascimbeni\inst{\ref{inst:13}}\and
R.~Ottensamer\inst{\ref{inst:40}}\and
I.~Pagano\inst{\ref{inst:50}}\and
E.~Pallé\inst{\ref{inst:21},\ref{inst:22}}\and
G.~Peter\inst{\ref{inst:27}}\and
D.~Piazza\inst{\ref{inst:51}}\and
G.~Piotto\inst{\ref{inst:13},\ref{inst:52}}\and
D.~Pollacco\inst{\ref{inst:17}}\and
D.~Queloz\inst{\ref{inst:6},\ref{inst:53}}\and
R.~Ragazzoni\inst{\ref{inst:13},\ref{inst:52}}\and
N.~Rando\inst{\ref{inst:23}}\and
H.~Rauer\inst{\ref{inst:54},\ref{inst:55}}\and
I.~Ribas\inst{\ref{inst:56},\ref{inst:57}}\and
N.~C.~Santos\inst{\ref{inst:15},\ref{inst:16}}\and
G.~Scandariato\inst{\ref{inst:50}}\and
D.~Ségransan\inst{\ref{inst:9}}\and
A.~E.~Simon\inst{\ref{inst:11},\ref{inst:5}}\and
A.~M.~S.~Smith\inst{\ref{inst:27}}\and
S.~Sulis\inst{\ref{inst:30}}\and
Gy.~M.~Szabó\inst{\ref{inst:33},\ref{inst:58}}\and
S.~Udry\inst{\ref{inst:9}}\and
S.~Ulmer-Moll\inst{\ref{inst:59},\ref{inst:19}}\and
V.~Van~Grootel\inst{\ref{inst:19}}\and
J.~Venturini\inst{\ref{inst:9}}\and
F.~Verrecchia\inst{\ref{inst:60},\ref{inst:61}}\and
E.~Villaver\inst{\ref{inst:21},\ref{inst:22}}\and
N.~A.~Walton\inst{\ref{inst:62}}\and
S.~Wolf\inst{\ref{inst:51}}\and
D.~Wolter\inst{\ref{inst:27}}\and
T.~Zingales\inst{\ref{inst:52},\ref{inst:13}}
}

\institute{
\label{inst:1} Space Research Institute, Austrian Academy of Sciences, Schmiedlstrasse 6, A-8042 Graz, Austria \\
\email{andrea.bonfanti@oeaw.ac.at}
\and
\label{inst:2} Dipartimento di Fisica, Università degli Studi di Torino, via Pietro Giuria 1, I-10125, Torino, Italy \and
\label{inst:3} Dipartimento di Fisica, Università di Trento, Via Sommarive 14, 38123 Povo \and
\label{inst:4} Dipartimento di Fisica e Astronomia, Università degli Studi di Padova, Vicolo dell’Osservatorio 3, 35122 Padova, Italy \and
\label{inst:5} Center for Space and Habitability, University of Bern, Gesellschaftsstrasse 6, 3012 Bern, Switzerland \and
\label{inst:6} ETH Zurich, Department of Physics, Wolfgang-Pauli-Strasse 2, CH-8093 Zurich, Switzerland \and
\label{inst:7} Institut d'astrophysique de Paris, UMR7095 CNRS, Universit\'e Pierre \& Marie Curie, 98bis boulevard Arago, 75014 Paris, France \and
\label{inst:8} Observatoire de Haute-Provence, CNRS, Universit\'e d'Aix-Marseille, 04870 Saint-Michel-l'Observatoire, France \and
\label{inst:9} Observatoire astronomique de l'Université de Genève, Chemin Pegasi 51, 1290 Versoix, Switzerland \and
\label{inst:10} Department of Astronomy, Stockholm University, AlbaNova University Center, 10691 Stockholm, Sweden \and
\label{inst:11} Space Research and Planetary Sciences, Physics Institute, University of Bern, Gesellschaftsstrasse 6, 3012 Bern, Switzerland \and
\label{inst:12} Dipartimento di Fisica e Astronomia "Galileo Galilei", Universita degli Studi di Padova, Vicolo dell'Osservatorio 3, 35122 Padova, Italy \and
\label{inst:13} INAF, Osservatorio Astronomico di Padova, Vicolo dell'Osservatorio 5, 35122 Padova, Italy \and
\label{inst:14} Centro di Ateneo di Studi e Attività Spaziali "G. Colombo", Università degli Studi di Padova, Via Venezia 15, 35131 Padova, Italy \and
\label{inst:15} Instituto de Astrofisica e Ciencias do Espaco, Universidade do Porto, CAUP, Rua das Estrelas, 4150-762 Porto, Portugal \and
\label{inst:16} Departamento de Fisica e Astronomia, Faculdade de Ciencias, Universidade do Porto, Rua do Campo Alegre, 4169-007 Porto, Portugal \and
\label{inst:17} Department of Physics, University of Warwick, Gibbet Hill Road, Coventry CV4 7AL, United Kingdom \and
\label{inst:18} CFisUC, Departamento de Física, Universidade de Coimbra, 3004-516 Coimbra, Portugal \and
\label{inst:19} Space sciences, Technologies and Astrophysics Research (STAR) Institute, Université de Liège, Allée du 6 Août 19C, 4000 Liège, Belgium \and
\label{inst:20} Astrobiology Research Unit, Université de Liège, Allée du 6 Août 19C, B-4000 Liège, Belgium \and
\label{inst:21} Instituto de Astrofísica de Canarias, Vía Láctea s/n, 38200 La Laguna, Tenerife, Spain \and
\label{inst:22} Departamento de Astrofísica, Universidad de La Laguna, Astrofísico Francisco Sanchez s/n, 38206 La Laguna, Tenerife, Spain \and
\label{inst:23} European Space Agency (ESA), European Space Research and Technology Centre (ESTEC), Keplerlaan 1, 2201 AZ Noordwijk, The Netherlands \and
\label{inst:24} Admatis, 5. Kandó Kálmán Street, 3534 Miskolc, Hungary \and
\label{inst:25} Depto. de Astrofísica, Centro de Astrobiología (CSIC-INTA), ESAC campus, 28692 Villanueva de la Cañada (Madrid), Spain \and
\label{inst:26} Centre for Exoplanet Science, SUPA School of Physics and Astronomy, University of St Andrews, North Haugh, St Andrews KY16 9SS, UK \and
\label{inst:27} Institute of Space Research, German Aerospace Center (DLR), Rutherfordstrasse 2, 12489 Berlin, Germany \and
\label{inst:28} INAF, Osservatorio Astrofisico di Torino, Via Osservatorio, 20, I-10025 Pino Torinese To, Italy \and
\label{inst:29} Centre for Mathematical Sciences, Lund University, Box 118, 221 00 Lund, Sweden \and
\label{inst:30} Aix Marseille Univ, CNRS, CNES, LAM, 38 rue Frédéric Joliot-Curie, 13388 Marseille, France \and
\label{inst:31} Universit\'e Grenoble Alpes, CNRS, IPAG, 38000 Grenoble, France \and
\label{inst:32} ARTORG Center for Biomedical Engineering Research, University of Bern, Bern, Switzerland \and
\label{inst:33} ELTE Gothard Astrophysical Observatory, 9700 Szombathely, Szent Imre h. u. 112, Hungary \and
\label{inst:34} LIRA, Observatoire de Paris, Universit\'e PSL, CNRS, Sorbonne Universit\'e, Universit\'e de Paris, 5 place Jules Janssen, 92195 Meudon, France \and
\label{inst:35} SRON Netherlands Institute for Space Research, Niels Bohrweg 4, 2333 CA Leiden, Netherlands \and
\label{inst:36} Centre Vie dans l’Univers, Faculté des sciences, Université de Genève, Quai Ernest-Ansermet 30, 1211 Genève 4, Switzerland \and
\label{inst:37} Leiden Observatory, University of Leiden, PO Box 9513, 2300 RA Leiden, The Netherlands \and
\label{inst:38} Department of Space, Earth and Environment, Chalmers University of Technology, Onsala Space Observatory, 439 92 Onsala, Sweden \and
\label{inst:39} National and Kapodistrian University of Athens, Department of Physics, University Campus, Zografos GR-157 84, Athens, Greece \and
\label{inst:40} Department of Astrophysics, University of Vienna, Türkenschanzstrasse 17, 1180 Vienna, Austria \and
\label{inst:41} Aix-Marseille Université, CNRS, CNES, Institut Origines, LAM, Marseille, France \and
\label{inst:42} Institute for Theoretical Physics and Computational Physics, Graz University of Technology, Petersgasse 16, 8010 Graz, Austria \and
\label{inst:43} Konkoly Observatory, Research Centre for Astronomy and Earth Sciences, 1121 Budapest, Konkoly Thege Miklós út 15-17, Hungary \and
\label{inst:44} ELTE E\"otv\"os Lor\'and University, Institute of Physics, P\'azm\'any P\'eter s\'et\'any 1/A, 1117 Budapest, Hungary \and
\label{inst:45} IMCCE, UMR8028 CNRS, Observatoire de Paris, PSL Univ., Sorbonne Univ., 77 av. Denfert-Rochereau, 75014 Paris, France \and
\label{inst:46} Institut d'astrophysique de Paris, UMR7095 CNRS, Université Pierre \& Marie Curie, 98bis blvd. Arago, 75014 Paris, France \and
\label{inst:47} Astrophysics Group, Lennard Jones Building, Keele University, Staffordshire, ST5 5BG, United Kingdom \and
\label{inst:48} Department of Astrophysics, University of Vienna, Tuerkenschanzstrasse 17, 1180 Vienna, Austria \and
\label{inst:49} European Space Agency, ESA - European Space Astronomy Centre, Camino Bajo del Castillo s/n, 28692 Villanueva de la Cañada, Madrid, Spain \and
\label{inst:50} INAF, Osservatorio Astrofisico di Catania, Via S. Sofia 78, 95123 Catania, Italy \and
\label{inst:51} Weltraumforschung und Planetologie, Physikalisches Institut, University of Bern, Gesellschaftsstrasse 6, 3012 Bern, Switzerland \and
\label{inst:52} Dipartimento di Fisica e Astronomia "Galileo Galilei", Università degli Studi di Padova, Vicolo dell'Osservatorio 3, 35122 Padova, Italy \and
\label{inst:53} Cavendish Laboratory, JJ Thomson Avenue, Cambridge CB3 0HE, UK \and
\label{inst:54} German Aerospace Center (DLR), Markgrafenstrasse 37, 10117 Berlin, Germany \and
\label{inst:55} Institut fuer Geologische Wissenschaften, Freie Universitaet Berlin, Malteserstrasse 74-100,12249 Berlin, Germany \and
\label{inst:56} Institut de Ciencies de l'Espai (ICE, CSIC), Campus UAB, Can Magrans s/n, 08193 Bellaterra, Spain \and
\label{inst:57} Institut d'Estudis Espacials de Catalunya (IEEC), 08860 Castelldefels (Barcelona), Spain \and
\label{inst:58} HUN-REN-ELTE Exoplanet Research Group, Szent Imre h. u. 112., Szombathely, H-9700, Hungary \and
\label{inst:59} Leiden Observatory, University of Leiden, Einsteinweg 55, 2333 CA Leiden, The Netherlands \and
\label{inst:60} Space Science Data Center, ASI, via del Politecnico snc, 00133 Roma, Italy \and
\label{inst:61} INAF, Osservatorio Astronomico di Roma, via Frascati 33, 00078 Monte Porzio Catone (RM), Italy \and
\label{inst:62} Institute of Astronomy, University of Cambridge, Madingley Road, Cambridge, CB3 0HA, United Kingdom
}

   \date{}

 
  \abstract
   {Following the 2022 alert of a TESS object of interest transiting TOI-5624 (a G7\,V star $\sim$100\,pc away), a CHEOPS campaign in 2023 detected four planetary signals at $P_b$\,$\approx$\,3.4, $P_c$\,$\approx$\,7.9, $P_d$\,$\approx$\,13.7, and $P_e$\,$\approx$\,21.5 days. These signals were later confirmed by additional TESS and CHEOPS photometry in 2024--2025.}
   {By using TESS and CHEOPS photometry, along with HARPS-N and SOPHIE high-resolution spectra, we determined the planet properties and performed a dynamical analysis of the system.} 
   {After analysing the photometric data, we extracted and modelled the radial velocity (RV) time series using two independent methodologies, both within a Markov chain Monte Carlo framework. We further integrated the N-body equations of motion, while simultaneously fitting the transit times and the detrended RVs, to dynamically characterise the system.}
   {We present the discovery of four transiting sub-Neptunes with radii of $R_b=2.314\pm0.035\,R_{\oplus}$, $R_c=2.474\pm0.042\,R_{\oplus}$, $R_d=3.584_{-0.050}^{+0.051}\,R_{\oplus}$, and $R_e=3.247_{-0.043}^{+0.042}\,R_{\oplus}$, along with masses of $M_b=9.4\pm1.4\,M_{\oplus}$, $M_c=4.8\pm1.9\,M_{\oplus}$, $M_d=4.9\pm2.2\,M_{\oplus}$, and $M_e=8.9_{-3.0}^{+2.9}\,M_{\oplus}$. Our photometric analysis revealed that the outermost transiting planet TOI-5624\,e shows significant transit-timing variations (TTVs). Indeed, we found a robust Keplerian signal in the RV time series close to the 2:1 period commensurability with TOI-5624\,e, which explains the TTV pattern exhibited by TOI-5624\,e according to our dynamical analysis. We labelled this non-transiting planet as TOI-5624\,f and determined its minimum mass to be $M_f\sin{i_f}=13.0\pm3.7\,M_{\oplus}$.}
   {Among the known systems hosting more than four planets, the remarkable precision with which the radii have been measured ($<1.7\%$) and the firm assessment (>\,$3\sigma$) of the mass for at least three planets has previously been reached only for TRAPPIST-1. Additional photometric observations will enable a better sample of the TTV modulation and a more robust dynamical determination of the masses.}
   
   \titlerunning{The multi-planet system TOI-5624}
   \maketitle
   \nolinenumbers

\section{Introduction}
Multi-planet systems are useful sources for setting valuable constraints on formation and evolution models by reducing their degrees of freedom, based on the fact that the planets in each system had formed within the same protoplanetary disc and currently orbit the same star \citep[e.g.,][]{fortier2013,alibert2013,raymond2022,armitage2024}.
Knowledge of the planetary parameters is essential to facilitate the exploration of mutual correlations and the identification of differences and similarities that shape the architecture of multi-planet systems \citep[e.g.,][]{weiss2018,mishra2023}. 

The transit method \citep{Winn2010} and the radial velocity (RV) technique \citep{hatzes2019} allow for the measurement of both the planetary radius and mass, from which the mean density, $\rho_p$, can be derived. Although $\rho_p$ is a bulk parameter, combining it with dedicated internal structure and atmospheric evolution models offers a first indication of the possible planetary structure and presence of an atmosphere \citep[e.g.,][]{wang2019,wordsworth2020,korshid2022,egger2024,haldemann2024}.

The present-day location and mutual distance between planets shed light on the migration mechanisms that may have occurred either during the formation processes or after protoplanetary disc dispersal \citep[e.g.,][]{goldreich1980,malhotra1995,hidekazu2002,mandell2011,johansen2019}. Significant dynamical interactions between planets are further expected under specific scenarios; for instance, in the presence of an orbital period commensurability between adjacent planets. In this case, the planets are close to (or potentially even trapped) in mean-motion resonances \citep[MMRs;][]{lee2002,correia2018} and, thus, they are likely to exhibit strong transit-timing variations \citep[TTVs;][]{agol2005,AgolFabrycky2018,leleu2021}. This phenomenon enables the mass determination of the interacting bodies, offering an alternative to the RV method for measuring planetary masses \citep[e.g.,][]{lithwick2012,borsato2014,deck2015,hadden2017,leleu2023}. 

The Transiting Exoplanet Survey Satellite \citep[TESS;][]{ricker2015} observed TOI-5624 in 2020 (Sector\,22), 2022 (Sectors\,48 and 49), and 2024 (Sectors\,75 and 76). 
A TESS object of interest (TOI-5624.01) was detected in 2022 by the automated QLP TESS pipeline \citep{fausnaugh2018} with a period of 13.73\,d. In summer 2022, we inspected Sectors 22, 48, and 49, whereby a combination of periodic searches with transit least squares \citep{hippke2019} revealed three additional planet candidates, with periods of 3.39, 7.89, and 21.49\,d.
To reassess the identified signals, we retrieved their ephemeris using \texttt{MonoTools} \citep{osborn2022} and carried out dedicated transit observations with the Characterising Exoplanet Satellite \citep[CHEOPS;][]{benz2021,fortier2024} between January and April 2023 through a Guaranteed Time Observation (GTO) programme (CH\_PR110045).
In December 2023, the 3.39\,d planet was also independently identified as a community TOI by M.~Jassim Munavar Hussain. Finally, all four planets were designated TOIs by the TESS team after collecting Sectors\,75 and 76 in April 2024 \citep[see also][]{jassimMunavarHussain2025}.

In this work, we present the discovery and characterisation of the four transiting planets around TOI-5624 by combining 5 TESS sectors and 36 CHEOPS transit observations with 96 HARPS-N \citep{cosentino2012} and 26 SOPHIE \citep{perruchot2008} high-precision RV measurements. We found that the 21.5~d planet (TOI-5624\,e) exhibits significant TTVs induced by a fifth non-transiting planet (TOI-5624\,f) most likely on a $\sim$45~d orbit. This paper is organised as follows. Sections~\ref{sec:photometry} and~\ref{sec:spectroscopy} present the photometric and spectroscopic data. Sect.~\ref{sec:hostStar} outlines the properties of the host star. Sect.~\ref{sec:dataAnalysis} describes the photometric, RV, and TTV analyses. Finally Sect.~\ref{sec:conclusions} reports the conclusions.

\section{Photometric data}\label{sec:photometry}
\subsection{TESS}
TESS observed TOI-5624 in five different sectors from 2020 to 2024: Sector 22 from 19 February to 17 March 2020 (120~s cadence), Sectors 48 and 49 from 28 January to 25 March 2022 (600~s cadence), and Sectors 75 and 76 from 30 January 2024 to 25 March 2024 (120~s cadence).
After downloading the TESS data product from the Mikulski Archive for Space Telescopes (MAST) portal\footnote{\url{https://mast.stsci.edu/portal/Mashup/Clients/Mast/Portal.html}}, we analysed the Pre-search Data Conditioning Simple Aperture Photometry \citep[PDCSAP;][]{jenkins2016}, where long-term trends and instrumental artefacts have been corrected for \citep{smith2012,stumpe2012,stumpe2014} by the Science Processing Operation Centre (SPOC), and applied a 5-median-absolute-deviation (MAD) clipping to reject outliers. For each transit event, we selected a temporal window spanning the transit duration plus $\sim$8~hours of out-of-transit data to ensure an adequate baseline for the following detrending. Then, we built up a light curve that lists not only the time (\ttt) and the PDCSAP flux along with its errors, but also ancillary detrending parameters\footnote{See \url{https://tasoc.dk/docs/EXP-TESS-ARC-ICD-TM-0014-Rev-F.pdf} for further details about the TESS data products.}, such as \textsc{mom\_centr1} and \textsc{mom\_centr2} (hereafter \ttx{} and \tty, respectively), and \textsc{pos\_corr1} and \textsc{pos\_corr2} (hereafter \ttdx{} and \ttdy, respectively). Following this procedure, we ended up with 44 TESS light curves (LCs).

\subsection{CHEOPS}
We collected 19 CHEOPS LCs in 2023 (GTO programme CH\_PR110045) and additional 17 CHEOPS LCs in 2025 (GTO programme CH\_PR110083).
The raw data of each visit were automatically processed by the CHEOPS Data Reduction Pipeline \citep[DRP;][]{hoyer2020}, which performs an instrumental calibration (bias, gain, linearisation, and flat-fielding correction) and environmental correction (cosmic ray hits, background, and smearing correction) before extracting the photometric signal of the target in various apertures. The log of the CHEOPS observations is reported in Table~\ref{tab:cheopsLog}, along with the file keys of each visit to query the mission archive or the Data Analysis Center for Exoplanets (DACE\footnote{\url{https://dace.unige.ch/cheopsDatabase/?}}).

We used the LCs obtained with the \texttt{DEFAULT} aperture and applied a 5-MAD clipping to reject flux outliers. Each LC is provided with ancillary parameters, such as the background level due to zodiacal and Earth stray light (\ttbg), the flux contamination from nearby stars entering the Field of View (\ttco), the smearing effect (\ttsm), the target location on the CCD (\ttx{} \& \tty), and the telescope roll angle \citep[\texttt{roll}; see][for further details]{bonfanti2021}. The flux vs. \texttt{roll} pattern was pre-detrended via Gaussian processes \citep[GPs;][]{rasmussen05} with a Matérn-3/2 kernel using the \texttt{celerite} package \citep{foreman17} and photometric uncertainties were inflated according to the GP covariance matrix.

\section{Spectroscopic data}\label{sec:spectroscopy}
\subsection{HARPS-N}\label{sec:HARPSN}
We spectroscopically monitored TOI-5624 with the High Accuracy Radial velocity Planet Searcher for the Northern hemisphere \citep[HARPS-N;][]{cosentino2012} echelle spectrograph, which is mounted at the 3.58\,m Italian Telescopio Nazionale Galileo (TNG) located at Roque de los Muchachos Observatory, La Palma, Spain. We acquired 98 high-resolution spectra ($R$\,=\,115\,000) as part of our HARPS-N programme dedicated to the Doppler follow-up of TOI-5624 (programme ID: A47TAC\_45; PI: Serrano). The observations cover a baseline of $\sim$426~d, between 3 February 2023 and 4 April 2024 (UTC). We used the second fibre of the spectrograph to monitor the sky background and set the exposure time to 1800~s, which led to a median signal-to-noise ratio (S/N) per pixel of 60 at 550~nm. Two spectra (6 June 2023 and 27 March 2024 UTC) were omitted because their S/N are below 15, resulting in a final sample of 96 usable HARPS-N spectra.

We employed the HARPS-N data reduction software \citep[\texttt{DRS};][]{pepe2002,Lovis2007} to extract the echelle spectra and computed absolute RV measurements via a multi-order cross-correlation with a G2 numerical mask \citep[][]{baranne1996}. The \texttt{DRS} was further used to extract three profile diagnostics of the cross-correlation functions (CCFs); namely, the bisector inverse slope (BIS), the full width at half maximum (FWHM), and the contrast ($A$). We also utilised the Template-Enhanced Radial velocity Re-analysis Application \citep[\texttt{TERRA;}][]{Anglada2012} to extract relative radial velocities, along with the following activity indicators: H$\alpha$, Na\,D1, Na\,D2, and \ion{Ca}{ii} H\,\&\,K S-index. From the latter, we computed the mean chromospheric activity index $\log{R'_{\mathrm{HK}}}=-4.50\pm0.03$.

Following \citet{simola2019}, we also extracted an independent RV time series by fitting skew-Normal (SN) functions onto the HARPS-N CCFs. Skew-Normal functions are intrinsically asymmetric, enabling the retrieval of the CCF-related activity indicators (i.e., the full-width-half-maximum FWHM$_{\mathrm{SN}}$, the contrast $A_{\mathrm{SN}}$, and the skewness $\gamma$) within the fitting procedure \citep[see, e.g.,][for further details]{bonfanti2023,bonfanti2025}. The subsequent HARPS-N RV time series, along with the CCF profile diagnostics and the activity indicators, are listed in Table~\ref{tab:RVdataHARPS-N}.

\subsection{SOPHIE}\label{sec:SOPHIE}
We also observed TOI-5624 with SOPHIE, the stabilized echelle spectrograph dedicated to high-precision RV measurements at the 1.93-m telescope of the Observatoire de Haute-Provence, France \citep{perruchot2008,bouchy2009,bouchy2013}. Using its high-resolution mode (resolving power $R\,=\,75\,000$) and the fast CCD readout, we secured 26 measurements between February 2023 and March 2025. Exposure times ranged between 11 and 53~minutes (typically 20~minutes), leading to an S/N per pixel between 22 and 41 (typically 30) at 550\,nm. The RVs were extracted with the SOPHIE pipeline \citep{bouchy2009,heidari2024,heidari2025}. In addition to an accurate wavelength calibration, the procedure includes corrections for bad pixels, cosmic rays, and charge transfer inefficiency of the CCD, as well as sky background and instrumental drifts. Sky background, in particular moonlight, is estimated and corrected for using the second SOPHIE fibre aperture that targets the sky 2$\arcmin$ away from the first fibre pointing to the star. 

The RVs and their uncertainties are derived from CCFs with a G2 numerical mask and fitted by Gaussians \citep{baranne1996,pepe2002}. The measurements are reported in Table~\ref{tab:RVdataSOPHIE}, together with the FWHM, contrast, and bisectors of the CCFs, as well as the $\log{R'_\mathrm{HK}}$ index measured on the SOPHIE spectra following \citet{boisse2010}.

\section{Host star properties}\label{sec:hostStar}
TOI-5624 is a late G-type star \citep{upgren1962} located $\sim$100\,pc away with a $G$-band magnitude of 10.5 \citep{GaiaCollaboration2023}. Both its $\log{R'_{\mathrm{HK}}}$ and the peak-to-peak amplitude of the normalised TESS sectors ($\sim$15 ppt) point to a moderately active star \citep{henry1996} and suggest an activity-induced RV root mean square (RMS) of\,$\sim$10\,m\,s$^{-1}$ \citep{hojjatpanah2020}. Given the scatter of a few m\,s$^{-1}$ inherent to the relation, this is in line with the stellar activity induced RV variation of $\sim$7\,m\,s$^{-1}$ we measured while fitting the planets to the RV time series without applying any noise modelling.

For the estimation of the spectroscopic parameters ($T_{\mathrm{eff}}$, $\log g$, [Fe/H]) we followed the \texttt{ARES+MOOG} method described by \citealt[][]{Santos-13, Sousa-14, Sousa-21}. We applied the \texttt{ARES} code\footnote{The last version, \texttt{ARES} v2, can be downloaded at \url{https://github.com/sousasag/ARES}} \citep{Sousa-07, Sousa-15} to the co-added HARPS-N spectrum of TOI-5624 (S/N\,$\approx$\,520 per pixel at 550\,\AA) to measure the equivalent widths (EWs) for the list of \ion{Fe}{i} and \ion{Fe}{ii} lines presented in \citet[][]{Sousa-08}. The best set of spectroscopic parameters was determined by using a minimisation process to find the ionisation and excitation equilibrium. This process makes use of a grid of Kurucz model atmospheres \citep{Kurucz-93} and the latest version of the radiative transfer code \texttt{MOOG} \citep{Sneden-73}. We also measured $v\sin{i_{\star}}$ from isolated metal lines, assuming a macroturbulence velocity of $v_{\mathrm{mac}}=v_{\mathrm{mac},\odot}=2.5$\,\kms. We further derived a more accurate trigonometric surface gravity using Gaia DR3 \citep{GaiaCollaboration2023} data following the same procedure as described in \citet[][]{Sousa-21}, which provides a value consistent with the spectroscopic one.

We computed the stellar radius of TOI-5624 using a Markov chain Monte Carlo (MCMC) modified infrared flux method \citep[IRFM;][]{Blackwell1977,Schanche2020}. From spectral energy distributions (SEDs) built using two sets of stellar atmospheric models \citep{Kurucz1993,Castelli2003} that were constrained by our spectroscopically derived stellar parameters, we produced broadband synthetic photometry that were compared to the observed fluxes within the MCMC in the following bandpasses: 2MASS $J$, $H$, and $K$, WISE $W1$ and $W2$, and Gaia $G$, $G_\mathrm{BP}$, and $G_\mathrm{RP}$ \citep{Skrutskie2006,Wright2010,GaiaCollaboration2023}. Thus, we were able to determine the stellar bolometric flux from which the angular diameter, $\theta$, of the target was obtained via the Stefan-Boltzmann law. By combining  the offset-corrected Gaia parallax $\pi$ \citep{lindegren2021} with $\theta$, we derived the stellar radius for each set of stellar atmospheres. Finally, we combined the two stellar radius posteriors via a Bayesian model averaging. This approach weighs the posterior probability distributions by the Bayesian information criterion of the MCMC analysis of both sets to account for model uncertainties.

We used two different sets of stellar evolutionary models to derive the mass $M_{\star}$ and age $t_{\star}$ of TOI-5624 starting from the derived effective temperature $T_{\mathrm{eff}}$, metallicity [Fe/H], and radius~$R_{\star}$. 
A first pair of mass and age estimates was computed via the isochrone placement routine \citep{bonfanti2015,bonfanti2016}, which interpolates the input parameters within pre-computed grids of \texttt{PARSEC}\footnote{\textsl{PA}dova and T\textsl{R}ieste \textsl{S}tellar \textsl{E}volutionary \textsl{C}ode: \url{https://stev.oapd.inaf.it/cgi-bin/cmd}} v1.2S \citep{marigo2017} tracks and isochrones. A second pair of mass and age estimates was inferred from the \texttt{CLES} \citep[Code Liègeois d'Évolution Stellaire;][]{scuflaire2008} code that computes the best-fit evolutionary track following a Levenberg-Marquadt minimisation scheme \citep[e.g.][]{salmon2021}.
The consistency of both mass and age was successfully checked via a $\chi^2$-based criterion and the distributions of the two respective pairs of outcomes were finally merged (i.e. summed), obtaining $M_{\star}=0.858_{-0.029}^{+0.033}\,M_{\odot}$ and $t_{\star}=5.7_{-2.3}^{+2.6}$ Gyr. See \citet{bonfanti2021} for a broader description of the statistical treatment and of the derivation process. The fundamental stellar properties are listed in Table~\ref{tab:star}.

\begin{table}
\caption{Stellar properties of TOI-5624.}
\label{tab:star}
\centering
\begin{tabular}{l c c}
\hline\hline   
\noalign{\smallskip}
\multicolumn{1}{l}{\multirow{4}*{Star names}} & \multicolumn{2}{l}{TOI-5624} \\
\multicolumn{1}{l}{}               & \multicolumn{2}{l}{TIC 53498154} \\
\multicolumn{1}{l}{}               & \multicolumn{2}{l}{BD+50\,1888} \\
\multicolumn{1}{l}{}               & \multicolumn{2}{l}{Gaia DR3 1546352569189373952} \\
\noalign{\smallskip}
\hline
\noalign{\smallskip}
Parameter & Value & Source \\
\noalign{\smallskip}
\hline
\noalign{\smallskip}
  RA\; [h:min:s] & 12:03:22.64 & Gaia DR3 \\
  DEC\; [$^\circ$:$\arcmin$:$\arcsec$] & +49:15:4.41 & Gaia DR3 \\
  $G$ mag & $10.5435\pm0.0028$ & Gaia DR3 \\
  $\pi$\; [mas] & $9.840\pm0.016$ & Gaia DR3\tablefootmark{(a)} \\
\noalign{\smallskip}
\hline
\noalign{\smallskip}
  $T_{\mathrm{eff}}$\; [K] & $5327\pm64$ & Spectroscopy \\\relax
  [Fe/H]                   & $-0.023\pm0.040$ & Spectroscopy \\
  $\log{g}$\; [cgs]        & $4.49\pm0.03$ & Spec \& Gaia DR3 \\
  $\log{R'_{\mathrm{HK}}}$ & $-4.50\pm0.03$ & Spectroscopy \\
  $v\sin{i_{\star}}$\; [km\,s$^{-1}$] & $2.0\pm0.5$ & Spectroscopy \\
  $R_{\star}$\; [$R_{\odot}$] & $0.820\pm0.005$ & IRFM \\
  $M_{\star}$\; [$M_{\odot}$] & $0.858_{-0.029}^{+0.033}$ & Isochrones \\
  $t_{\star}$\; [Gyr] & $5.7_{-2.3}^{+2.6}$ & Isochrones \\
  $L_{\star}$\; [$L_{\odot}$] & $0.486\pm0.024$ & From $T_{\mathrm{eff}}$ \& $R_{\star}$ \\
  $\rho_{\star}$\; [$\rho_{\odot}$] & $1.556\pm0.063$ & From $M_{\star}$ \& $R_{\star}$ \\
\noalign{\smallskip}
\hline                        
\end{tabular}
\tablefoot{RA \& DEC are reported in the J2000 reference frame. \\ \tablefoottext{a}{Zero-point correction from \citet{lindegren2021} applied.}}
\end{table}

\section{Data analysis}\label{sec:dataAnalysis}
\subsection{Photometric analysis with four planets}\label{sec:LConly_4pla}
First, we modelled the TESS and CHEOPS LCs assuming a four-planet scenario and investigated whether the four transiting planets exhibit TTVs. To this end, we used the \texttt{MCMCI} code \citep{bonfanti2020}, that performs an MCMC analysis, while detrending the time series with low-order polynomials.
In detail, we proceeded as follows. We launched several \texttt{MCMCI} mini-runs (10\,000 steps each) to investigate the best set of polynomial orders to be attributed to the various detrending parameters of each LC, according to the Bayesian information criterion \citep[BIC;][]{schwarz1978} minimisation (see Table~\ref{tab:polyDetrendingTE} and \ref{tab:polyDetrendingCH}). 
We then performed an \texttt{MCMCI} run (200\,000 steps) to properly adjust the photometric error bars accounting for both white and red noise \citep{pont2006,bonfanti2020}. Finally, we launched three independent \texttt{MCMCI} runs (200\,000 steps each, 20\% burn-in) to obtain the system's parameters after checking the convergence of the posterior distributions via the Gelman-Rubin (GR) statistic \citep[$\hat{R}$;][]{gelman1992}. For each planet, we adopted uniform priors (unbounded, except for the physical limits) on the transit depth, d$F$, the impact parameter, $b$, and the transit timing of each transit event, $T_{\mathrm{tr}}$. Normal priors were imposed on $T_{\mathrm{eff}}$, [Fe/H], $R_{\star}$, and $M_{\star}$ according to the values listed in Table~\ref{tab:star} with a twofold aim: (i) constrain the transit shape model via $\rho_{\star}$, as inferred from $M_{\star}$ and $R_{\star}$; (ii) evaluate the quadratic limb darkening (LD) coefficients following interpolation of the LD table pre-computed with the \texttt{get\_lds.py} routine\footnote{\url{https://github.com/nespinoza/limb-darkening}} \citep{espinoza2015}.

\begin{figure}
\centering
\includegraphics[width=\hsize]{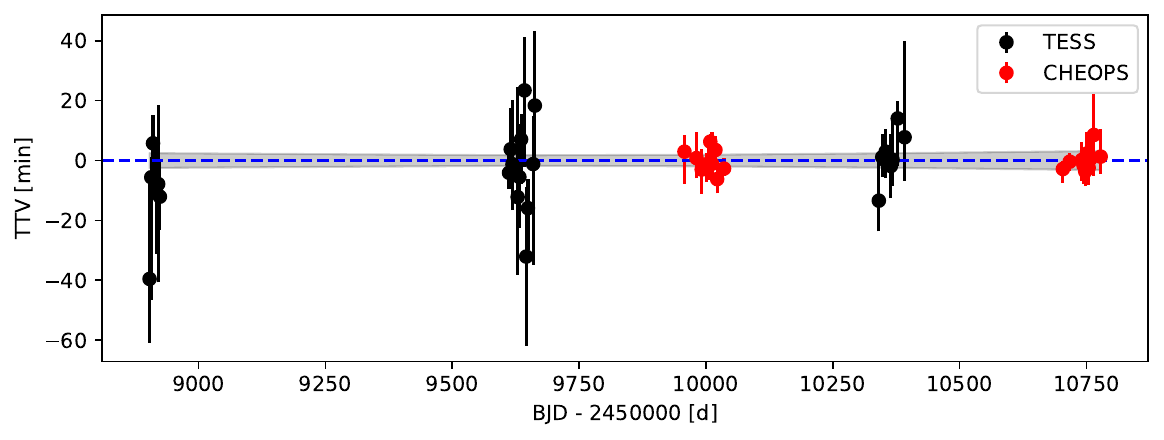} \\
\includegraphics[width=\hsize]{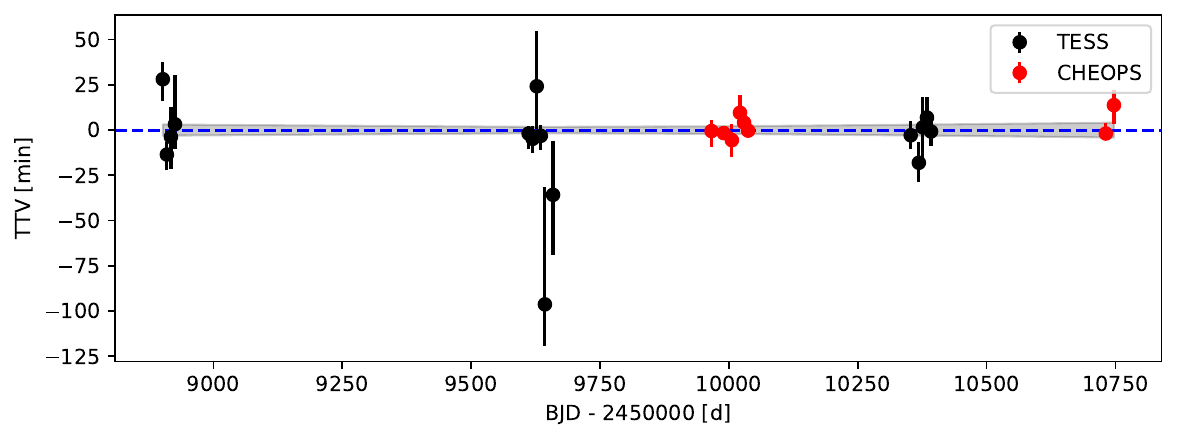} \\
\includegraphics[width=\hsize]{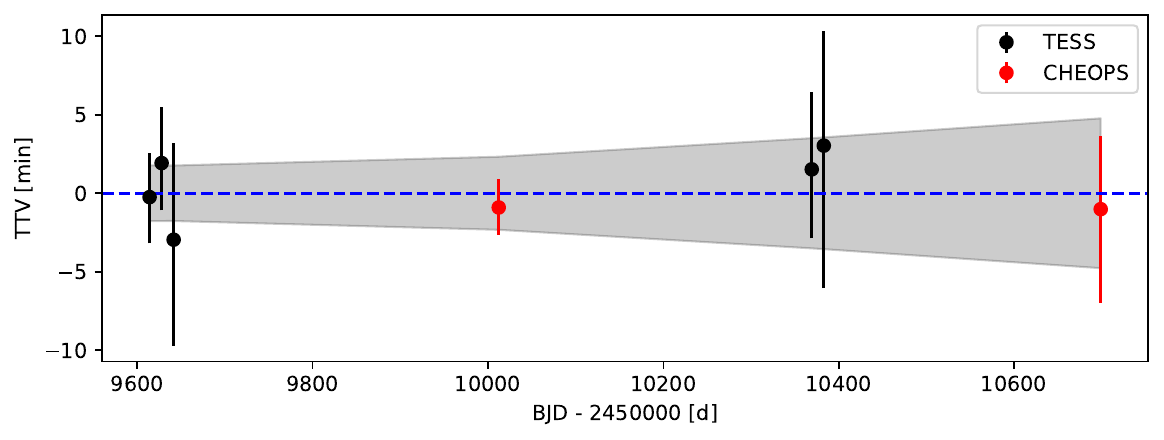} \\
\includegraphics[width=\hsize]{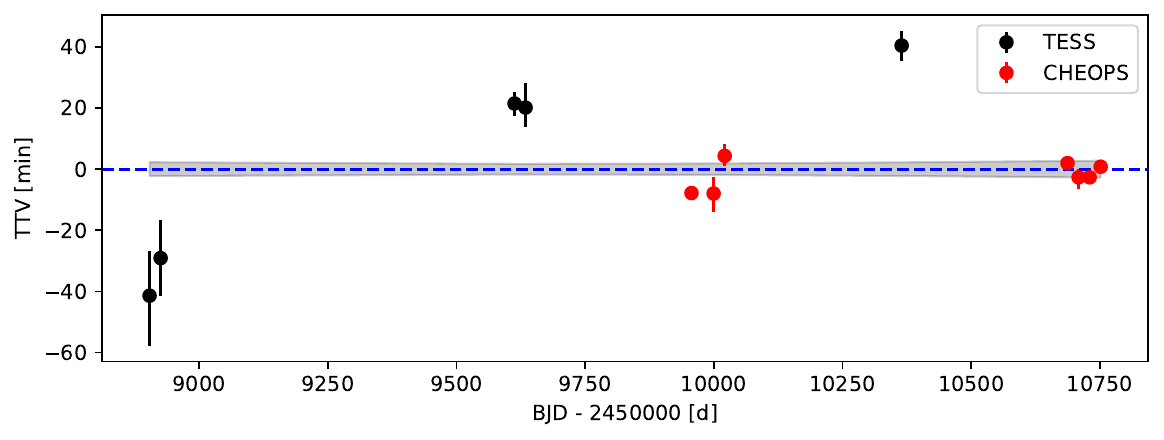}
\caption{TTVs with respect to linear ephemerides for TOI-5624\,b, c, d, and e (top to bottom). The grey shaded area marks the 1$\sigma$ uncertainty region of the linear ephemerides' model.}
\label{fig:TTVs}
\end{figure}

\begin{figure*}
\centering
\includegraphics[height=0.235\textheight]{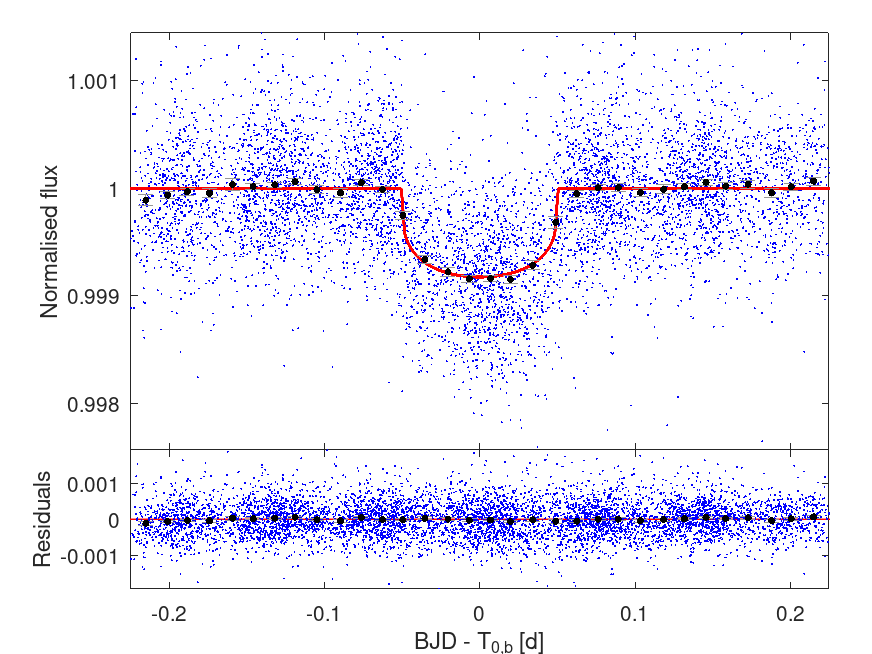}
\includegraphics[height=0.235\textheight]{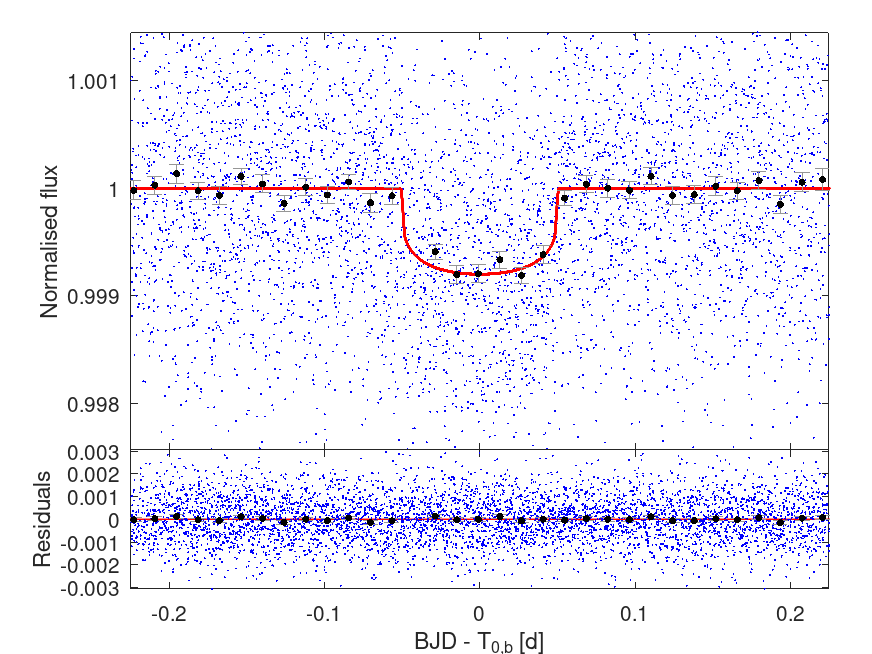} \\
\includegraphics[height=0.235\textheight]{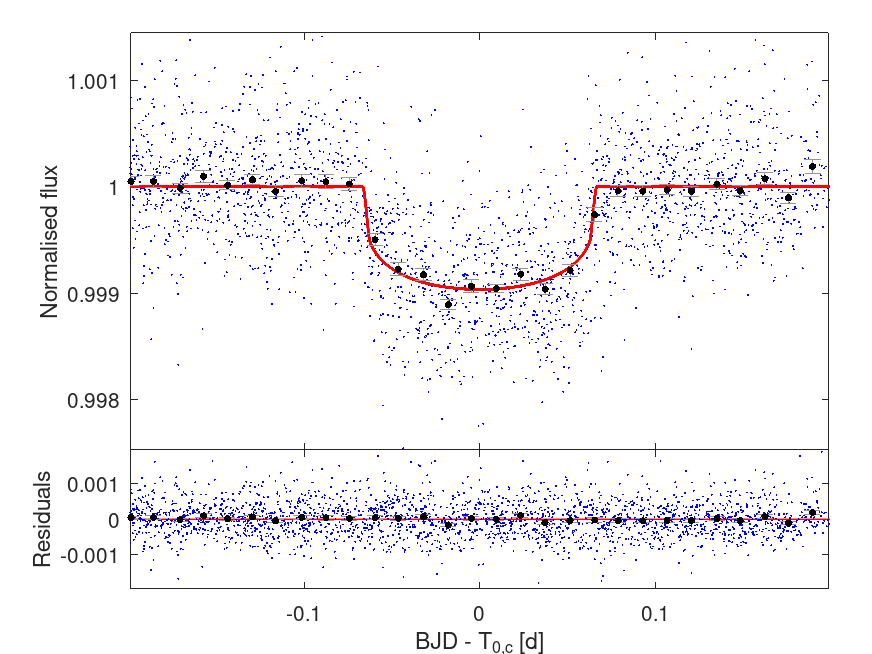}
\includegraphics[height=0.235\textheight]{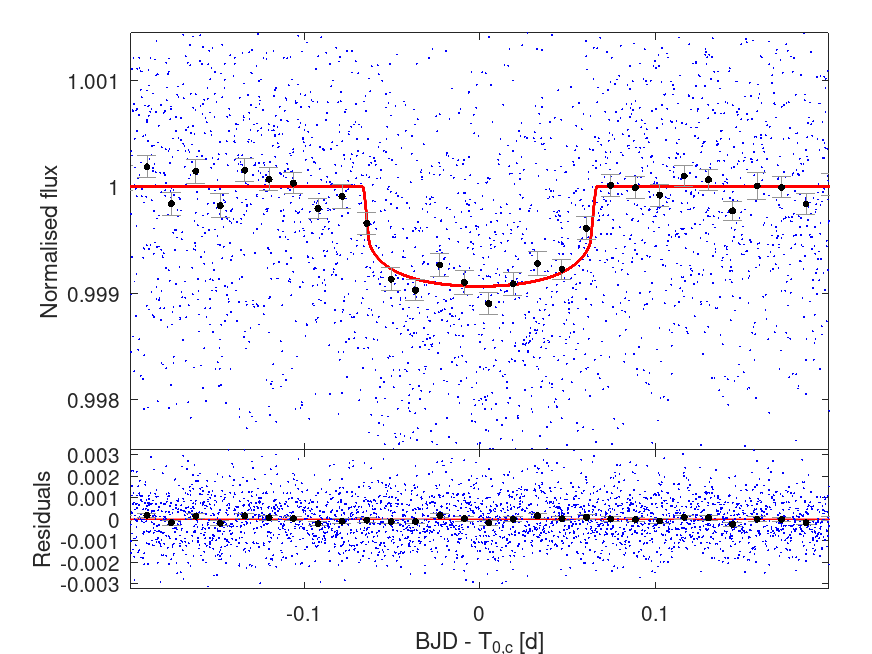} \\
\includegraphics[height=0.235\textheight]{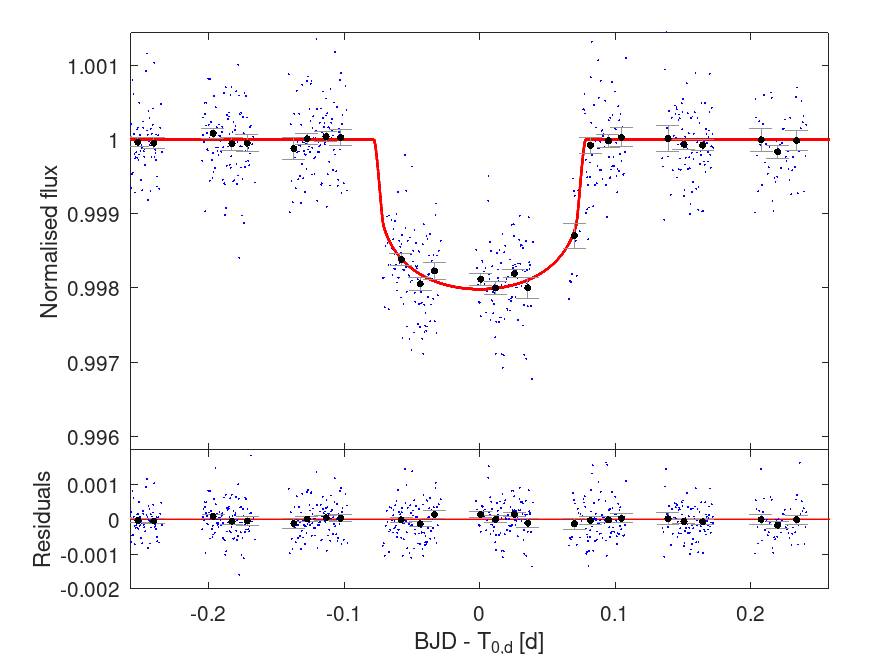}
\includegraphics[height=0.235\textheight]{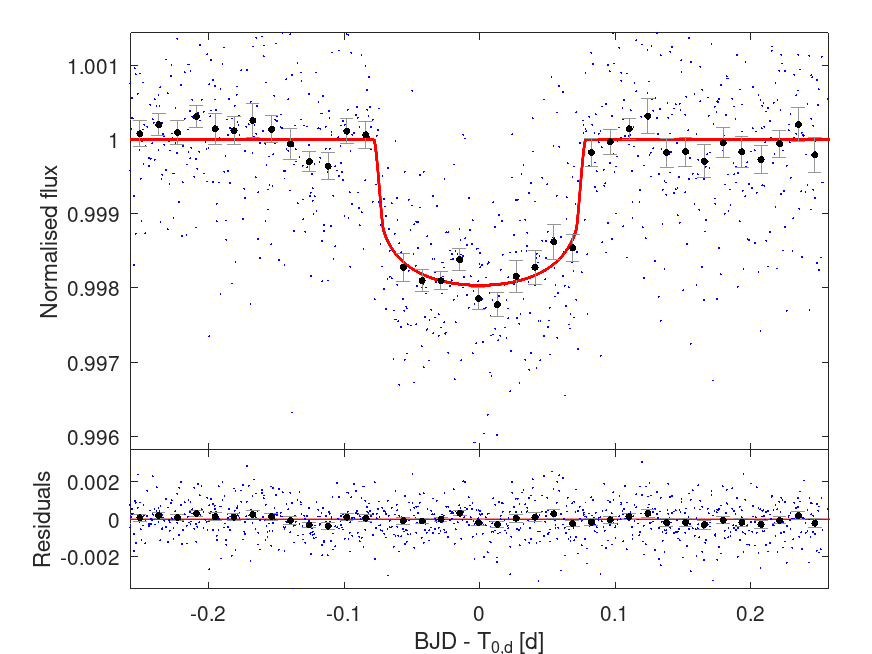} \\
\includegraphics[height=0.235\textheight]{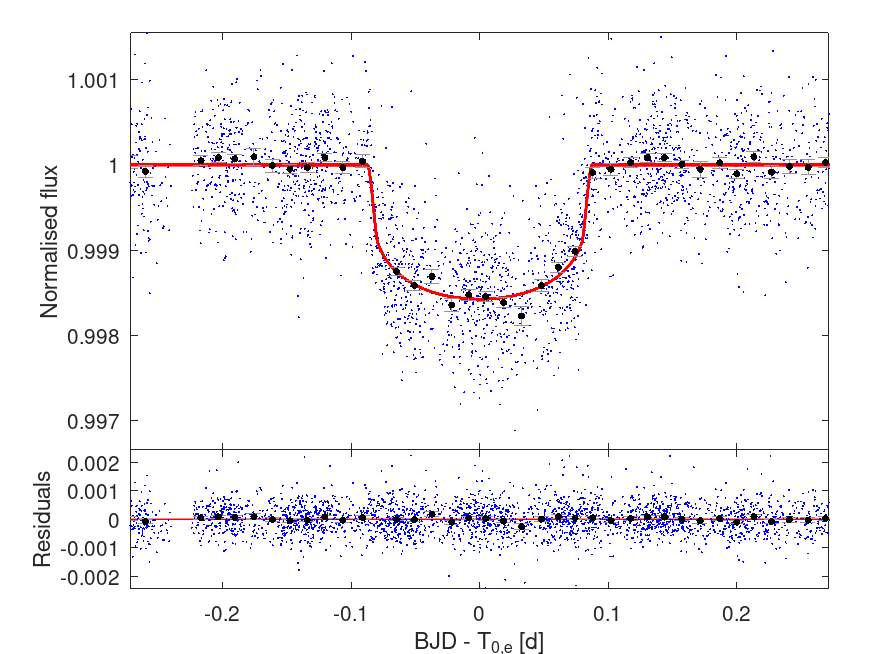}
\includegraphics[height=0.235\textheight]{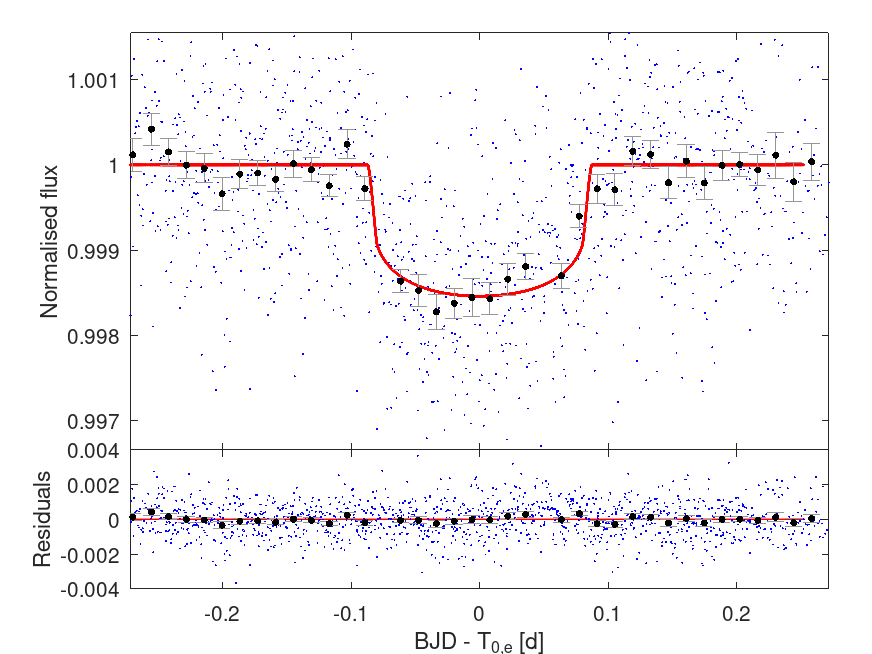}
\caption{Folded and detrendend LCs of planets b to e (top to bottom) as observed by CHEOPS (left) and TESS (right). Blue dots are the original data points, while black dots show the 20-minute binned data. The best-fit transit model is superimposed in red and the corresponding residuals are shown in the bottom panel of each plot.}
\label{fig:LCs}
\end{figure*}

The results ($\hat{R}\lesssim1.01$ for all the jump parameters) are summarised in Table~\ref{tab:planetParameters}. The four transiting planets are sub-Neptunes with radii of $R_b=2.314\pm0.035\,R_{\oplus}$, $R_c=2.474\pm0.042\,R_{\oplus}$, $R_d=3.584_{-0.050}^{+0.051}\,R_{\oplus}$, and $R_e=3.247_{-0.043}^{+0.042}\,R_{\oplus}$ (relative uncertainties~$\lesssim$\,1.7\%). We used the $T_{\mathrm{tr}}$ set of each planet to compute its least-square-based linear ephemerides and then we produced the TTV plots in Fig.~\ref{fig:TTVs}. As quantified by the reduced chi-square ($\hat{\chi}^2$), none of the planets shows statistically significant TTVs ($\hat{\chi}^2_b$\,$\sim$\,0.5, $\hat{\chi}^2_c$\,$\sim$\,1, and $\hat{\chi}^2_d$\,$\sim$\,0.2), except for TOI-5624\,e, whose $\hat{\chi}^2$ sensibly deviates from unity ($\hat{\chi}^2_e$\,$\sim$\,11). The TTV significance of TOI-5624\,e with a peak-to-peak variation of $\sim$\,80 minutes hints at the presence of an outer perturber, whose orbital period is likely commensurable with $P_e$.
After accounting for TTVs, we produced the folded detrended LCs of the four planets shown in Fig.~\ref{fig:LCs}.

\subsection{Searching for a fifth transiting planet}\label{sec:search5pla_LC}

We looked for the putative fifth planet both in the available TESS data and in the RV time series. The only transit-like signal we could detect in the photometry is in Sector 22 at the epoch $T_{\mathrm{ExoFOP}} = 2\,458\,924.4$ BJD, which appears on the ExoFOP webpage\footnote{\url{https://exofop.ipac.caltech.edu/tess/target.php?id=53498154}} as well.
To check whether this transit-like signal is indeed genuine, we further extracted its TESS LC from the Full Frame Image \citep[FFI;][]{sullivan2015}. Briefly, we utilised the \texttt{PATHOS} pipeline \citep{2019MNRAS.490.3806N}, which uses FFIs, empirical point spread functions (PSFs), and the Gaia DR3 catalogue, to extract the LCs of the target star with five different methods (PSF-fitting, 1-px, 2-px, 3-px, and 4-px radius aperture photometry) after subtracting all the neighbouring stars within a radius of 10 pixels ($\sim$3.5$\arcmin$) from TOI-5624. Raw light curves were then corrected with co-trending basis vectors \citep{2021MNRAS.505.3767N}, to minimise the RMS and preserve the stellar variability (see Appendix A of \citealt{2025A&A...693A..32N} for further details).

Although we were able to fit a transit model onto the PDCSAP flux (Fig.~\ref{fig:transitPla5}, left) with a depth of $1072_{-330}^{+350}$\,ppm, two similar \texttt{MCMCI} analyses using the SAP and \texttt{PATHOS}-extracted fluxes led to no transit (no\_tr) detections (Fig.~\ref{fig:transitPla5}, middle and right). If we force a planet to transit within the SAP and \texttt{PATHOS} LCs (tr scenario) by fixing the planet parameters to those inferred from the PDCSAP analysis, run \texttt{MCMCI}, and compute $\Delta\mathrm{BIC} = \mathrm{BIC}_{\mathrm{tr}}-\mathrm{BIC}_{\mathrm{no\_tr}}$, we get $\Delta\mathrm{BIC} = +10$ and +16 for the SAP and \texttt{PATHOS} photometry, respectively.
As the outcomes depend on the photometric extraction pipelines, with the BIC that does not favour the transit scenario, we rejected the hypothesis that a fifth planet transits within the available photometry and we focussed on the RV time series.

\subsection{Searching for a fifth planet in the RV data}\label{sec:search5pla_RV}
We first computed the generalised Lomb-Scargle (GLS) periodograms \citep{zechmeister2009} of different RV-related activity indicators (see Sect.~\ref{sec:HARPSN}). As shown in Fig.~\ref{fig:RVactivity}, there is always a prominent peak at $\sim$\,20 days, which is likely the rotation period of the star, $\Prot$. This is consistent with the upper limit $P_{\mathrm{rot,}\star}^{\mathrm{up}}=20.7\pm5.2$~d, as inferred from $R_{\star}$ and $v\sin{i_{\star}}$; we note that $\Prot$\,$\sim$\,20 d is close to the orbital period of TOI-5624\,e. To corroborate our conclusions, for each of the five TESS sectors extracted by the SAP, we removed all the temporal windows containing transit events and we produced the corresponding GLS periodogram plus a second one after removing the most prominent peak from the first one (Fig.~\ref{fig:glsLC}). All these pairs of periodograms peak at $\Prot$\,$\sim$\,20 d and at its first harmonic ($\Prot/2$), except for Sector~75 whose most prominent peaks are the second ($\Prot/3$) and first ($\Prot/2$) harmonic of $\Prot$.

We then performed an RV-only analysis by fitting the four transiting planets on circular orbits using the \texttt{MCMCI} code that allows the simultaneous detrending of the RV time series against time and four ancillary parameters via low-order polynomials. After executing several \texttt{MCMCI} mini-runs, where we alternated different combinations of four ancillary parameters extracted from the seven activity indicators available for the detrending of the SN-based HARPS-N data, it turned out that the set (FWHM$_{\mathrm{SN}}$, $\gamma$, NaD2, H$\alpha$) displays the best performance in removing the stellar activity, as quantified by the BIC. For SOPHIE data, instead, we used the RV-related activity indicators given by the official pipeline described in Sect.~\ref{sec:SOPHIE}. Then, we launched two independent \texttt{MCMCI} runs and after checking the convergence of the jump parameters (i.e. the time of inferior conjunction, $T_0$, the orbital period, $P$, and the RV semi-amplitude, $K$) via the GR statistic ($\hat{R}<1.003$), we computed the GLS periodogram of the HARPS-N RV residuals (obtained after subtracting the Keplerians of the four transiting planets to the detrended RV time series: Fig.~\ref{fig:glsResi4}, top panel) that peaks at $\sim$\,6.4\,d. After removing this signal (Fig.~\ref{fig:glsResi4}, bottom panel), the highest peak is at $\sim$\,45\,d followed by a peak at $\sim$\,10\,d. Fig.~\ref{fig:ProtHarmonics} shows the pre-whitened GLS periodograms of the CCF FWHM (first row) and $\gamma$ (second row) that peak at $\sim$10\,d and $\sim$6.4\,d. A similar behaviour is seen in the other RV-related activity indicators (not shown here). In addition, peaks at the same frequencies are found in the GLS periodograms of the SAP TESS LCs as mentioned above (Fig.~\ref{fig:glsLC}). Thus, we ascribed the $\sim$10\,d and $\sim$6.4\,d signals to the first and second harmonic of $\Prot$, respectively. The signal at $\sim$45\,d seen in Fig.~\ref{fig:glsResi4} has no counterpart in any of the activity indicators, hinting at the presence of an outer planet (TOI-5624\,f). Planet\,f could be the perturber inducing the TTVs on TOI-5624\,e, being their period ratio $P_f$\,:\,$P_e$ close to the 2:1 first-order commensurability. We successfully explored the five-planet scenario in Sect.~\ref{sec:MCMCI_analysis}, \ref{sec:pyaneti_analysis}, and \ref{sec:TRADES}.

\subsection{RV analysis with MCMCI: Five-planet fit}
\label{sec:MCMCI_analysis}

Using the \texttt{MCMCI} code, we modelled the HARPS-N RVs as extracted from the SN-fit (Sect.~\ref{sec:HARPSN}) and the SOPHIE RVs (Sect.~\ref{sec:SOPHIE}) by fitting five planets on circular orbits, while imposing Gaussian priors on ($P$, $T_0$) of the four transiting planets according to the linear ephemerides we derived in Section~\ref{sec:LConly_4pla}.
The orbital period, $P_f$, and the time of inferior conjunction $T_{0,f}$ of TOI-5624\,f as well as the RV semi-amplitudes of all the five planets were subject to uniform unbounded priors.

The stellar activity in the HARPS-N and SOPHIE time series ($RV_{\star}^{\mathrm{HN}}$ and $RV_{\star}^{\mathrm{SO}}$, respectively) was simultaneously modelled within the MCMC framework via%
\begin{gather}
\begin{multlined}
RV_{\star}^{\mathrm{HN}} = \beta_0^{\mathrm{HN}} + \beta_{1,t}^{\mathrm{HN}} t + \beta_{1,F}^{\mathrm{HN}} \mathrm{FWHM_{SN}} + \displaystyle\sum_{k=1}^{3}{\beta_{k,\gamma}^{\mathrm{HN}} \gamma^k} + \\
+ \beta_{k,H\alpha}^{\mathrm{HN}} H\alpha + \displaystyle\sum_{k=1}^{3}{\beta_{k,\mathrm{NaD2}}^{\mathrm{HN}} \mathrm{NaD2}^k} - \beta_{\mathrm{h2}}\sin{\left[\frac{2\pi}{P_{\mathrm{h2}}}(t-T_{0,\mathrm{h2}})\right]},
\label{eq:RVactivityHN}    
\end{multlined} \\
\begin{multlined}
RV_{\star}^{\mathrm{SO}} = \beta_0^{\mathrm{SO}} + \displaystyle\sum_{k=1}^{2}{\beta_{k,t}^{\mathrm{SO}} t^k} + \displaystyle\sum_{k=1}^{4}{\beta_{k,\mathrm{BIS}}^{\mathrm{SO}} \mathrm{BIS}^k} + \displaystyle\sum_{k=1}^{3}{\beta_{k,A}^{\mathrm{SO}} A^k} + \\
+ \displaystyle\sum_{k=1}^{4}{\beta_{k,R}^{\mathrm{SO}} \log{R'_{\mathrm{HK}}}^k} - \beta_{\mathrm{h2}}\sin{\left[\frac{2\pi}{P_{\mathrm{h2}}}(t-T_{0,\mathrm{h2}})\right]}\,,
\label{eq:RVactivitySO}    
\end{multlined}
\end{gather}
where the $\beta$ parameters are the polynomial coefficients, the polynomial orders were assessed following the BIC minimisation criterion and the sinusoid's parameters, $P_{\mathrm{h2}}$ and $T_{0,\mathrm{h2}}$, were subject to the following Gaussian priors\footnote{$\mathcal{N}(\mu,\sigma)$ denotes a gaussian density function with mean $\mu$ and standard deviation $\sigma$.} $P_{\mathrm{h2}}=\mathcal{N}(6.391,0.011)$ and $T_{0,\mathrm{h2}}=\mathcal{N}(2\,459\,979.29,0.21)$ as inferred from the periodogram in Fig.~\ref{fig:glsResi4} that shows the second harmonic of $\Prot$.

\begin{figure*}
    \centering
    \includegraphics[width=0.495\textwidth]{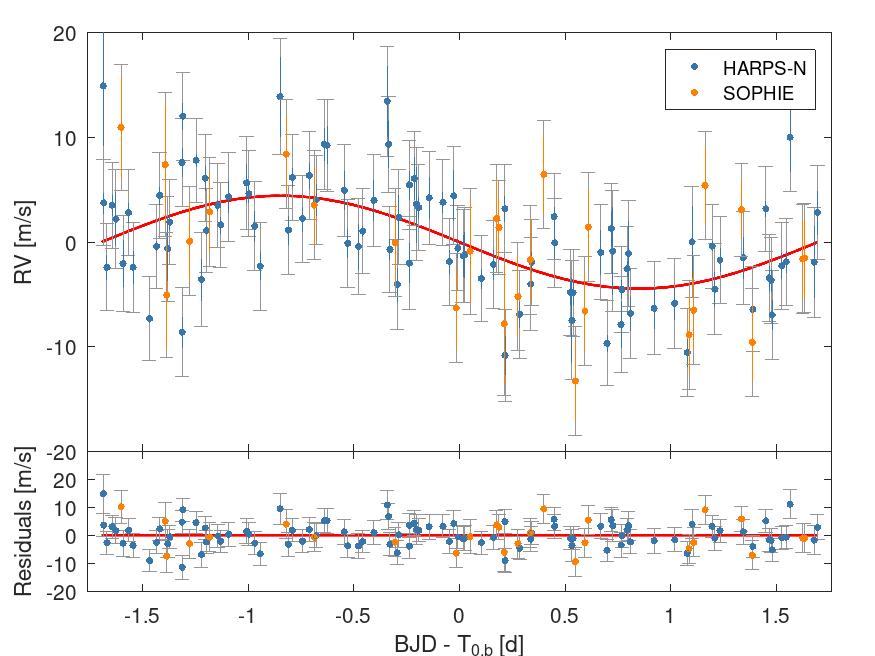}
    \includegraphics[width=0.495\textwidth]{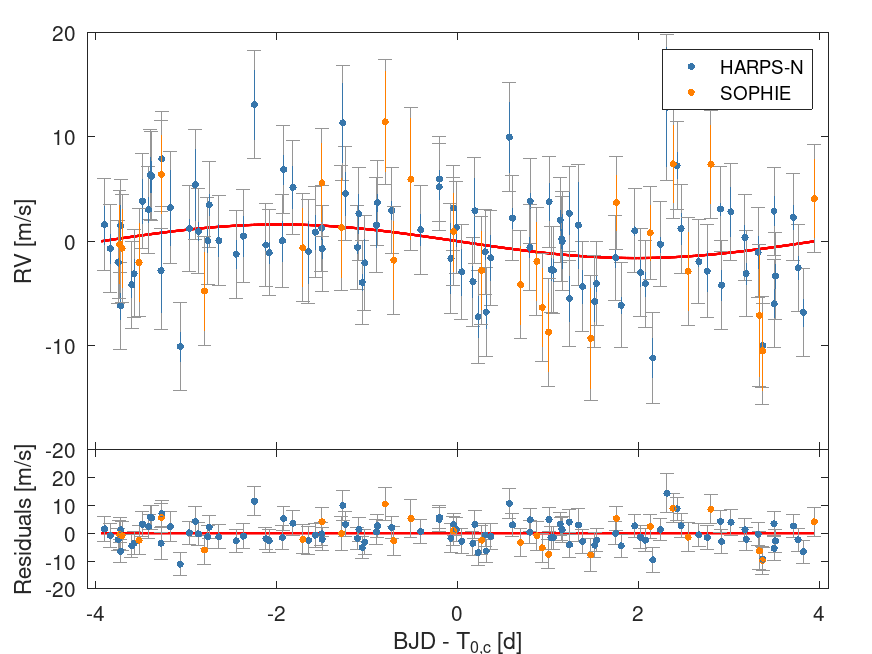} \\
    \includegraphics[width=0.495\textwidth]{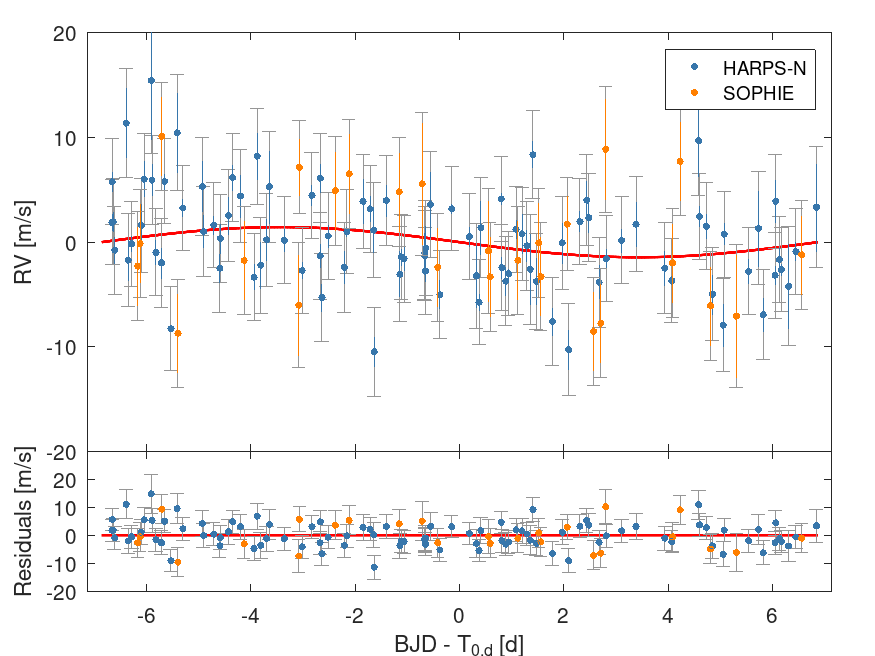}
    \includegraphics[width=0.495\textwidth]{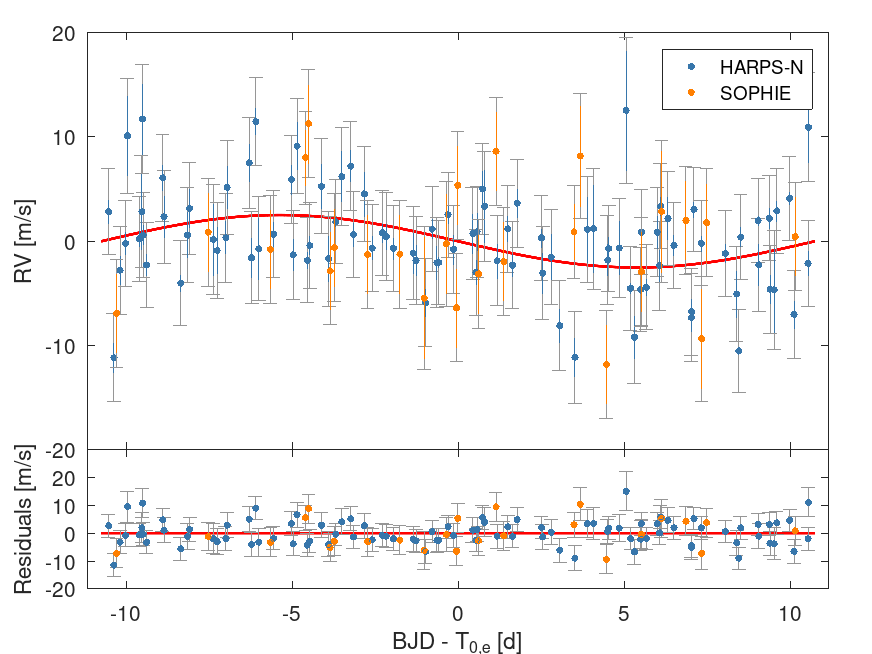} \\
    \sidecaption
    \includegraphics[width=0.495\textwidth]{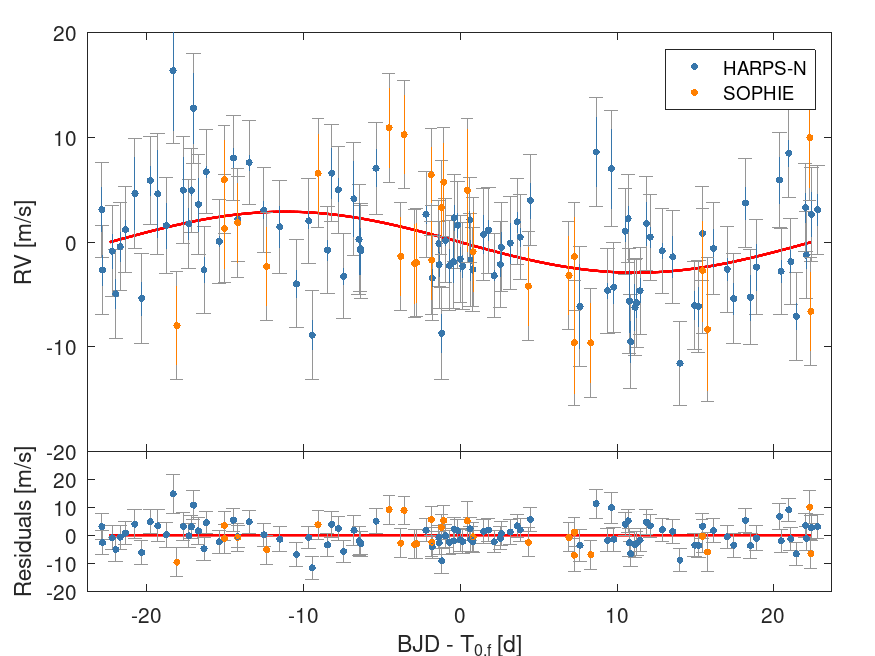}
    \caption{Detrended RV time series folded according to the period of TOI-5624\,b (top-left), TOI-5624\,c (top-right), TOI-5624\,d (middle-left), TOI-5624\,e (middle-right), and TOI-5624\,f (bottom-left) after removing the Keplerian signals of the other planets as inferred from the \texttt{MCMCI} analysis. The marker colour distinguishes the two different instruments (light blue for HARPS-N and orange for SOPHIE), while the red line represents the inferred model. The jitter contribution in the error bars is highlighted in grey. The RV semi-amplitudes are measured at the 6.9, 2.5, 2.2, 3.1, and 3.5$\sigma$ level for planets b, c, d, e, and f, respectively.}
    \label{fig:RVphaseFolded}
\end{figure*}

We performed four independent \texttt{MCMCI} runs made of 100\,000 steps each (burn-in 20\%) that consistently converged according to the GR statistic ($\hat{R}<1.027$). As listed in Table~\ref{tab:planetParameters}, we obtained robust mass estimates for TOI-5624\,b, e, and f (detection >\,3\,$\sigma$), while measuring the mass of TOI-5624\,c and d at the 2.5 and 2.2$\sigma$ level, respectively. We obtained $M_b=9.4\pm1.4\,M_{\oplus}$, $M_c=4.8\pm1.9\,M_{\oplus}$, $M_d=4.9\pm2.2\,M_{\oplus}$, $M_e=8.9_{-3.0}^{+2.9}\,M_{\oplus}$ for planets b to e, and a planet minimum mass $M_f\sin{i_f}=13.0\pm3.7\,M_{\oplus}$ for TOI-5624\,f. Each panel in Fig.~\ref{fig:RVphaseFolded} shows the detrended RV time series phase-folded to the period of that planet, after removing the Keplerian signals of the other planets detected in the system.

We further performed a similar \texttt{MCMCI} analysis, but allowing the planetary eccentricities to vary. For each planet, we assumed ($e\cos{\omega}$; $e\sin{\omega}$) as jump parameters, where $e$ is the orbital eccentricity and $\omega$ the argument of periastron \citep{anderson2011}.
It turned out that all the $e$ values do not significantly differ from zero, being less than 1.5$\sigma$ away from zero. In addition, when comparing the circular analysis ($e=0$) with the eccentric one ($e\neq0$), we obtained $\Delta\mathrm{BIC}=\mathrm{BIC}_{e\neq0}-\mathrm{BIC}_{e=0}=+37$, which states that assuming circular orbits is a reasonable assumption given the RV data in our hands. Moreover, the TTV dynamical analysis performed in Sect.~\ref{sec:TRADES} confirms low orbital eccentricities with $e$ values that again do not significantly differ from zero. The eccentricity value exhibiting the maximum $\sigma$-deviation from zero ($\sim$\,2.7$\sigma$) is $e_e=0.0165_{-0.0062}^{+0.0110}$ (Sect.~\ref{sec:TRADES}; Tab.~\ref{Table:TRADES_results}). \citet{kipping2008} shows that in low-eccentricity systems ($e$ lower than a few hundredths), the transit light curve remains effectively indistinguishable from a circular orbit ($e=0$) model, which justifies the assumption for $e=0$ both in the RV analysis and in the photometric analysis of Sect.~\ref{sec:LConly_4pla}.

By propagating the coarse ephemerides of TOI-5624\,f obtained from the RVs, there is a full sequence of epochs not covered by the available photometric data in which the planet could have transited its host star. While the RV data alone cannot rule out the hypothesis that TOI-5624\,f is a transiting planet, the dynamical analysis (Sect.~\ref{sec:TRADES}) finds TOI-5624\,f as a non-transiting planet. Future photometric surveys may help clarify this, though TESS is not currently foreseen to re-observe this system\footnote{\url{https://heasarc.gsfc.nasa.gov/wsgi-scripts/TESS/TESS-point_Web_Tool/TESS-point_Web_Tool/wtv_v2.0.py/TICID_result/ticid=53498154}}. 

Finally, we note that all the transiting planets are consistent with being coplanar given their orbital inclinations $i$ (see Table~\ref{tab:planetParameters}). By combining $a/R_{\star}$ of TOI-5624\,f and that the planet does not transit, we inferred an orbital inclination $i_f<89^{\circ}$, which differs at least by $\sim$\,$0.5^{\circ}$ from the average orbital inclination of the transiting planets in the system. On the one hand, this difference may be small if compared with the typical mutual inclination dispersion among Kepler \citep{borucki2010} multi-planet systems, which amounts to a few degrees \citep[see e.g.][and references therein]{millholland2021}. On the other hand, this difference is large with respect to the standard deviation of $0.12^{\circ}$ as computed from the orbital inclinations of the planets transiting TOI-5624, which might challenge the present-day dynamical models in the interpretation of the system architecture.

\begin{table*}
\caption{Planetary parameters derived from LC and RV analyses with the \texttt{MCMCI} code.}
\label{tab:planetParameters}
\centering
\begin{tabular}{llllll}
\hline
\hline
\noalign{\smallskip}
Parameter & TOI-5624\,b & TOI-5624\,c & TOI-5624\,d & TOI-5624\,e & TOI-5624\,f \\
\noalign{\smallskip}
\hline
\noalign{\smallskip}
$P$\,\tablefootmark{(a)} \;[d] & $3.3903473\pm0.0000054$ & $7.885385\pm0.000018$ & $13.731468_{-0.000041}^{+0.000042}$ & $21.489936\pm0.000029$ & $45.37_{-0.90}^{+0.74}$ \\
$T_0$\,\tablefootmark{(a)} \;[BJD] & $9649.1284\pm0.0012$ & $9650.9241\pm0.0012$ & $9655.2041\pm0.0012$ & $9655.6215\pm0.0012$ & $10338.3_{-3.0}^{+2.6}$ \\
d$F$\,\tablefootmark{(b)} \;[ppm] & $667_{-18}^{+19}$ & $762_{-24}^{+25}$ & $1599_{-38}^{+40}$ & $1312\pm29$ & \\
$b$ \; & $0.114_{-0.079}^{+0.097}$ & $0.143_{-0.093}^{+0.089}$ & $0.153_{-0.097}^{+0.088}$ & $0.391_{-0.029}^{+0.027}$ & \\
$K$ \;[m\,s$^{-1}$] & $4.46\pm0.65$ & $1.70_{-0.68}^{+0.69}$ & $1.45\pm0.65$ & $2.27_{-0.75}^{+0.74}$ & $2.59\pm0.73$ \\
\noalign{\smallskip}
\hline
\noalign{\smallskip}
$T_{14}$ \;[h] & $2.392_{-0.031}^{+0.029}$ & $3.165_{-0.041}^{+0.038}$ & $3.850_{-0.046}^{+0.041}$ & $4.180_{-0.029}^{+0.031}$ & \\
$a$ \;[AU] & $0.04201_{-0.00043}^{+0.00041}$ & $0.07374_{-0.00075}^{+0.00072}$ & $0.1067_{-0.0011}^{+0.0010}$ & $0.1439_{-0.0015}^{+0.0014}$ & $0.2366_{-0.0041}^{+0.0040}$\\
$i$ \;[$^{\circ}$] & $89.41_{-0.51}^{+0.41}$ & $89.58_{-0.27}^{+0.28}$ & $89.69_{-0.18}^{+0.20}$ & $89.405_{-0.047}^{+0.049}$ & \\
$R_p/R_{\star}$ \; & $0.02582\pm0.00036$ & $0.02760\pm0.00044$ & $0.03999_{-0.00048}^{+0.00049}$ & $0.03622\pm0.00040$ & \\
$a/R_{\star}$ \; & $11.01_{-0.12}^{+0.11}$ & $19.32_{-0.21}^{+0.19}$ & $27.97_{-0.30}^{+0.27}$ & $37.70_{-0.41}^{+0.36}$ & $62.0\pm1.1$ \\
$T_{\mathrm{eq}}$\,\tablefootmark{(c)} \;[K] & $1136\pm15$ & $857\pm11$ & $712.6_{-9.4}^{+9.5}$ & $613.8_{-8.1}^{+8.2}$ & $478.5_{-7.0}^{+7.1}$ \\
$e$ \; & 0 (fixed) & 0 (fixed) & 0 (fixed) & 0 (fixed) & 0 (fixed) \\
$\omega$ \;[$^{\circ}$] & 90 (fixed) & 90 (fixed) & 90 (fixed) & 90 (fixed) & 90 (fixed) \\
$R_p$ \;[$R_{\oplus}$] & $2.314\pm0.035$ & $2.474\pm0.042$ & $3.584_{-0.050}^{+0.051}$ & $3.247_{-0.043}^{+0.042}$ \\
$M_p$ \;[$M_{\oplus}$] & $9.4\pm1.4$ & $4.8\pm1.9$ & $4.9\pm2.2$ & $8.9_{-3.0}^{+2.9}$ & $13.0\pm3.7$\,\tablefootmark{(d)} \\
$\rho_p$ \;[g\,cm$^{-3}$] & $4.00_{-0.59}^{+.60}$ & $1.72_{-0.69}^{+0.70}$ & $0.59_{-0.26}^{+0.27}$ & $1.46\pm0.48$ &  \\
\noalign{\smallskip}
\hline
\noalign{\smallskip}
$u_{1,TE}$ \; & $0.370\pm0.012$ & \\
$u_{2,TE}$ \; & $0.2384\pm0.0070$ & \\
$u_{1,CH}$ \; & $0.503\pm0.014$ & \\
$u_{2,CH}$ \; & $0.200\pm0.010$ & \\
$\sigma_{\text{jit,H-N}}$ \;[m\,s$^{-1}$] & $3.98_{-0.10}^{+0.13}$ & \\
$\sigma_{\text{jit,SO}}$ \;[m\,s$^{-1}$] & $3.54\pm0.32$ & \\
\noalign{\smallskip}
\hline
\end{tabular}
\tablefoot{The planetary jump (i.e. fitted) parameters are listed in the top part of the table. All jump parameters but the LD coefficients ($u_{1,\mathrm{TE}}=\mathcal{N}(0.371,0.011)$, $u_{2,\mathrm{TE}}=\mathcal{N}(0.2387,0.0066)$, $u_{1,\mathrm{CH}}=\mathcal{N}(0.503,0.013)$, $u_{2,\mathrm{CH}}=\mathcal{N}(0.1991,0.0096)$) were subject to uniform unbounded priors (except for the physical limits) following the parameterisation detailed in \citet{bonfanti2020}. The HARPS-N and SOPHIE RV jitter terms are referred to as $\sigma_{\text{jit,H-N}}$ and $\sigma_{\text{jit,SO}}$, respectively. \\
\tablefoottext{a}{The orbital period $P$ and the time of inferior conjunction $T_0$ for planets b to e are computed following a linear ephemerides model. The period of planet f inferred from the TTV+RV dynamical analysis (Sect.~\ref{sec:TRADES}) is $P_f=44.092_{-0.057}^{+0.015}$. All the $T_0$s are shifted by $-2\,450\,000$.} \\
\tablefoottext{b}{d$F\equiv \left(\frac{R_p}{R_{\star}}\right)^2$} \;
\tablefoottext{c}{$T_{\mathrm{eq}}$ is computed assuming albedo equal to zero and full heat recirculation.} \;
\tablefoottext{d}{It is the planetary minimum mass $M_p\sin{i}$.}
}
\end{table*}

\subsection{RV analysis with \texttt{pyaneti}}
\label{sec:pyaneti_analysis}

As a sanity check, we jointly modelled the HARPS-N \texttt{TERRA} RVs and activity indicators following a multi-dimensional GP approach \citep{Rajpaul2015} as implemented in the MCMC-based code \texttt{pyaneti} \citep{Barragan2019,Barragan2022}. This approach assumes that the same function $G(t)$ and its time derivative $\dot{G}(t)$ can describe the activity-induced stellar signal seen both in the Doppler measurements and in the activity indicators. For this purpose, alongside the RVs, we modelled the FWHM of the \texttt{DRS} cross-correlation functions, as its periodogram shows a strong signal at the stellar rotation period ($\Prot$\,$\approx$\,20\,d). 
Following \citet{Rajpaul2015}, we used a 2D GP defined as
\begin{subequations}
\label{eq:GP}
\begin{align}
\mathrm{RV}(t) &= A_{\mathrm{RV}} + B_{\mathrm{RV}}\,G(t) + C_{\mathrm{RV}}\,\dot{G}(t), \label{eq:RV_t} \\
\mathrm{FWHM}(t) &= A_{\mathrm{FWHM}} + B_{\mathrm{FWHM}}\,G(t), \label{eq:FWHM_t}
\end{align}
\end{subequations}
where $A_{\mathrm{RV}}$ and $A_{\mathrm{FWHM}}$ are offset terms, while $B_{\mathrm{RV}}$, $C_{\mathrm{RV}}$, and $B_{\mathrm{FWHM}}$ are amplitudes that relate the function $G(t)$ and $\dot{G}(t)$ to the RV and FWHM time series. For the GP, we chose the following quasi-periodic covariance function,
\begin{equation}
\kappa(t_{i},t_{j}) = \exp\left[- \frac{\sin{^2}\left[\pi(t_{i}-t_{j})/P_\mathrm{GP}\right]}{2\lambda_\mathrm{p}^{2}} - \frac{(t_{i}-t_{j})^{2}}{2\lambda_\mathrm{e}^{2}}\right]\,,
\label{eq:Kernel}
\end{equation}
where $P_\mathrm{GP}$ is the characteristic period of the GP (interpreted as $\Prot$), $\lambda_\mathrm{p}$ is the inverse of the harmonic complexity (accounting for the distribution of active regions on the stellar surface), and $\lambda_\mathrm{e}$ is the long-term evolution timescale of spots and plages.

To account for the Doppler reflex motions induced by the five planets orbiting TOI-5624, we added five Keplerians to the RV model in Eq.~(\ref{eq:RV_t}). We assumed the orbits to be circular and adopted Gaussian priors on periods ($P$) and times of reference transit ($T_0$) for the four transiting planets (TOI-5624~b, c, d, and e), as derived from the modelling presented in Sect.~\ref{sec:LConly_4pla}. RV and FWHM jitter terms were also modelled to account for any instrumental noise that was not already captured in the nominal uncertainties. We used uniform priors for the orbital period and time of inferior conjunction of the non-transiting outer planet (TOI-5624~f), as well as for the RV semi-amplitudes of all five planets (Table~\ref{tab:PyanetiResults}). For the GP parameters, we adopted a uniform prior on $P_\mathrm{GP}$ centred on the stellar rotation period found by the frequency analysis of the activity indicators, and wide uniform priors for the remaining hyperparameters and amplitudes, as listed in Table~\ref{tab:PyanetiResults}. We sampled the parameter space with 500 chains and used the last 500 iterations of the converged chains, applying a thin factor of 10 to obtain a final sample of 250,000 independent points for each fitted parameter. 

Table~\ref{tab:PyanetiResults} reports the inferred values and uncertainties for both the model and the derived planetary parameters. They are defined as the median and 68.3\% region of the credible interval of the marginalised posterior distributions for each inferred parameter. The time series of the HARPS-N \texttt{TERRA} Doppler measurements and FWHM of the \texttt{DRS} cross-correlation functions are shown in Fig.~\ref{fig:RV_curves_pyaneti}, along with the inferred models, and the phase-folded Doppler reflex motions induced by each planet. We found that multi-dimensional GP-derived RV variations are in excellent agreement (within 1$\sigma$) with those presented in Sect.~\ref{sec:MCMCI_analysis}, corroborating our results.

\subsection{TTV+RV dynamical analysis}\label{sec:TRADES}
As shown in Figure~\ref{fig:TTVs}, TOI-5624\,e exhibits significant TTVs, indicative of strong gravitational interactions within the system. These variations could be driven by both the neighbouring planet~d, which lies close to a 3:2 commensurability with planet~e, and by a possible outer non-transiting companion, planet~f, whose signal is detected in the radial velocities and appears near a 2:1 commensurability with planet~e.

To investigate the dynamical origin of the observed TTVs and to test the plausibility of the fifth planet, we performed a joint N-body analysis of the system using the TRAnsits and Dynamics of Exoplanetary Systems \citep[\texttt{TRADES},][]{borsato2014, borsato_2019, borsato_2021, borsato_2024}\footnote{\url{https://github.com/lucaborsato/trades}.} code. This allowed us to integrate the full N-body equations of motion and to simultaneously fit both transit times and the detrended radial velocities taken from Tables~\ref{tab:RVdataHARPS-N}, \ref{tab:RVdataSOPHIE}. The analysis was carried out under two different hypotheses: (1) a four-planet system including only the transiting planets (from b to e) and (2) a five-planet configuration including an additional planet~f.

Given the extensive integration time required (five years) to reproduce the observational baseline and the short orbital periods of the two inner planets (b and c), we neglected their mutual gravitational interactions in the N-body model. This simplification is justified because the inner planets lie far from first-order resonances with the outer planet and, therefore, they are expected to have a negligible dynamical influence on its TTVs. The period ratio $P_d/P_c = 1.742$ places the pair $\sim$\,0.5\% interior to the 7:4 commensurability, while $P_d/P_b = 4.05$ lies $\sim$\,1.3\% exterior to the 4:1 commensurability. Both correspond to higher order resonances, which are expected to generate substantially weaker TTV signals compared to first-order resonances, as their strength scales with higher powers of the orbital eccentricity and therefore becomes negligible for near-circular orbits (e.g.\ \citealt{Murray_Dermott_1999}; \citealt{Agol_Deck_2016}). This interpretation is further supported by Fig.~\ref{fig:TTVs}, where the two inner planets do not show significant deviations from a linear ephemeris.

For each planet, we fit the orbital period $P$, the planet-to-star mass ratio, $M_\mathrm{p}/M_\star$, the eccentricity, $e$, the argument of periastron, $\omega$, and the mean longitude,~$\lambda$\footnote{Defined as $\lambda = \mathcal{M} + \omega + \Omega$, where $\mathcal{M}$ is the mean anomaly, $\omega$ the argument of periastron, and $\Omega$ the longitude of the ascending node.}.
To reduce correlations and improve sampling efficiency, we employed the parametrisation $\left(\sqrt{e}\cos\omega, \sqrt{e}\sin\omega\right)$ instead of fitting $e$ and $\omega$ directly \citep{anderson2011}.
The stellar parameters ($M_\star$, $R_\star$) and the planetary radii ($R_\mathrm{p}$), together with the orbital inclinations, $i$, were fixed to the values reported in Tables~\ref{tab:star} and~\ref{tab:planetParameters}.
The longitude of the ascending node was set to $\Omega = 180^\circ$ for planet d and e. For the outer planet~f, both the orbital inclination and the longitude of the ascending node were left as free parameters, with $\Omega_f$ sampled uniformly within $[170^{\circ},190^{\circ}]$ to explore possible small mutual inclinations with the inner system.
We adopted uniform priors on all fitted parameters (see Table~\ref{Table:TRADES_results}) and included an additional RV jitter term, $\sigma_{\mathrm{jitter}}$, parametrised as $\log_2 \sigma_{\mathrm{jitter}}$ to ensure efficient sampling.

The exploration of the parameter space was initially performed using the \pyde\footnote{\url{https://github.com/hpparvi/PyDE}} differential evolution optimiser \citep{Storn_97, Parviainen_2016}, with a population of 120 individuals evolved over 100\,000 generations.
The best-fitting solutions from this stage were adopted as starting points for the \emcee{} ensemble sampler \citep{Foremanmackey_2013}, which employed 120 walkers to explore the posterior distribution for 500\,000 steps.
Following the approach of \citet{Leonardi_2025}, the \emcee{} sampling combined two proposal schemes: a differential evolution proposal for 80\% of the walkers \citep{Nelson_2014} and a snooker update for the remaining 20\% \citep{terbraak_2008}, improving mixing efficiency in multi-modal posteriors.
Convergence of the chains was assessed using the Geweke \citep{Geweke_1991} and GR diagnostics, together with autocorrelation analysis \citep{Goodman_Weare_2010} and visual inspection, requiring the total chain length to exceed 50 integrated autocorrelation times.
The first 250\,000 steps were discarded as burn-in, and the chains were thinned by a factor of 100.

\begin{figure}
    \centering
    \includegraphics[width=0.9\columnwidth]{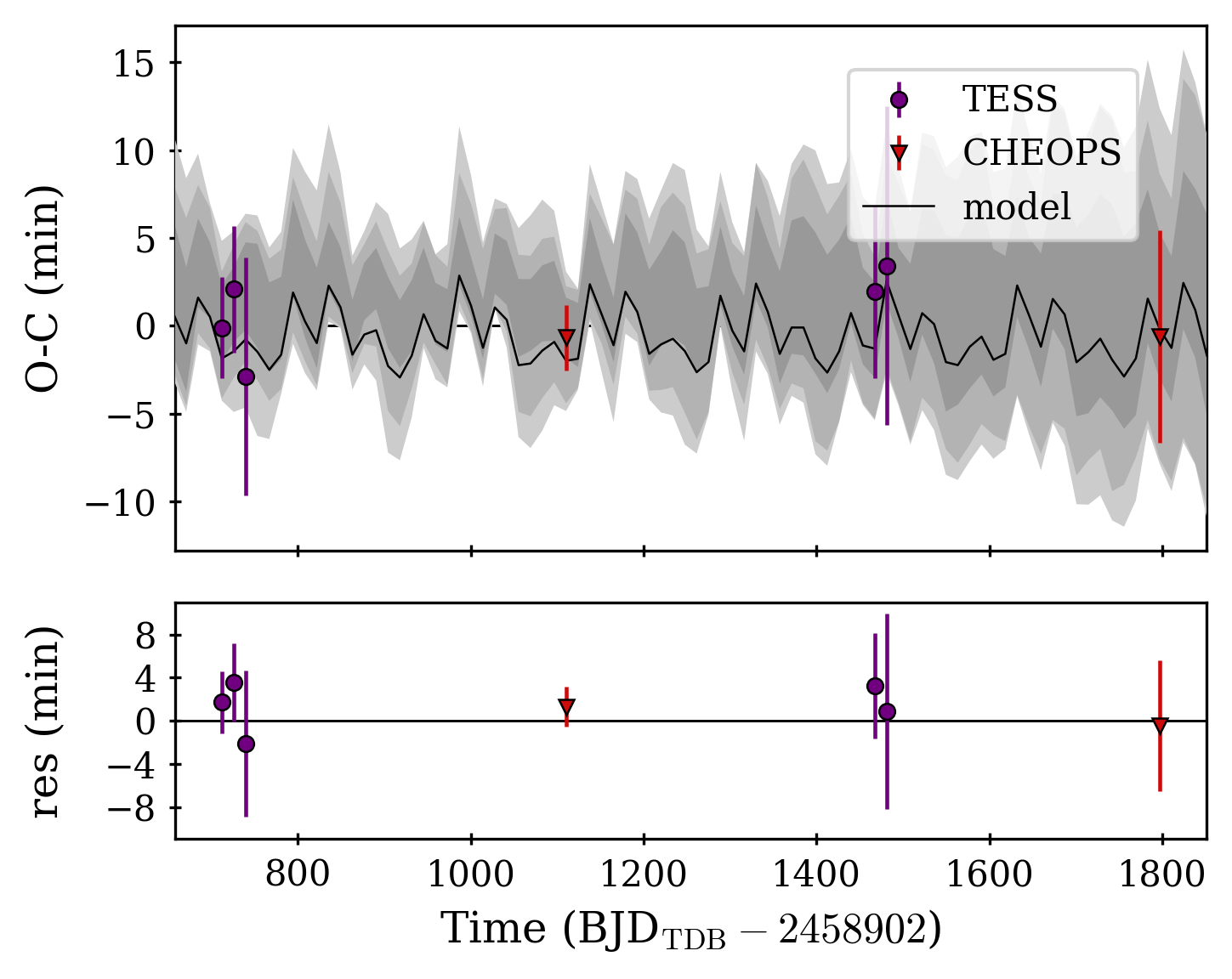} \\
    \includegraphics[width=0.9\columnwidth]{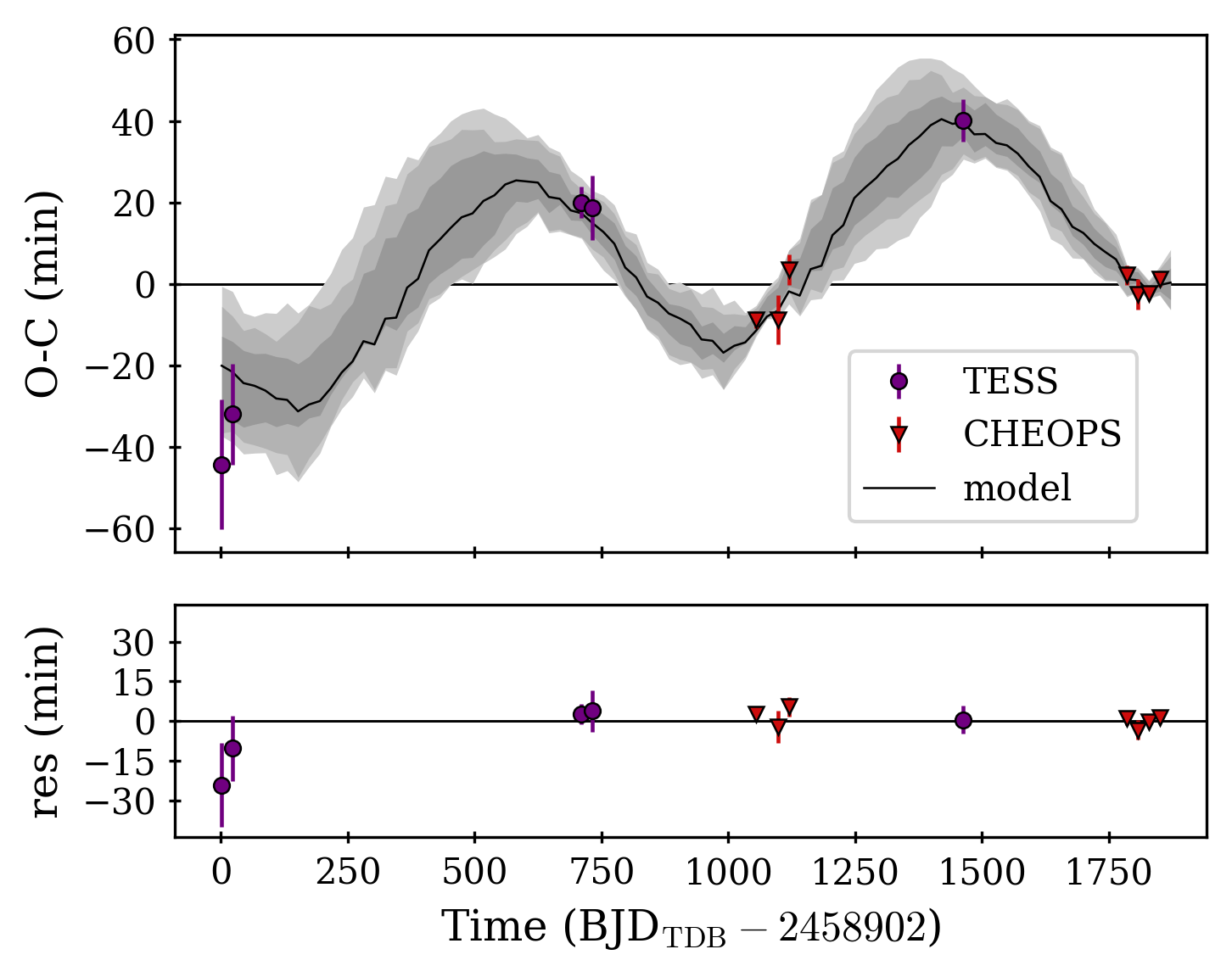}
    \caption{O$-$C diagrams of TOI-5624\,d  (top) \&\,e (bottom) along with the residuals (res) with respect to the best-fit model when assuming the five-planet configuration. The best-fit \trades{} model is shown as a black line, while the shaded areas display the $1\sigma$, $2\sigma$, and $3\sigma$ confidence intervals.}
    \label{fig:oc_2}
\end{figure}

The final parameter estimates along with the adopted priors are listed in Table~\ref{Table:TRADES_results}.
The observed-minus-calculated (O$-$C) diagrams for hypotheses (1) and (2) are shown in Fig.~\ref{fig:oc_1} and Fig.~\ref{fig:oc_2}, respectively. The fit to the transit timings of TOI-5624\,e is noticeably poorer under the four-planet configuration.
Indeed, model comparison between the four-planet (4p) and five-planet (5p) scenarios was performed via $\Delta\mathrm{BIC} = \mathrm{BIC}_{\text{5p}} - \mathrm{BIC}_{\text{4p}}$. The resulting value, $\Delta\mathrm{BIC} = -80$, strongly favours the inclusion of the fifth outer planet TOI-5624\,f. Likewise, adopting the approximate relation for the logarithm of the Bayes factor, $\log \mathcal{B} \simeq -\frac{1}{2}\Delta\mathrm{BIC}$ \citep{kass_raftery_95}, we obtained $\log \mathcal{B} = +40$, which further confirms an increased evidence for the five-planet configuration. While the five-planet model reproduces the overall timing pattern of TOI-5624\,d and e (see Fig.\ref{fig:oc_2}), the resulting posterior distributions do not yield significant improvements in the planetary mass estimates with respect to the RV-only analyses. In particular, the inclination of the outer planet~f remains largely unconstrained and tends to favour a non-coplanar, non-transiting configuration. The orbital period of TOI-5624\,f turned out to be $P_{f,\mathrm{dyn}}=44.092_{-0.057}^{+0.015}$\,d that differs from the strict $P_f$:$P_e$\,=\,2:1 commensurability by 2.6\%. It is consistent with the estimate from the RV analysis ($P_{f,\mathrm{RV}}=45.37_{-0.90}^{+0.74}$\,d) at the $\sim$\,1.4$\sigma$ level.

To further assess the robustness of the five-planet solution, we performed two additional dynamical analyses. First, we re-ran the TTV+RV analysis adopting a broader prior on the orbital period of the outer planet, allowing $P_f$ to vary within the interval $[1,50]$\,days. This range also includes configurations in which the additional planet would reside between TOI-5624\,d and TOI-5624\,e. Second, we repeated the dynamical fit using only the transit mid-times ($T_{\mathrm{tr}}$), excluding the RV measurements, but leaving the broader prior on the period of planet f. This test was designed to verify that the inferred signal of the fifth planet is not driven primarily by the RV dataset, where the Keplerian reflex motion associated with TOI-5624\,f is clearly detected. Both tests yield results consistent with the nominal five-planet solution (see Fig.~\ref{fig:oc_Pf_1_50}), recovering a planetary signal with orbital periods of $44.494_{-0.026}^{+0.055}$ and $42.308_{-0.71}^{+0.20}$ days, respectively. These sanity checks support the robustness of the inferred architecture and indicate that the detection of TOI-5624\,f is not sensitive to the adopted prior assumptions or to the inclusion of the RV dataset. In other words, only an outer perturber is capable of reproducing the observed TTVs of planet e.

Following \citet{correia2005,correia2010}, we also ran a stability analysis of the best-fit solution (Table~\ref{tab:planetParameters}). The system was integrated over $10^4$~yr for each initial condition and a stability indicator was calculated from the frequency analysis of the mean longitude \citep{laskar1990,laskar1993} of TOI-5624\,f. We found that the five-planet system is stable, the strict 2:1 MMR between TOI-5624\,e and f would be unstable, except for a small stability region around $P$\,$\sim$\,42.9 day and eccentricities lower than 0.01, with planet f that is not trapped in there in any case (Fig.~\ref{fig:stabilityTOI5624f}, left panel).
We also found that the inclination of TOI-5624\,f must be around $90^\circ \pm 60^\circ$ (Fig.~\ref{fig:stabilityTOI5624f}, right panel); thus, we can put an upper limit on its true mass $M_f^{\mathrm{up}}$\,$\sim$\,$26\,M_\oplus$.

Given the current temporal baseline and the limited sampling of the expected TTV super-period, the available data do not allow for a robust dynamical mass determination. Additional (e.g. CHEOPS) transit observations, ideally spanning a longer temporal window, are therefore required to better sample the TTV modulation, to reduce the degeneracies between mass and eccentricity, and to confirm the dynamical link between TOI-5624\,e and~f.
In the perspective of follow-up observations, we note that future transit timings of TOI-5624\,d predicted by the dynamical model are consistent with the linear ephemeris reported in Table~\ref{tab:planetParameters}. In case of TOI-5624\,e, instead, the uncertainties of the predicted transit timings would not aid with the scheduling of future transit events. Thus, we suggest to account for a $\sim$\,1.5\,h buffer (i.e. the peak-to-peak amplitude of the observed TTVs) when planning future observations of TOI-5624\,e.

\section{Conclusions}\label{sec:conclusions}
TOI-5624 hosts four transiting sub-Neptunes (orbital periods $P_b$\,$\approx$\,3.4, $P_c$\,$\approx$\,7.9, $P_d$\,$\approx$\,13.7, and $P_e$\,$\approx$\,21.5 days; $R_{\rm p}$: 2.3--3.6\,$R_{\oplus}$) that we characterised using TESS and CHEOPS photometry. The outermost planet (i.e. TOI-5624\,e) is affected by statistically significant TTVs ($\sim$80 min peak-to-peak). We thus searched for additional planets in the system that could explain the observed TTV pattern, but found none in the available photometry. We then conducted frequency analyses of the HARPS-N RV time series and found a signal hinting at the presence of an outer planet near a 2:1 period commensurability with TOI-5624\,e. Thus, we performed two fully independent RV analyses (both in terms of the extraction technique for retrieving the RV time series and of the MCMC statistical framework employed for the data analysis) in which we assumed a five-planet scenario. All jump parameters converged and the two independent analyses gave consistent results within 1$\sigma$, obtaining $M_b=9.4\pm1.4\,M_{\oplus}$ ($\rho_b=4.00_{-0.59}^{+0.60}$\,g\,cm$^{-3}$), $M_c=4.8\pm1.9\,M_{\oplus}$ ($\rho_c=1.72_{-0.69}^{+0.70}$\,g\,cm$^{-3}$), $M_d=4.9\pm2.2\,M_{\oplus}$ ($\rho_d=0.59_{-0.26}^{+0.27}$\,g\,cm$^{-3}$), $M_e=8.9_{-3.0}^{+2.9}\,M_{\oplus}$ ($\rho_e=1.46\pm0.48$\,g\,cm$^{-3}$) for planets b to e, and a planet minimum mass $M_f\sin{i_f}=13.0\pm3.7\,M_{\oplus}$ for TOI-5624\,f. Stability simulations suggest a mass upper limit $M_f^{\mathrm{up}}$\,$\sim$\,26\,$M_{\oplus}$ for this outer planet.

We finally investigated the dynamical origin of the TTV pattern exhibited by TOI-5624\,e by performing two N-body analyses of the system considering either a four or a five-planet configuration. We found that the five-planet scenario is much more favoured by the Bayesian evidence ($\Delta\mathrm{BIC}=-78$) and that the model is able to properly reproduce the transit times' pattern observed for TOI-5624\,e. This constitutes an independent confirmation of TOI-5624\,f. However, further photometric observations are required to properly map the phase of the TTV super period, to dynamically constrain the planet masses, and to definitely confirm the dynamical link between TOI-5624\,e and f.

Hosting at least five planets, TOI-5624 is a golden target in the context of multi-planet systems and we were able to characterise the radii of the four transiting planets at a precision level better than $\sim$1.7\%, while firmly detecting three planets in the RV time series (precision >\,3$\sigma$). The RV semi-amplitudes of the other two planets (i.e. TOI-5624\,c and d) were measured at the 2.5 and 2.2$\sigma$ level, respectively. As of today\footnote{\href{https://exoplanetarchive.ipac.caltech.edu/}{NASA Exoplanet Archive}, queried on 4 December 2025}, 76 multi-planet systems containing at least four transiting planets are known out of the $\sim$4500 validated exoplanet systems. Of these, only four systems, namely, TRAPPIST-1 \citep{agol2021}, HD\,110067 \citep{luque2023}, TOI-178 \citep{leleu2024}, and HIP\,41378 \citep{howard2025}, have been characterised well enough to measure the radii of at least four of their planets with a precision better than 2\%. Moreover, TOI-5624 is the only system for which at least three planet masses were firmly measured (>\,3$\sigma$) using RV data only.
Future photometric and RV data would enable the investigation of the presence of additional planets, making this system an even more compelling subject of study.

\begin{acknowledgements}
We thank the anonymous referee, as the comments provided have sensibly improved the quality of the manuscript.
CHEOPS is an ESA mission in partnership with Switzerland with important contributions to the payload and the ground segment from Austria, Belgium, France, Germany, Hungary, Italy, Portugal, Spain, Sweden, and the United Kingdom. The CHEOPS Consortium would like to gratefully acknowledge the support received by all the agencies, offices, universities, and industries involved. Their flexibility and willingness to explore new approaches were essential to the success of this mission. CHEOPS data analysed in this article will be made available in the CHEOPS mission archive (\url{https://cheops.unige.ch/archive_browser/}). 
DGa and LMSe acknowledge financial support from the CRT foundation under Grant No. 2018.2323 ‘Gaseous or rocky? Unveiling the nature of small worlds’. 
CBr and ASi acknowledge support from the Swiss Space Office through the ESA PRODEX program. 
S.G.S. acknowledge support from FCT through FCT contract nr. CEECIND/00826/2018 and POPH/FSE (EC). 
The Portuguese team thanks the Portuguese Space Agency for the provision of financial support in the framework of the PRODEX Programme of the European Space Agency (ESA) under contract number 4000142255. 
TWi acknowledges support from the UKSA and the University of Warwick. 
ACo, ADe, BEd, KGa, and JKo acknowledge their role as ESA-appointed CHEOPS Science Team Members. 
ABr was supported by the SNSA. 
M.G. is F.R.S.-FNRS Research Director. 
BAk and MLe acknowledge support of the Swiss National Science Foundation under grant number PCEFP2\_194576. 
YAl acknowledges support from the Swiss National Science Foundation (SNSF) under grant 200020\_192038. 
RAl, DBa, EPa, IRi, and EVi acknowledge financial support from the Agencia Estatal de Investigación of the Ministerio de Ciencia e Innovación MCIN/AEI/10.13039/501100011033 and the ERDF “A way of making Europe” through projects PID2021-125627OB-C31, PID2021-125627OB-C32, PID2021-127289NB-I00, PID2023-150468NB-I00 and PID2023-149439NB-C41. 
SCCB acknowledges the support from Fundação para a Ciência e Tecnologia (FCT) in the form of work contract through the Scientific Employment Incentive program with reference 2023.06687.CEECIND and DOI 10.54499/2023.06687.CEECIND/CP2839/CT0002. 
LBo, VNa, IPa, GPi, RRa, GSc, and TZi acknowledge support from CHEOPS ASI-INAF agreement n. 2019-29-HH.0. TZi also acknowledges support from NVIDIA Academic Hardware Grant Program for the use of the Titan V GPU card and the Italian MUR Departments of Excellence grant 2023-2027 “Quantum Frontiers”.
ACC acknowledges support from STFC consolidated grant number ST/V000861/1, and UKSA grant number ST/X002217/1. 
P.E.C. is funded by the Austrian Science Fund (FWF) Erwin Schroedinger Fellowship, program J4595-N. 
This project was supported by the CNES. 
ADe acknowledges financial support from the Swiss National Science Foundation (SNSF) for project 200021\_200726. 
This work was supported by FCT - Funda\c{c}\~{a}o para a Ci\^{e}ncia e a Tecnologia through national funds and by FEDER through COMPETE2020 through the research grants UIDB/04434/2020, UIDP/04434/2020, 2022.06962.PTDC. 
O.D.S.D. is supported in the form of work contract (DL 57/2016/CP1364/CT0004) funded by national funds through FCT. 
B.-O. D. acknowledges support from the Swiss State Secretariat for Education, Research and Innovation (SERI) under contract number MB22.00046. 
This project has received funding from the Swiss National Science Foundation for project 200021\_200726. It has also been carried out within the framework of the National Centre of Competence in Research PlanetS supported by the Swiss National Science Foundation under grants 51NF40\_182901 and 51NF40\_205606. The authors acknowledge the financial support of the SNSF. 
MF and CMP gratefully acknowledge the support of the Swedish National Space Agency (DNR 65/19, 174/18). 
MNG is the ESA CHEOPS Project Scientist and Mission Representative. BMM is the ESA CHEOPS Project Scientist. KGI was the ESA CHEOPS Project Scientist until the end of December 2022 and Mission Representative until the end of January 2023. All of them are/were responsible for the Guest Observers (GO) Programme. None of them relay/relayed proprietary information between the GO and Guaranteed Time Observation (GTO) Programmes, nor do/did they decide on the definition and target selection of the GTO Programme. 
CHe acknowledges financial support from the Österreichische Akademie der Wissenschaften and from the European Union H2020-MSCA-ITN-2019 under Grant Agreement no. 860470 (CHAMELEON). Calculations were performed using supercomputer resources provided by the Vienna Scientific Cluster (VSC). 
K.W.F.L. was supported by Deutsche Forschungsgemeinschaft grants RA714/14-1 within the DFG Schwerpunkt SPP 1992, Exploring the Diversity of Extrasolar Planets. 
This work was granted access to the HPC resources of MesoPSL financed by the Region Ile de France and the project Equip@Meso (reference ANR-10-EQPX-29-01) of the programme Investissements d'Avenir supervised by the Agence Nationale pour la Recherche. 
PM acknowledges support from STFC research grant number ST/R000638/1. 
This work was also partially supported by a grant from the Simons Foundation (PI Queloz, grant number 327127). 
NCSa acknowledges funding by the European Union (ERC, FIERCE, 101052347). Views and opinions expressed are however those of the author(s) only and do not necessarily reflect those of the European Union or the European Research Council. Neither the European Union nor the granting authority can be held responsible for them. 
GyMSz acknowledges the support of the Hungarian National Research, Development and Innovation Office (NKFIH) grant K-125015, a PRODEX Experiment Agreement No. 4000137122, the Lendület LP2018-7/2021 grant of the Hungarian Academy of Science and the support of the city of Szombathely. 
V.V.G. is an F.R.S-FNRS Research Associate. 
JV acknowledges support from the Swiss National Science Foundation (SNSF) under grant PZ00P2\_208945. 
NAW acknowledges UKSA grant ST/R004838/1.
This work uses observations secured with the SOPHIE spectrograph at the 1.93-m telescope of Observatoire Haute-Provence, France, with the support of its staff. This work was supported by the ``Programme National de Plan\'etologie'' (PNP) of CNRS/INSU, and CNES.
The stability maps were performed at the Oblivion Supercomputer at the University of \'Evora (\href{https://oblivion.hpc.uevora.pt}{https://oblivion.hpc.uevora.pt}).
\end{acknowledgements}

\bibliographystyle{aa}
\bibliography{biblio}

\begin{appendix}

\onecolumn
\section{RV time series}
\begin{table}[h!]
\caption{Radial velocities, $\overline{RV}$, as inferred from the SN fit onto the centred HARPS-N CCFs with their errors $\sigma_{\mathrm{RV}}$. They are followed by the ancillary parameters used for the detrending (i.e. $\mathrm{FWHM_{SN}}$, $H_{\alpha}$, $\gamma$, and NaD2) and by the detrended RV values ($RV_{\mathrm{det}}$) with their errors which also account for the jitter ($\sigma_{\mathrm{RV(det+jitter)}}$).}
\label{tab:RVdataHARPS-N}
\centering
\begin{tabular}{rrrrrrrrr}
\hline\hline
\noalign{\smallskip}
\multicolumn{1}{c}{$\mathrm{BJD_{TDB}}$} & \multicolumn{1}{c}{$\overline{RV}$} & \multicolumn{1}{c}{$\sigma_{\mathrm{RV}}$} & \multicolumn{1}{c}{$\mathrm{FWHM_{SN}}$} & \multicolumn{1}{c}{$H_{\alpha}$} & \multicolumn{1}{c}{$\gamma$} & \multicolumn{1}{c}{NaD2} & \multicolumn{1}{c}{$RV_{\mathrm{det}}$} & \multicolumn{1}{c}{$\sigma_{\mathrm{RV(det+jitter)}}$} \\
\multicolumn{1}{c}{[$\mathrm{JD-2\,450\,000}$]} & \multicolumn{1}{c}{[m\,s$^{-1}$]} & \multicolumn{1}{c}{[m\,s$^{-1}$]} & \multicolumn{1}{c}{[km\,s$^{-1}$]} & \multicolumn{1}{c}{\AA} & \multicolumn{1}{c}{} & \multicolumn{1}{c}{\AA} & \multicolumn{1}{c}{[m\,s$^{-1}$]} & \multicolumn{1}{c}{[m\,s$^{-1}$]} \\
\noalign{\smallskip}
\hline
\noalign{\smallskip}
  9978.709867 & $ 10.117$ & 2.121 & 7.121 &  0.848 & $-0.0225$ & $ 0.607$  & $ 0.682$ & 4.343 \\
  9983.642536 & $-10.244$ & 1.471 & 7.086 &  0.852 & $-0.0111$ & $ 0.615$  & $-4.457$ & 4.065 \\
  9994.608233 & $ 13.547$ & 2.578 & 7.109 &  0.854 & $-0.0317$ & $ 0.607$  & $13.190$ & 4.584 \\
  9996.638267 & $ 18.790$ & 2.467 & 7.130 &  0.846 & $-0.0214$ & $ 0.612$  & $ 7.150$ & 4.522 \\
\ldots & \ldots & \ldots & \ldots & \ldots & \ldots & \ldots & \ldots & \ldots \\
\noalign{\smallskip}
\hline
\end{tabular}
\tablefoot{The whole table is available in electronic form at the Strasbourg astronomical Data Center (CDS)}
\end{table}

\begin{table}[h!]
\caption{Radial velocities, $\overline{RV}$, as extracted from the centred SOPHIE CCFs with their errors $\sigma_{\mathrm{RV}}$. They are followed by the CCF-related parameters (i.e. FWHM, $A$, and BIS), the $\log{R\arcmin_{\mathrm{HK}}}$, and by the detrended RV values ($RV_{\mathrm{det}}$) with their errors which also account for the jitter ($\sigma_{\mathrm{RV(det+jitter)}}$).}
\label{tab:RVdataSOPHIE}
\centering
\begin{tabular}{rrrrrrrrr}
\hline\hline
\noalign{\smallskip}
\multicolumn{1}{c}{$\mathrm{BJD_{TDB}}$} & \multicolumn{1}{c}{$\overline{RV}$} & \multicolumn{1}{c}{$\sigma_{\mathrm{RV}}$} & \multicolumn{1}{c}{FWHM} & \multicolumn{1}{c}{$A$} & \multicolumn{1}{c}{BIS} & \multicolumn{1}{c}{$\log{R\arcmin_{\mathrm{HK}}}$} & \multicolumn{1}{c}{$RV_{\mathrm{det}}$} & \multicolumn{1}{c}{$\sigma_{\mathrm{RV(det+jitter)}}$} \\
\multicolumn{1}{c}{[$\mathrm{JD-2\,450\,000}$]} & \multicolumn{1}{c}{[m\,s$^{-1}$]} & \multicolumn{1}{c}{[m\,s$^{-1}$]} & \multicolumn{1}{c}{[km\,s$^{-1}$]} & \multicolumn{1}{c}{[\%]} & \multicolumn{1}{c}{} & \multicolumn{1}{c}{} & \multicolumn{1}{c}{[m\,s$^{-1}$]} & \multicolumn{1}{c}{[m\,s$^{-1}$]} \\
\noalign{\smallskip}
\hline
\noalign{\smallskip}
  9978.585740 & $ 13.000$ & 4.000 & 8.220 & 43.700 & $-0.0260$ & $-4.790$  & $-7.351$ & 5.186 \\
 10016.447890 & $ 15.000$ & 4.000 & 8.220 & 43.900 & $-0.0400$ & $-4.530$  & $ 5.947$ & 5.186 \\
 10042.392200 & $ -2.000$ & 4.000 & 8.230 & 43.600 & $-0.0420$ & $-4.540$  & $-6.133$ & 5.186 \\
 10110.402050 & $  1.000$ & 3.000 & 8.190 & 44.300 & $-0.0110$ & $-4.610$  & $-3.346$ & 4.460 \\
\ldots & \ldots & \ldots & \ldots & \ldots & \ldots & \ldots & \ldots & \ldots \\
\noalign{\smallskip}
\hline
\end{tabular}
\tablefoot{The whole table is available in electronic form at the Strasbourg astronomical Data Center (CDS)}
\end{table}

\section{Additional figures}
\begin{figure}[h] 
\centering
\includegraphics[width=0.33\textwidth]{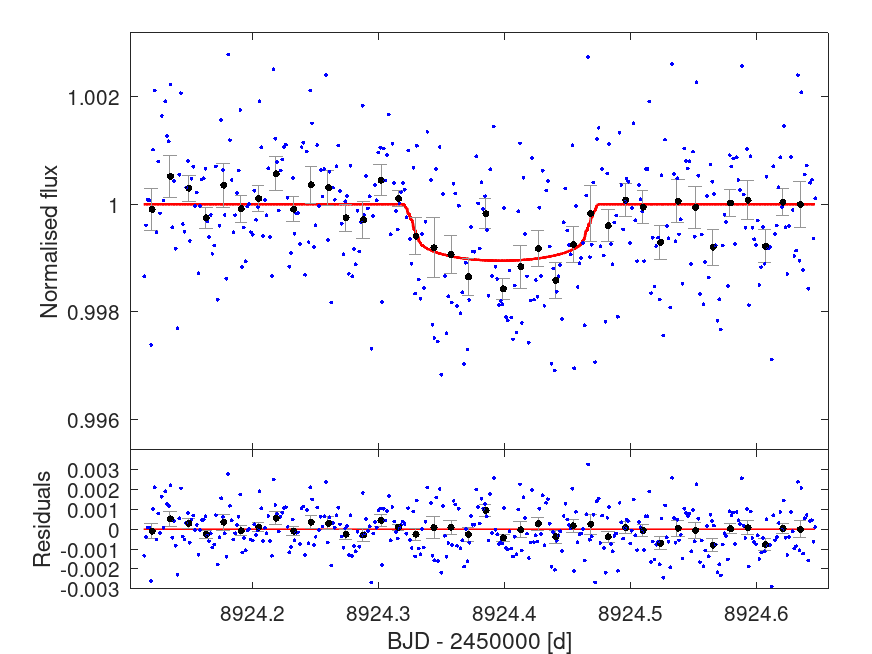}
\includegraphics[width=0.33\textwidth]{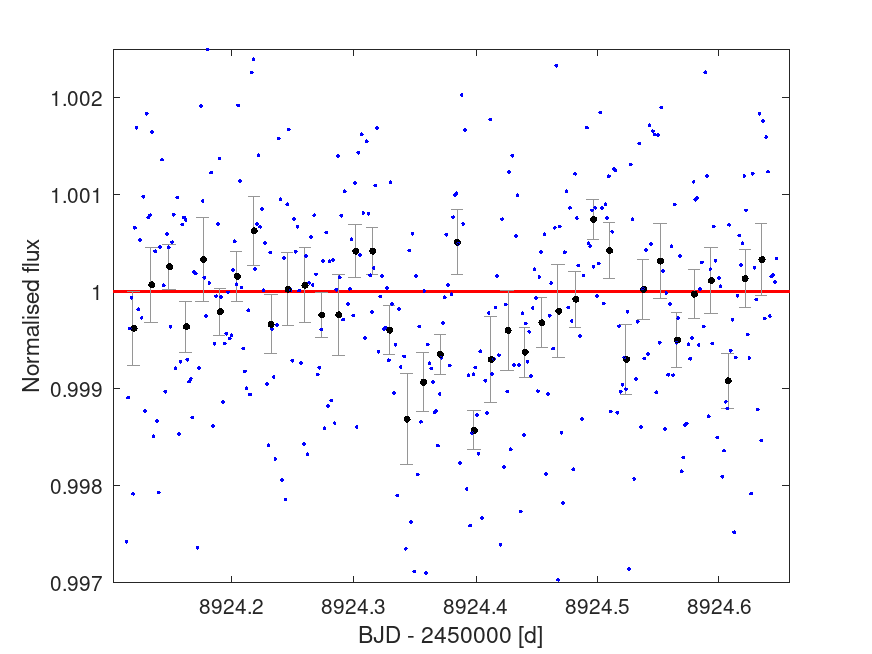}
\includegraphics[width=0.33\textwidth]{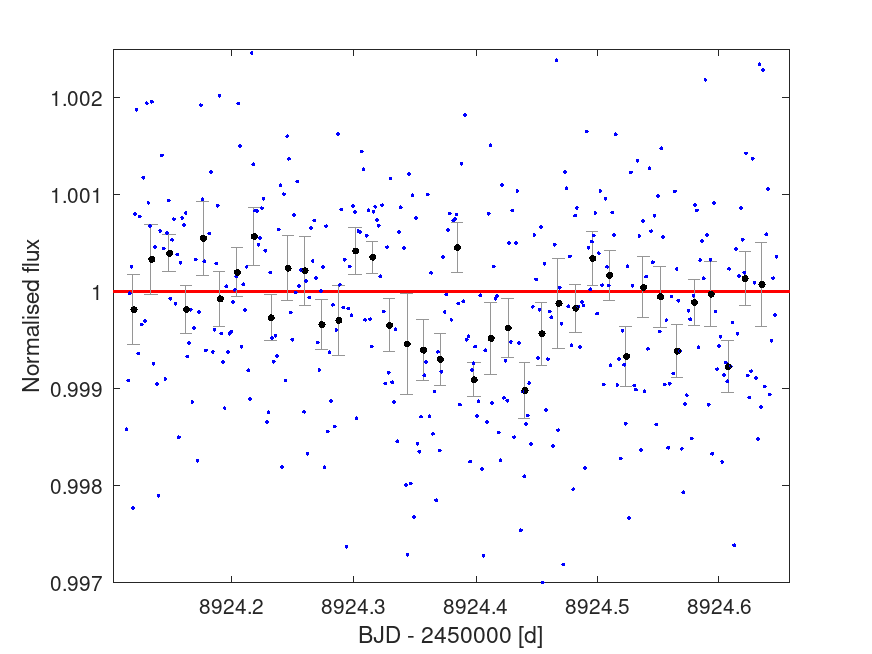}
\caption{Transit modelling of the wiggle in TESS Sector 22 based on the photometry as extracted from PDCSAP (left), SAP (mid), and \texttt{PATHOS} (right). The \texttt{MCMCI} analysis does not lead to any transit detection when attempting to fit a transit in the SAP and \texttt{PATHOS} photometry. The outcomes depend on the extraction pipeline, which plays against the genuineness of the putative transit event as confirmed by the BIC as well (see text for further details).}
\label{fig:transitPla5}
\end{figure}

\begin{figure}
    \centering
    \includegraphics[width=0.246\textwidth]{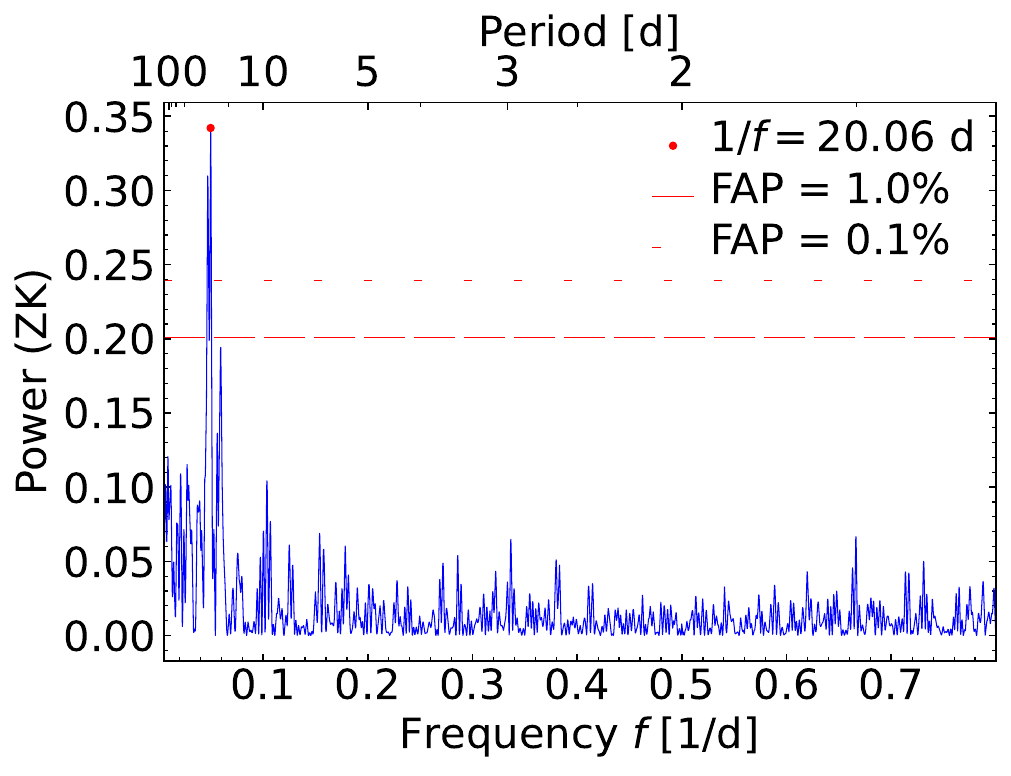}
    \includegraphics[width=0.246\textwidth]{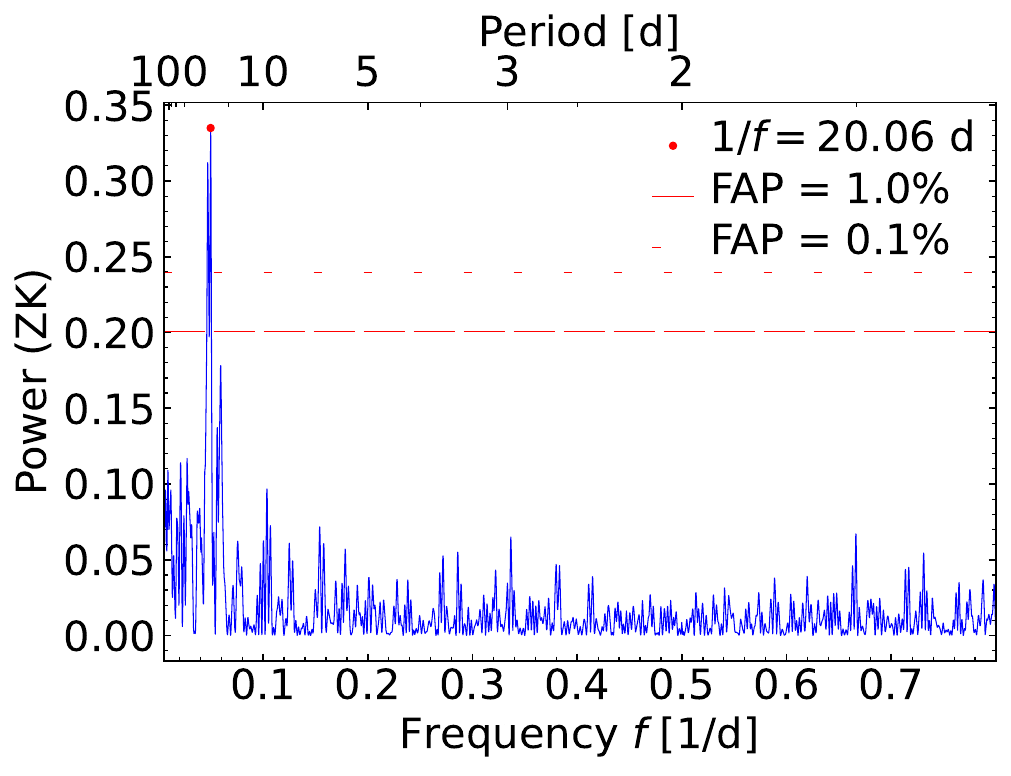}
    \includegraphics[width=0.246\textwidth]{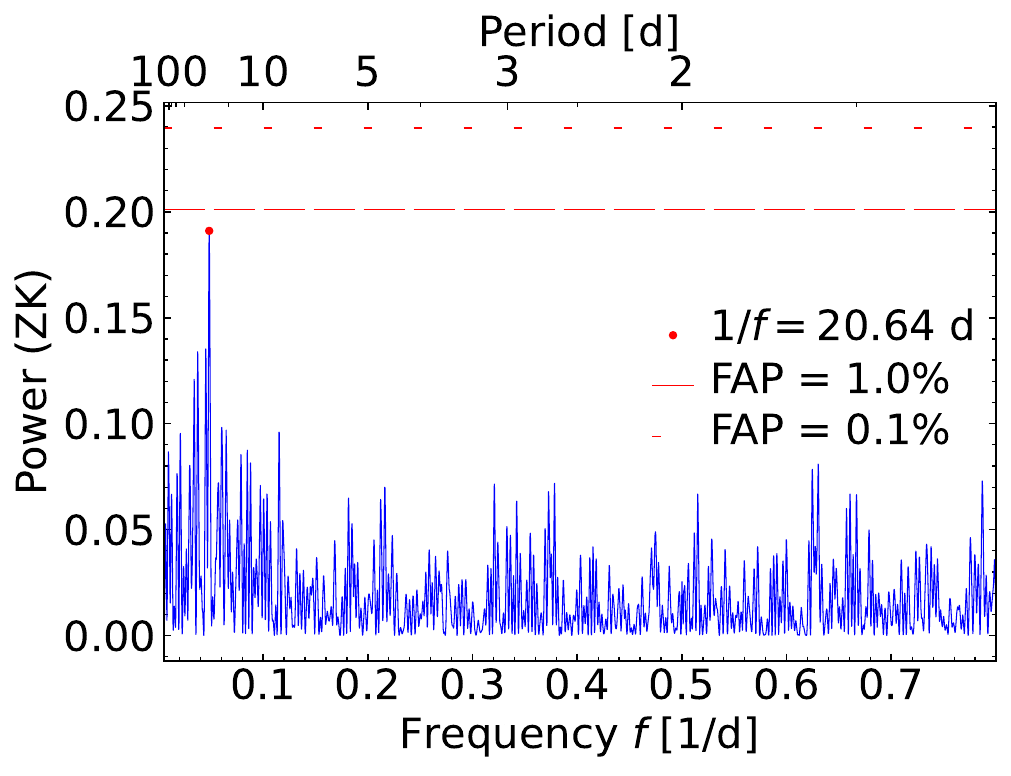}
    \includegraphics[width=0.246\textwidth]{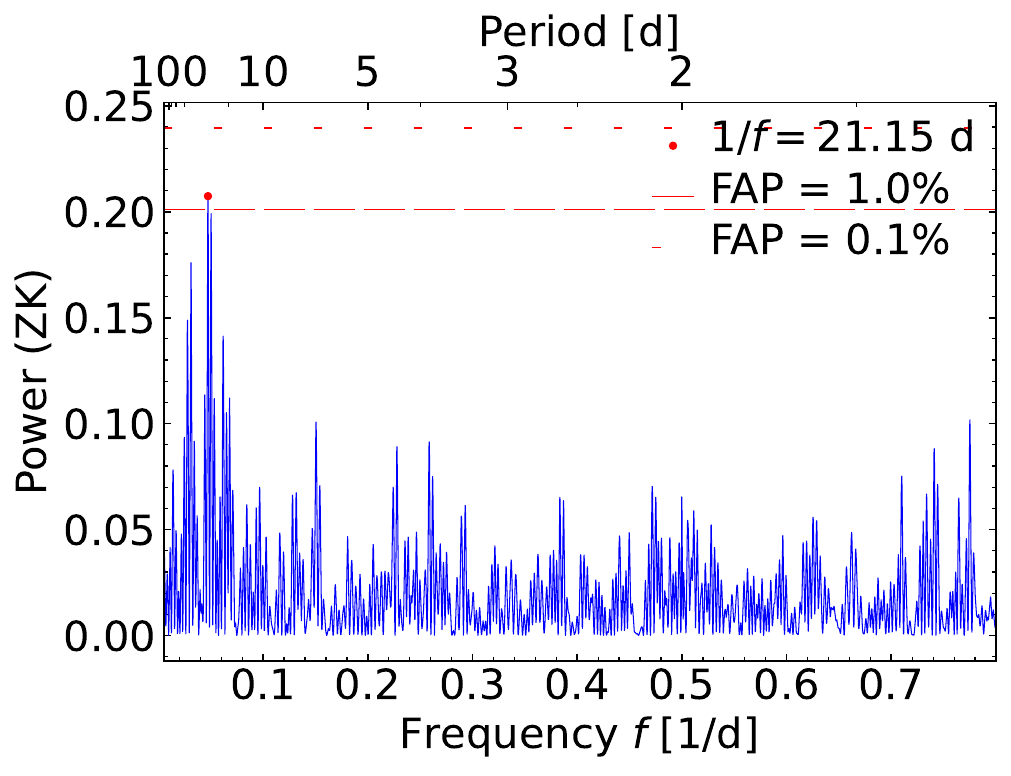} \\
    \includegraphics[width=0.246\textwidth]{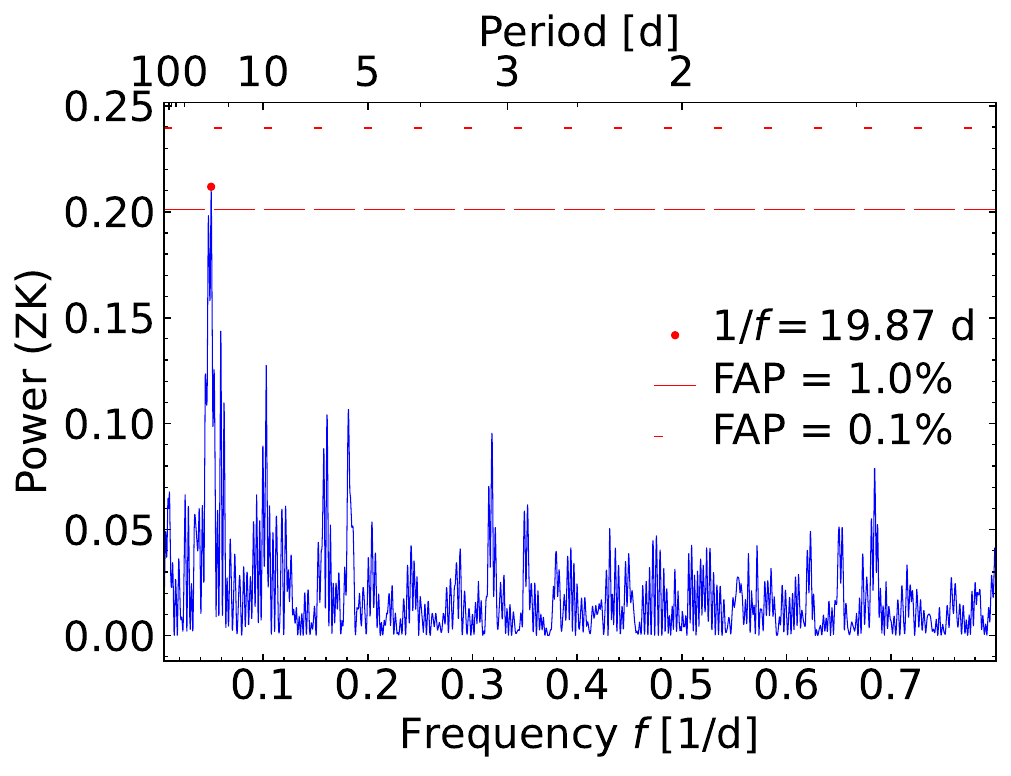}
    \includegraphics[width=0.246\textwidth]{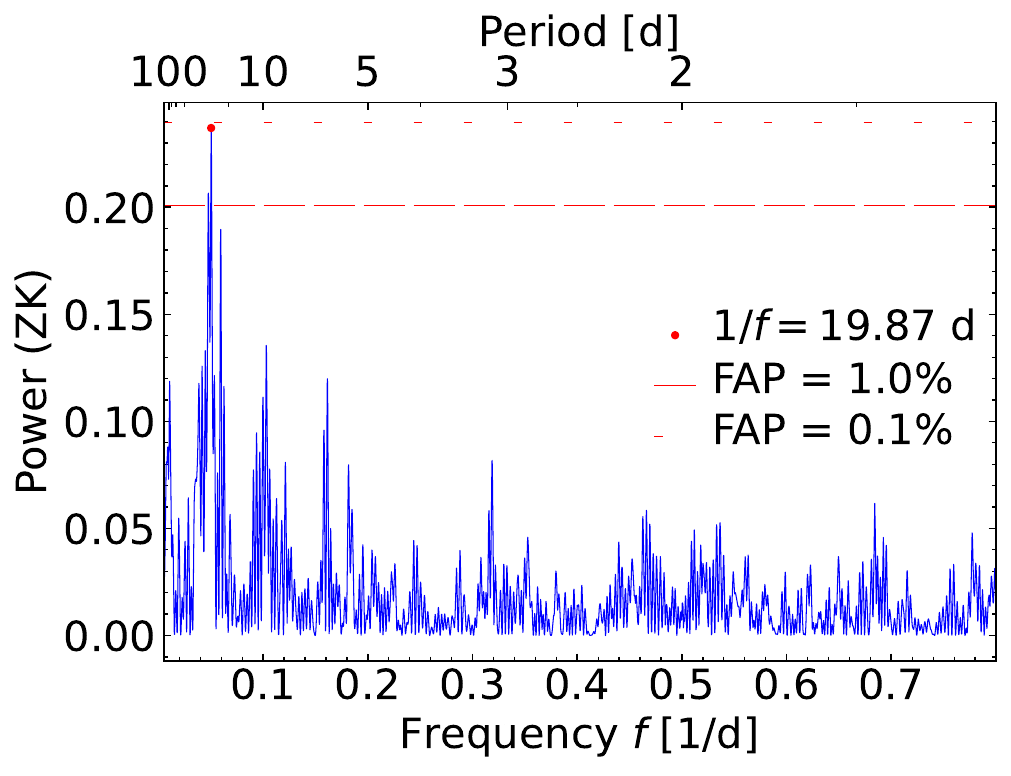}
    \includegraphics[width=0.246\textwidth]{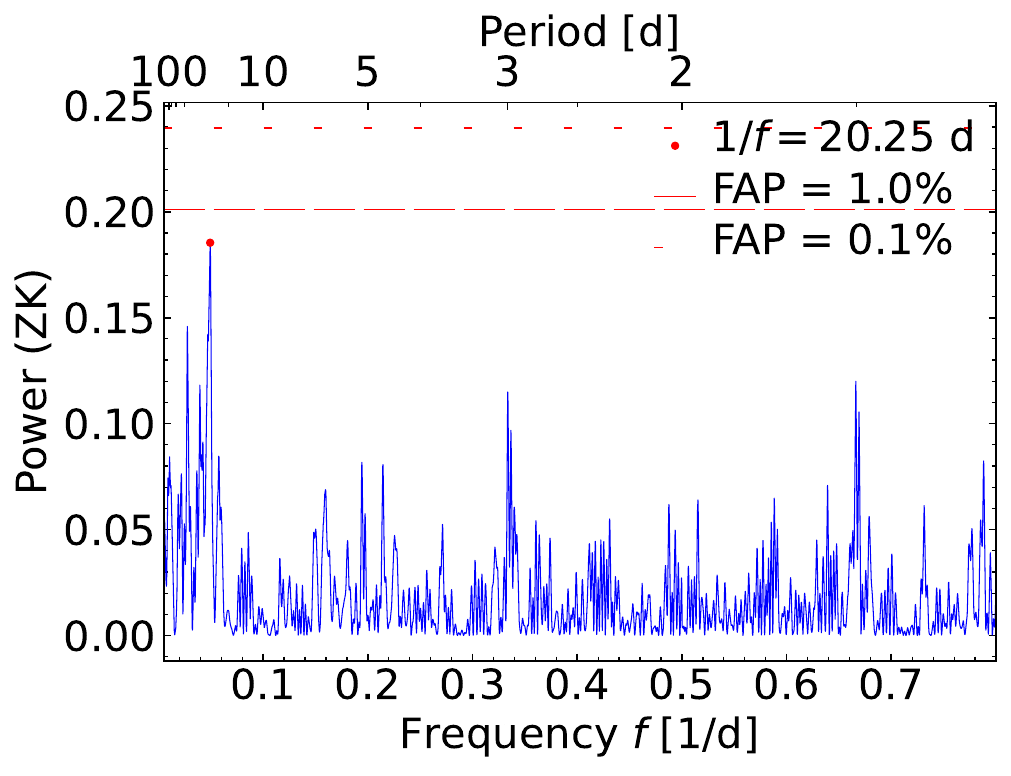}
    \includegraphics[width=0.246\textwidth]{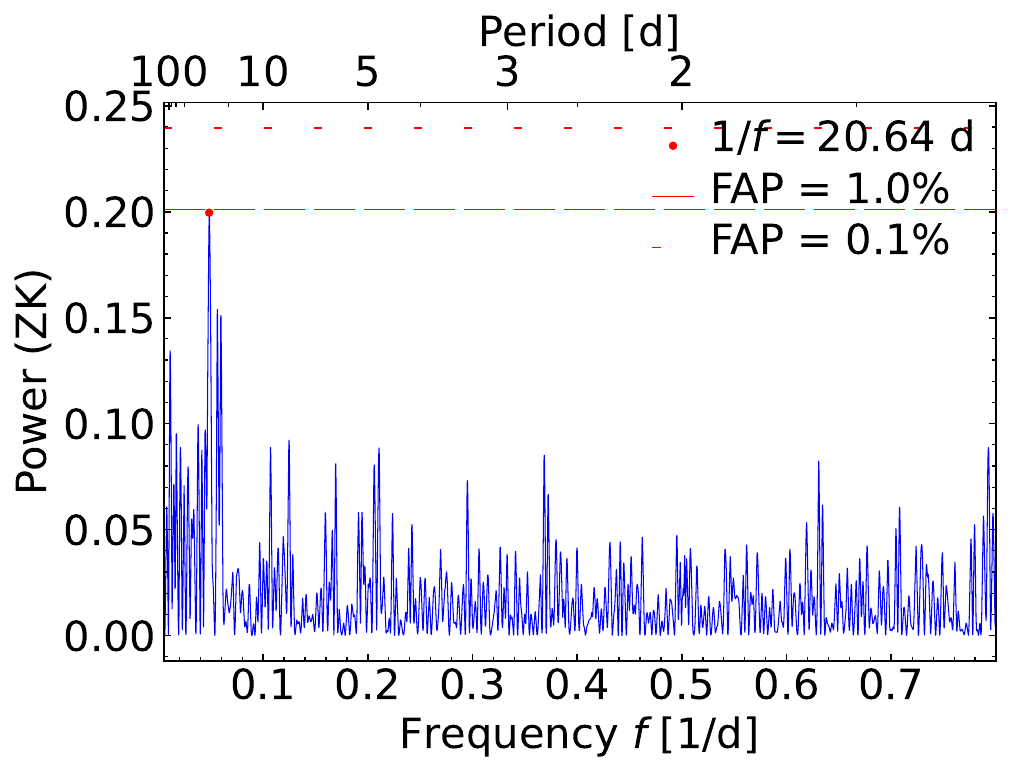}
    \caption{GLS periodograms of different activity indicators as inferred from the HARPS-N RV time series with the FAP levels at 1\% and 0.1\% marked as dashed and dotted red lines, respectively. First row: from left to right FWHM, FWHM$_\mathrm{SN}$, $A$, and $A_{\mathrm{SN}}$. Second row: from left to right BIS, $\gamma$, $\log{R'_{HK}}$, and $H_{\alpha}$, where the latter two were subject to a pre-whitening. The most prominent peak is always at $\sim$20 days, which is likely the stellar rotation period ($\Prot$).}
    \label{fig:RVactivity}
\end{figure}

\begin{figure}
    \includegraphics[width=0.246\textwidth]{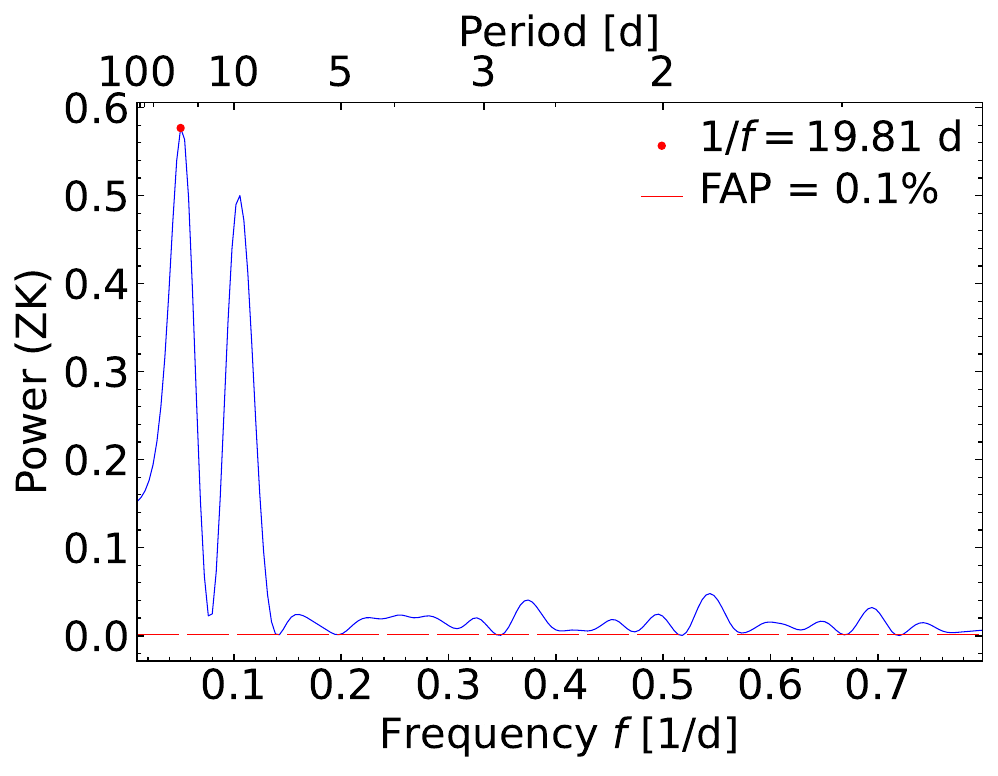}
    \includegraphics[width=0.246\textwidth]{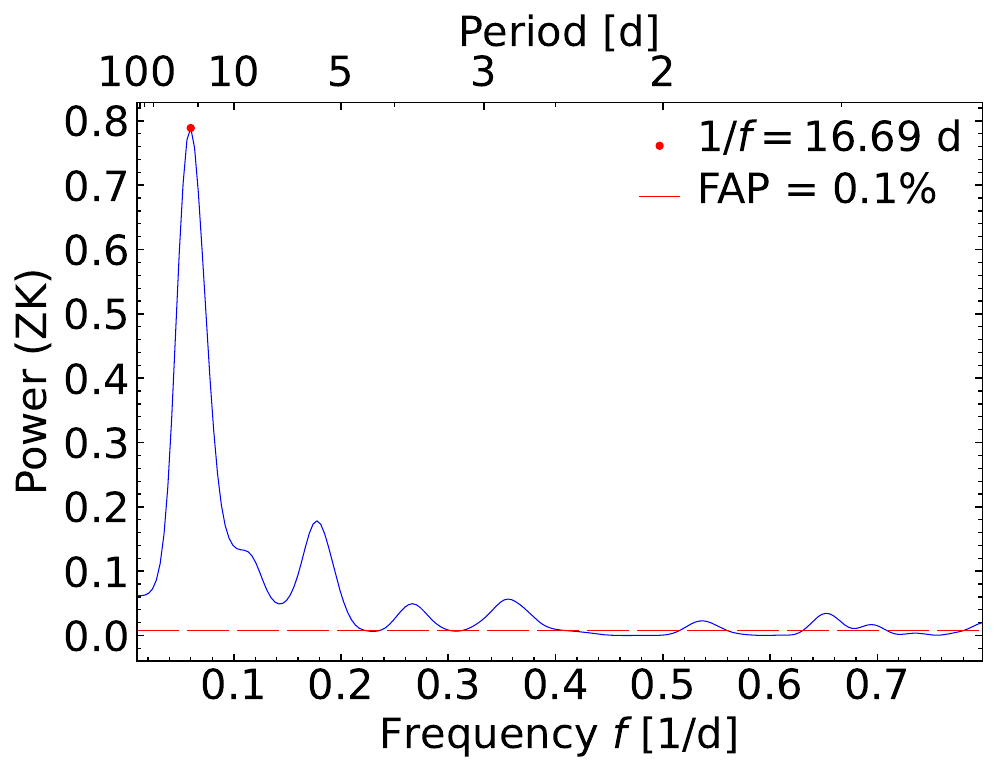}
    \includegraphics[width=0.246\textwidth]{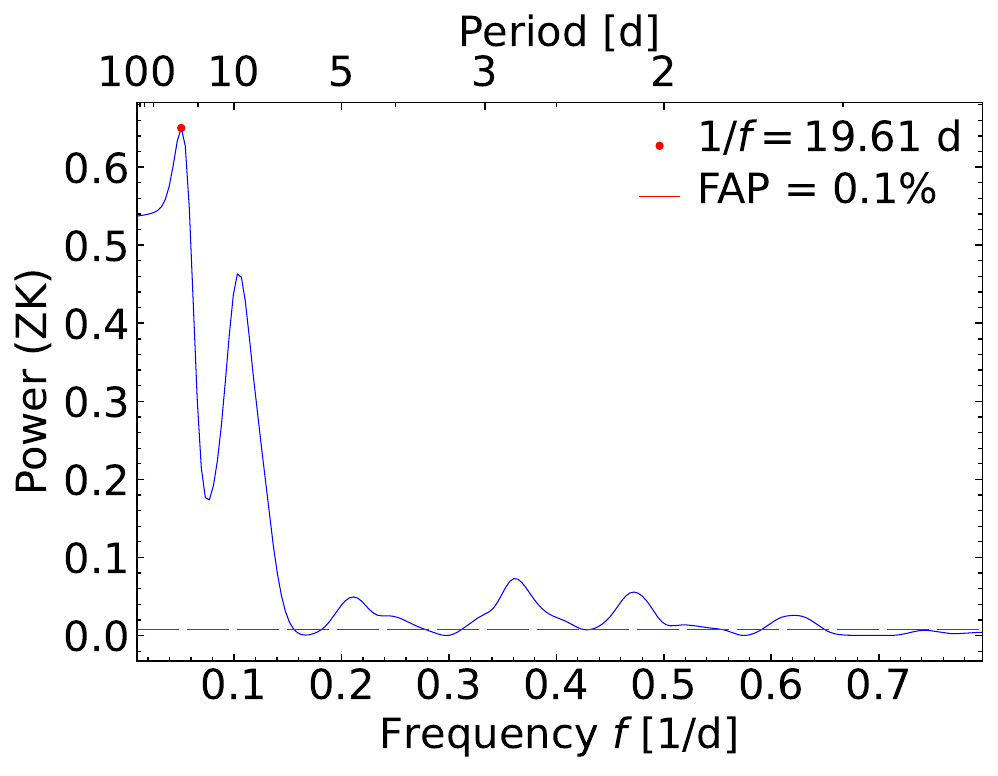}
    \includegraphics[width=0.246\textwidth]{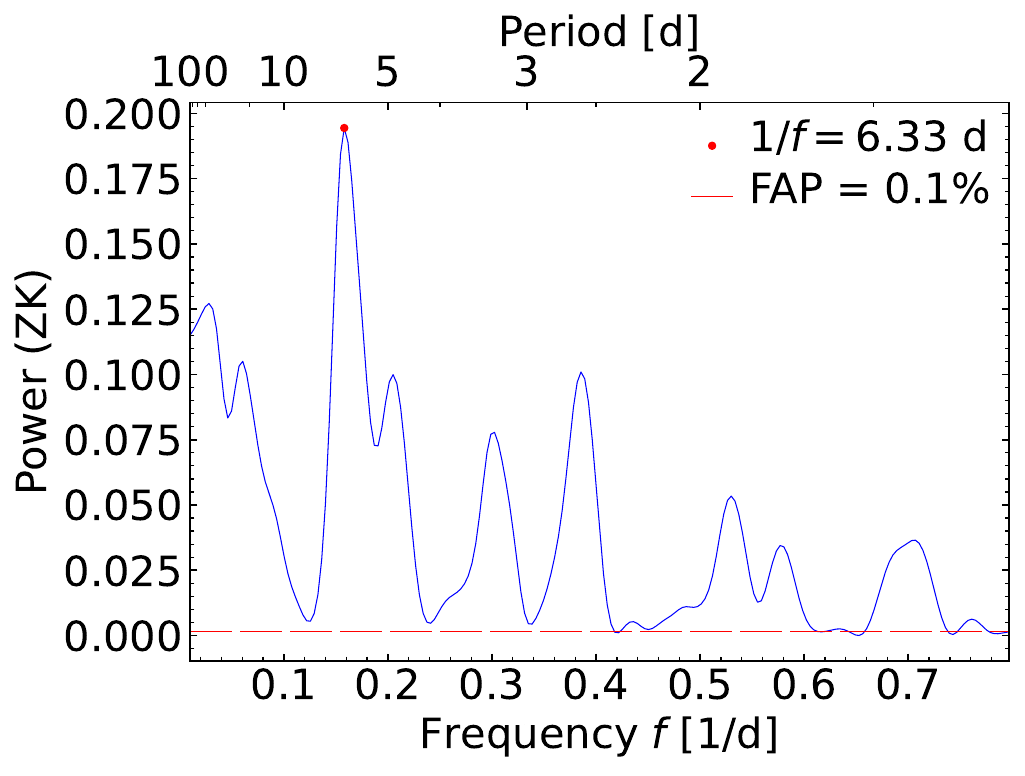} \\
    \includegraphics[width=0.246\textwidth]{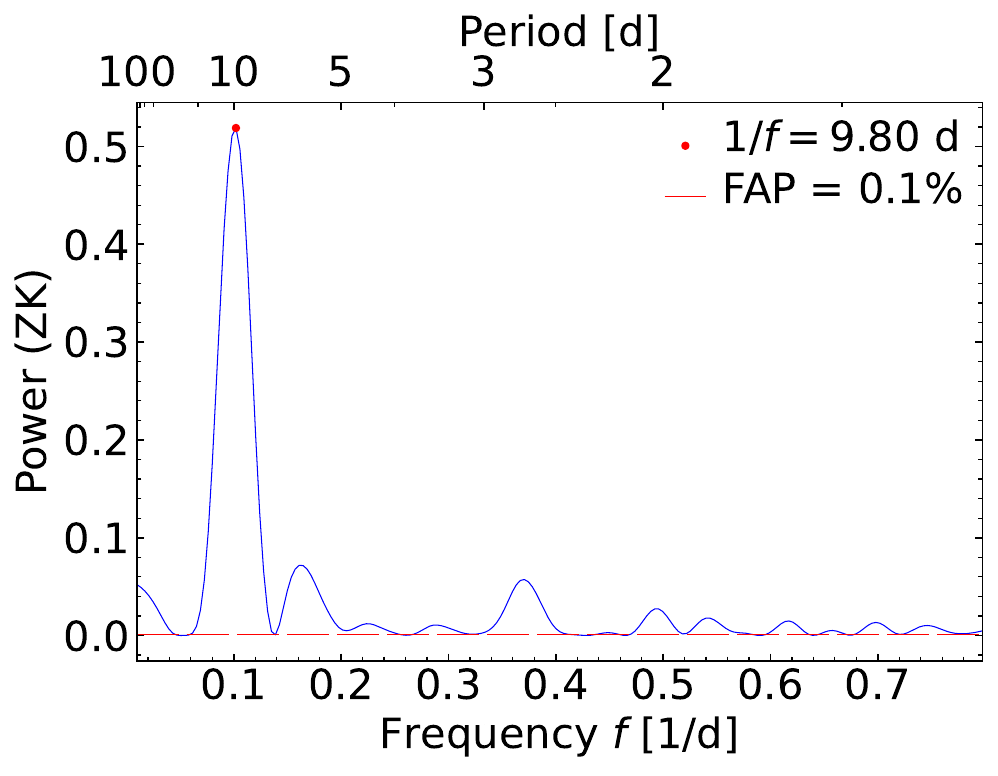}
    \includegraphics[width=0.246\textwidth]{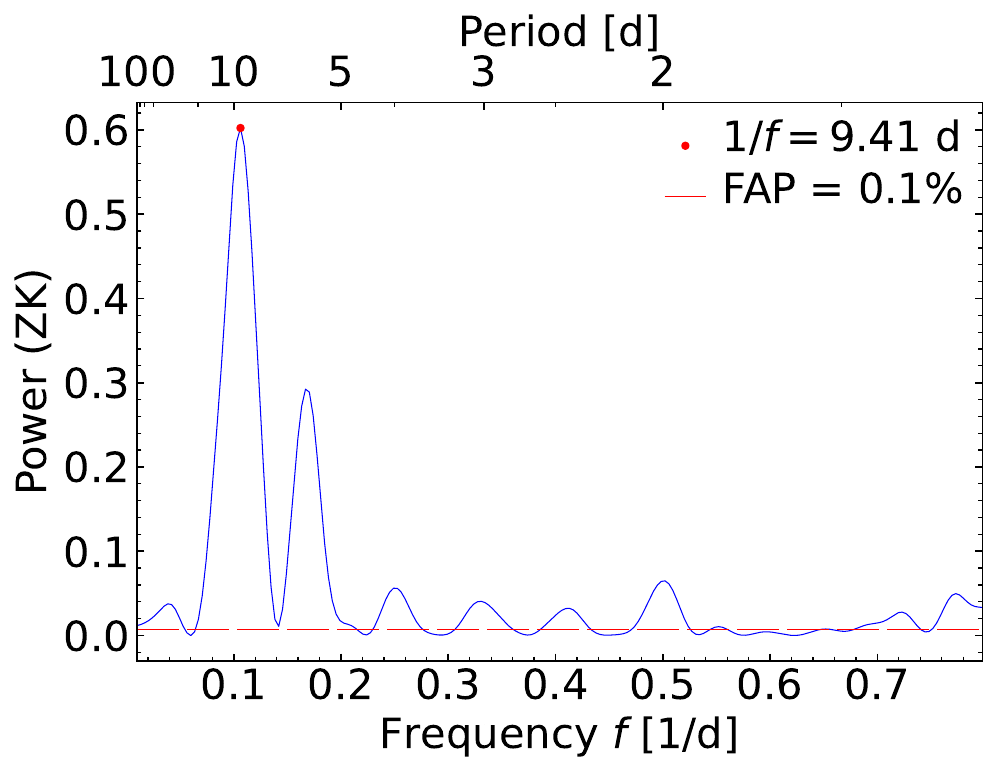}
    \includegraphics[width=0.246\textwidth]{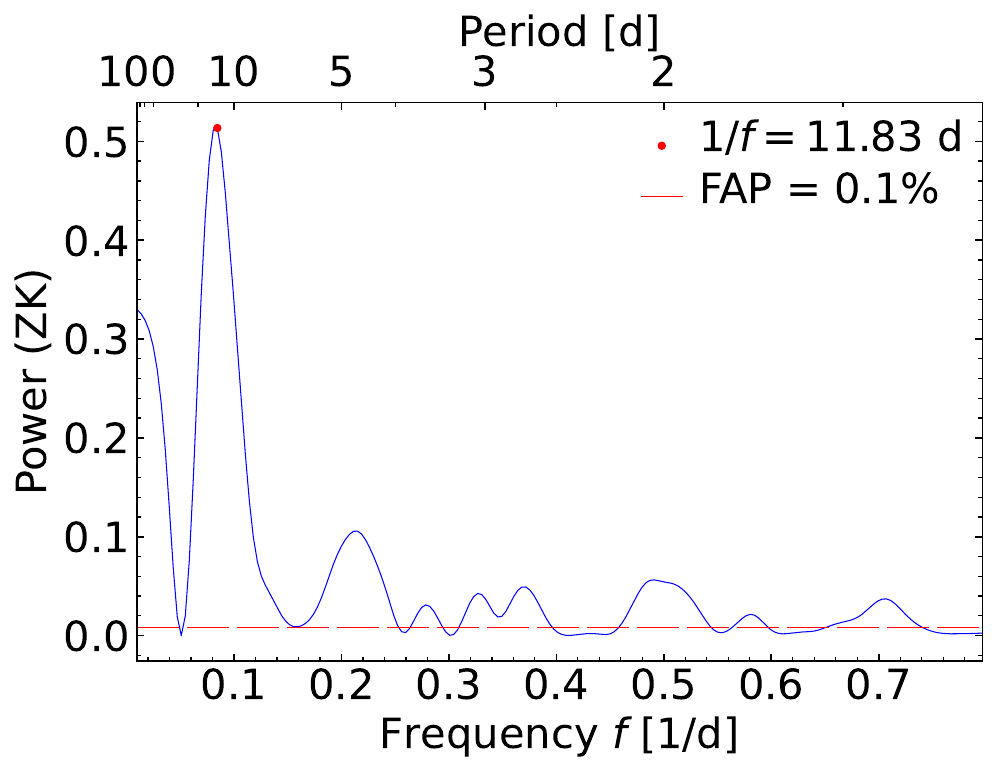}
    \includegraphics[width=0.246\textwidth]{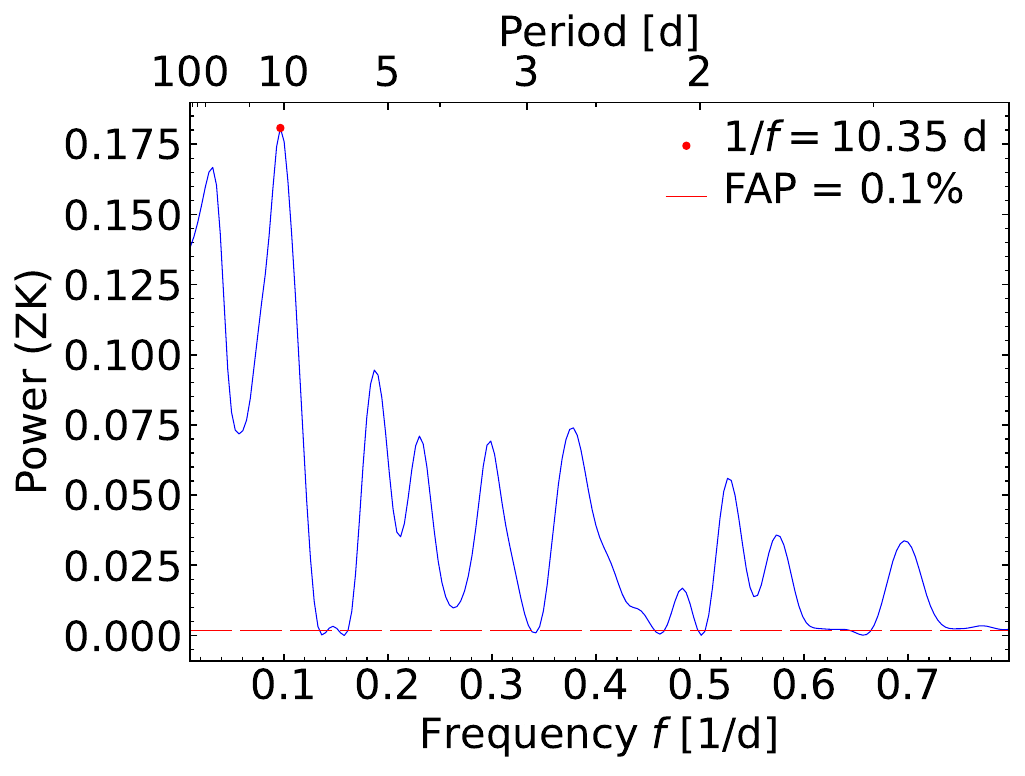} \\
    \includegraphics[width=0.246\textwidth]{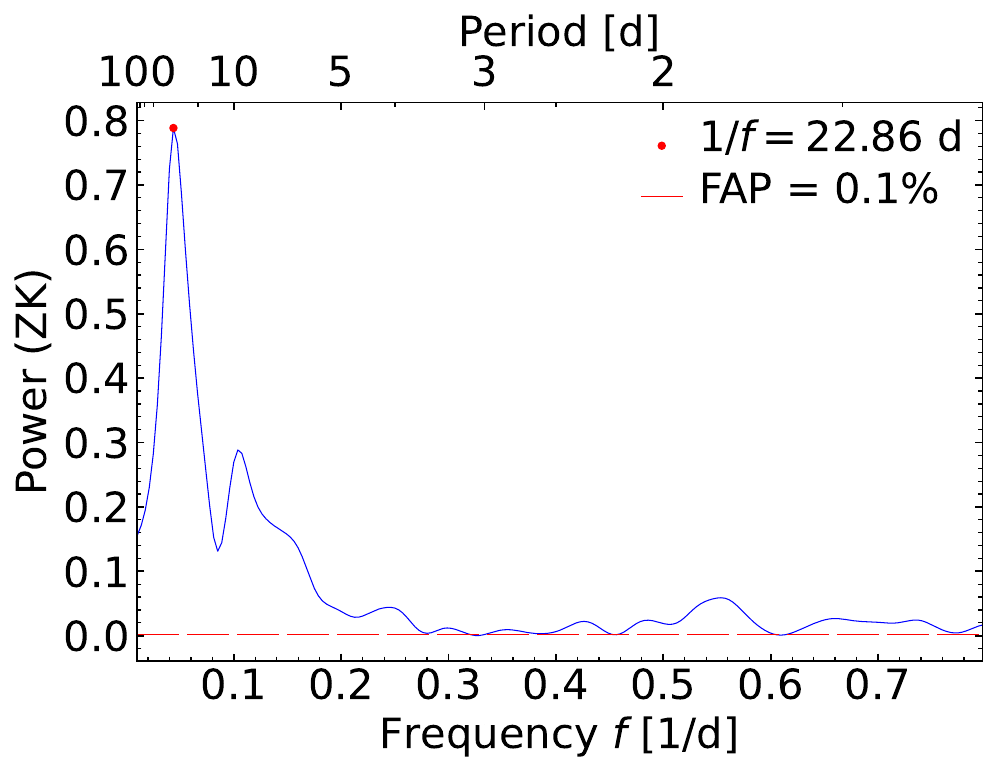} \\
    \sidecaption
    \includegraphics[width=0.246\textwidth]{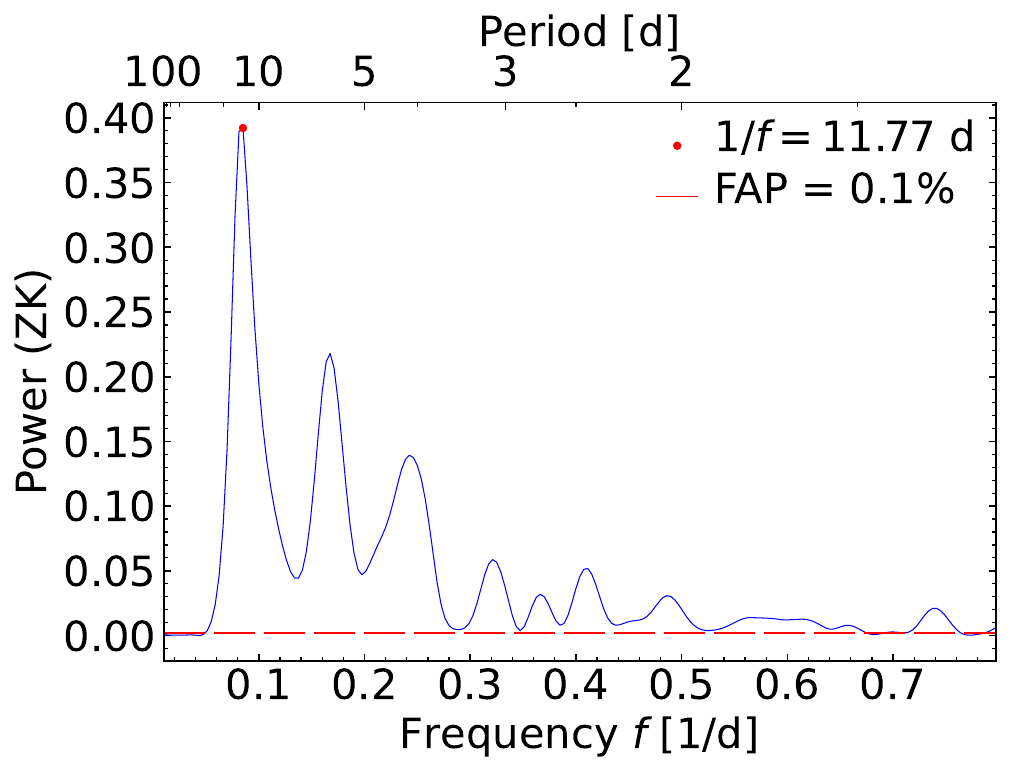}
    \caption{GLS periodograms of the five TESS sectors (SAP photometry) computed after removing all the transit windows. First row: from left to right sector 22, 48, 49, and 75. Second row: same as first row, but after subtracting the most prominent peak. Third row: sector 76. Fourth row: sector 76 after subtracting the most prominent peak. The most prominent peak is always at $\Prot$\,$\sim$\,20 days, except for Sector 75 (last panels in the first and second row) where the second ($\Prot$/3) and first ($\Prot$/2) harmonic of $\Prot$ appear.}
    \label{fig:glsLC}
\end{figure}

\begin{figure}
  \begin{minipage}[b]{0.48\linewidth} 
    \includegraphics[width=0.9\linewidth]{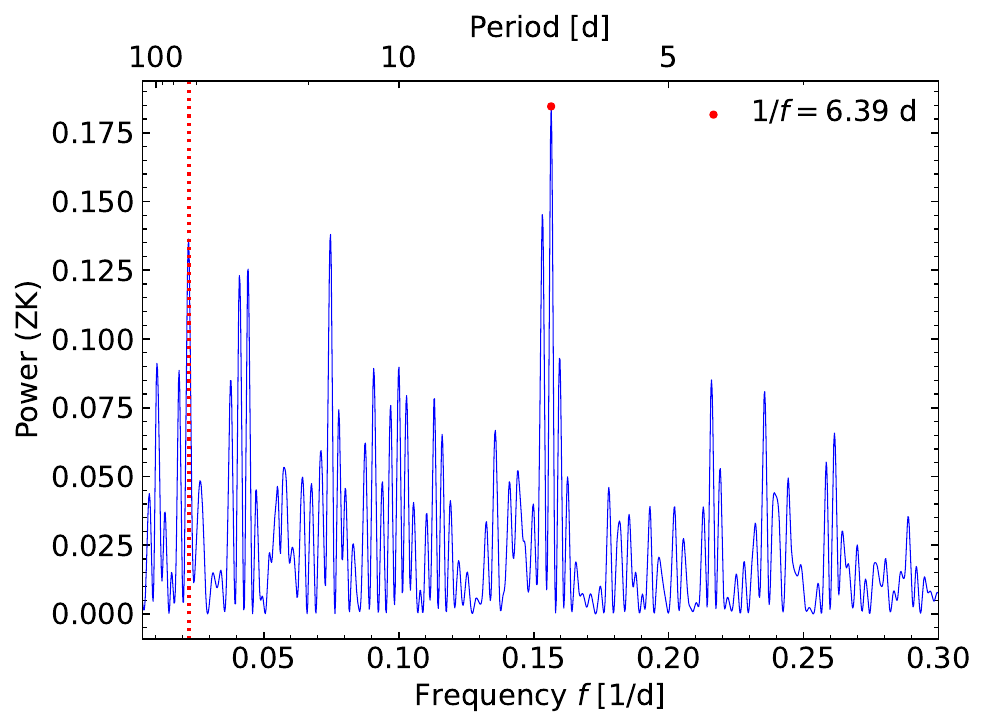} \\
    \includegraphics[width=0.9\linewidth]{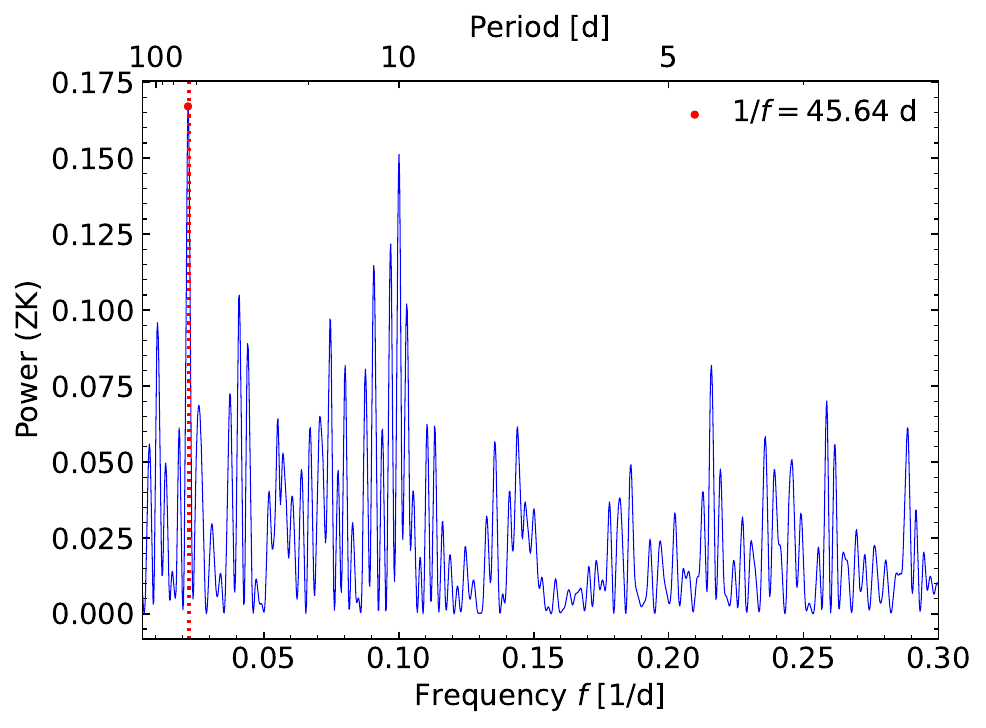}
    \caption{\textit{Top}. GLS periodogram of the RV residuals obtained after subtracting the Keplerian signals of the four transiting planets to the HARPS-N detrended RV time series (Sect.~\ref{sec:search5pla_RV}). The most prominent peak is the second harmonic of $\Prot$. The vertical dotted red line marks the signal of TOI-5624\,f at $\sim$\,45\,d. \textit{Bottom}. Same but after further subtracting the signal at $\sim$\,6.4\,d. The signal at $\sim$\,45\,d becomes the most prominent one with the second highest peak being $\Prot$/2.}
    \label{fig:glsResi4}
  \end{minipage}\hfill
  \begin{minipage}[b]{0.5\linewidth} 
    \begin{subfigure}{0.495\linewidth}
      \includegraphics[width=\linewidth]{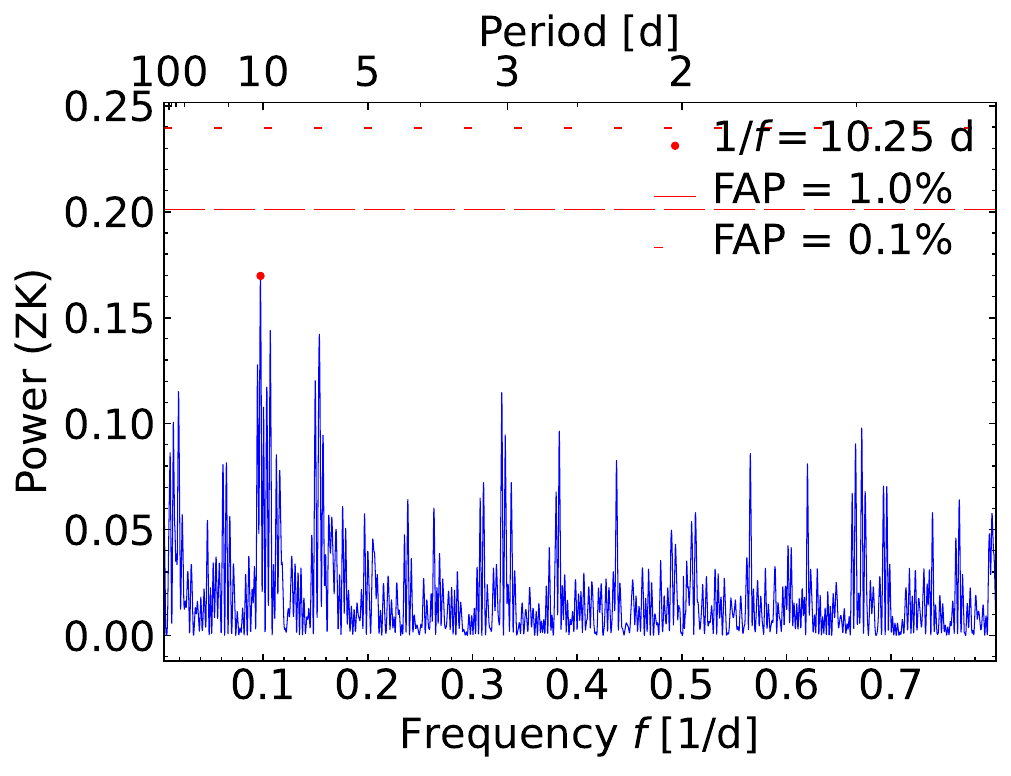}
    \end{subfigure}\hfill
    \begin{subfigure}{0.495\linewidth}
      \includegraphics[width=\linewidth]{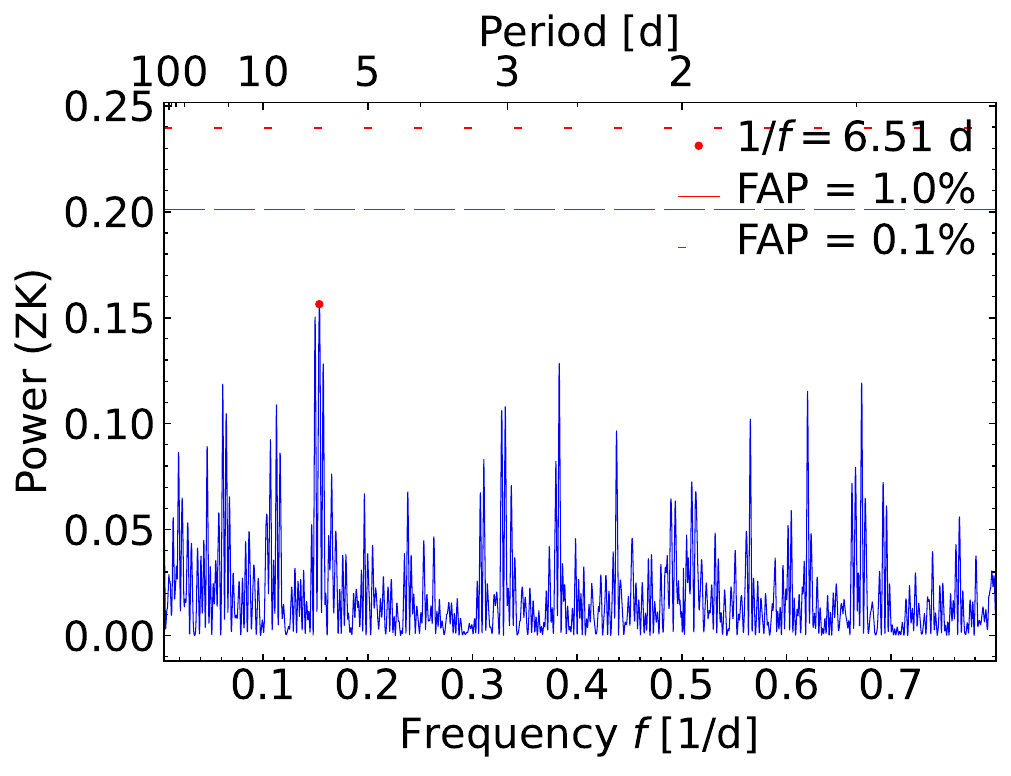}
    \end{subfigure}
    \begin{subfigure}{0.495\linewidth}
      \includegraphics[width=\linewidth]{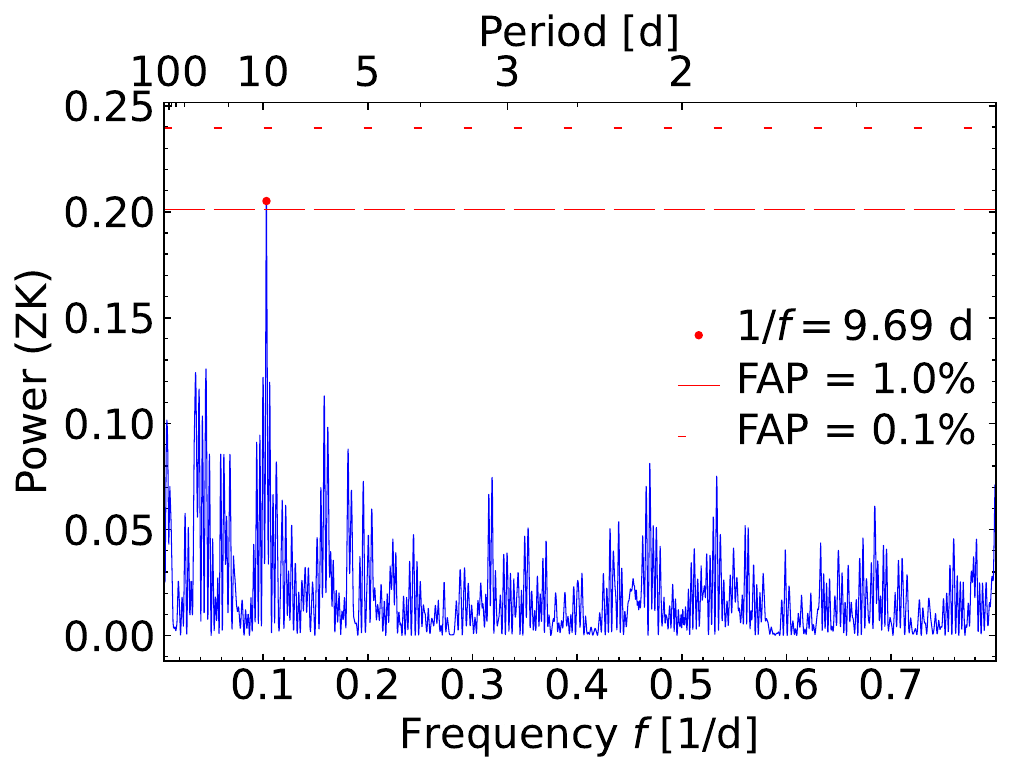}
    \end{subfigure}\hfill
    \begin{subfigure}{0.495\linewidth}
      \includegraphics[width=\linewidth]{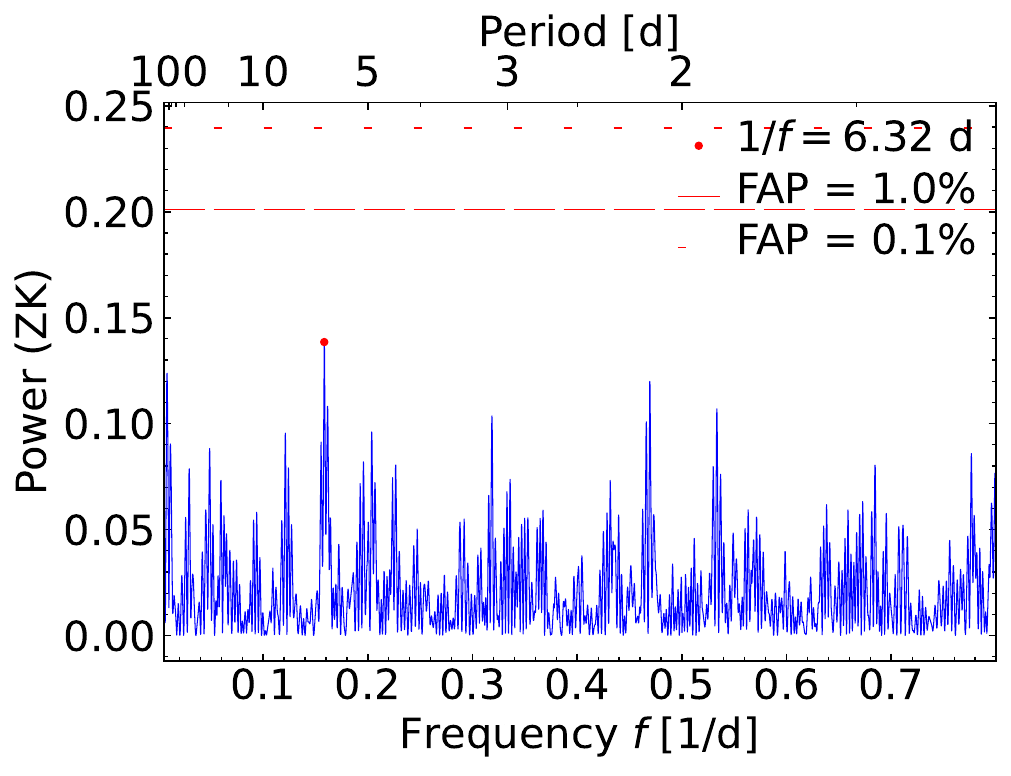}
    \end{subfigure}
    \caption{GLS periodograms of the FHWM (first row) and $\gamma$ (second row) as computed from the HARPS-N time series after a pre-whitening. Although not always significant, we systematically observe peaks at $\sim$10\,d (first column) and $\sim$6.4\,d (second column), which we ascribe to the first ($\Prot$/2) and second ($\Prot$/3) harmonics of the stellar rotation period ($\Prot$). A similar behaviour is seen in the other RV-related activity indicators (not shown here).}
    \label{fig:ProtHarmonics}
  \end{minipage}
\end{figure}

\begin{figure}[!ht]
    \centering
    \includegraphics[width=0.497\linewidth]{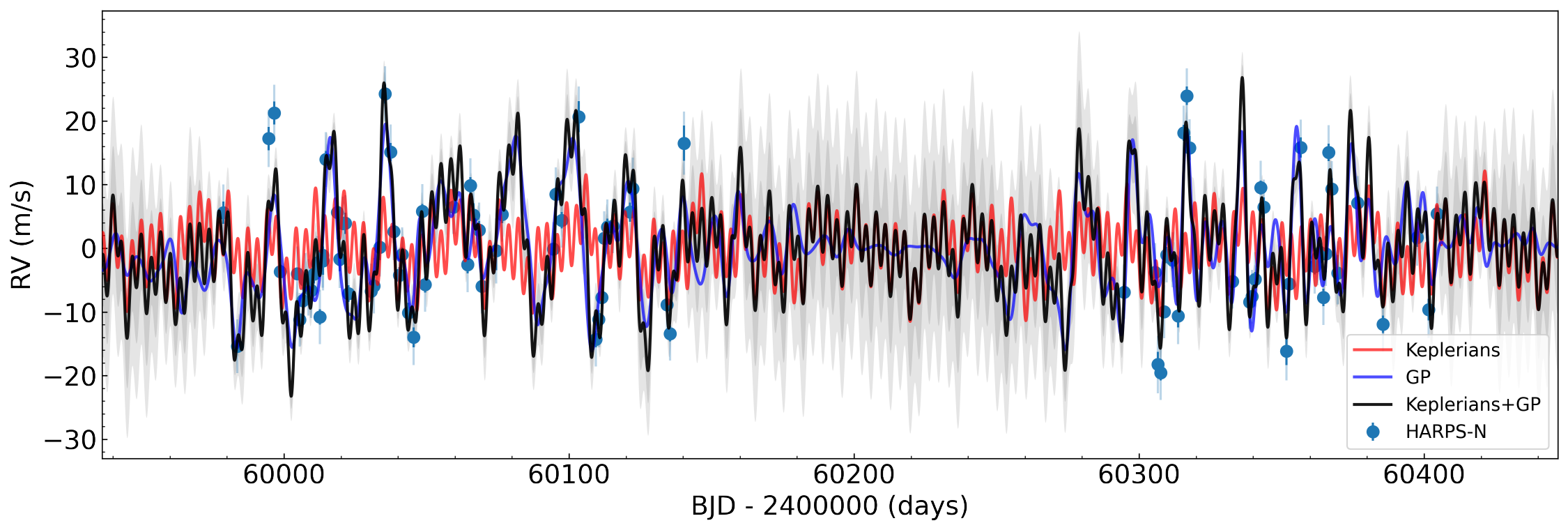}
    \includegraphics[width=0.497\linewidth]{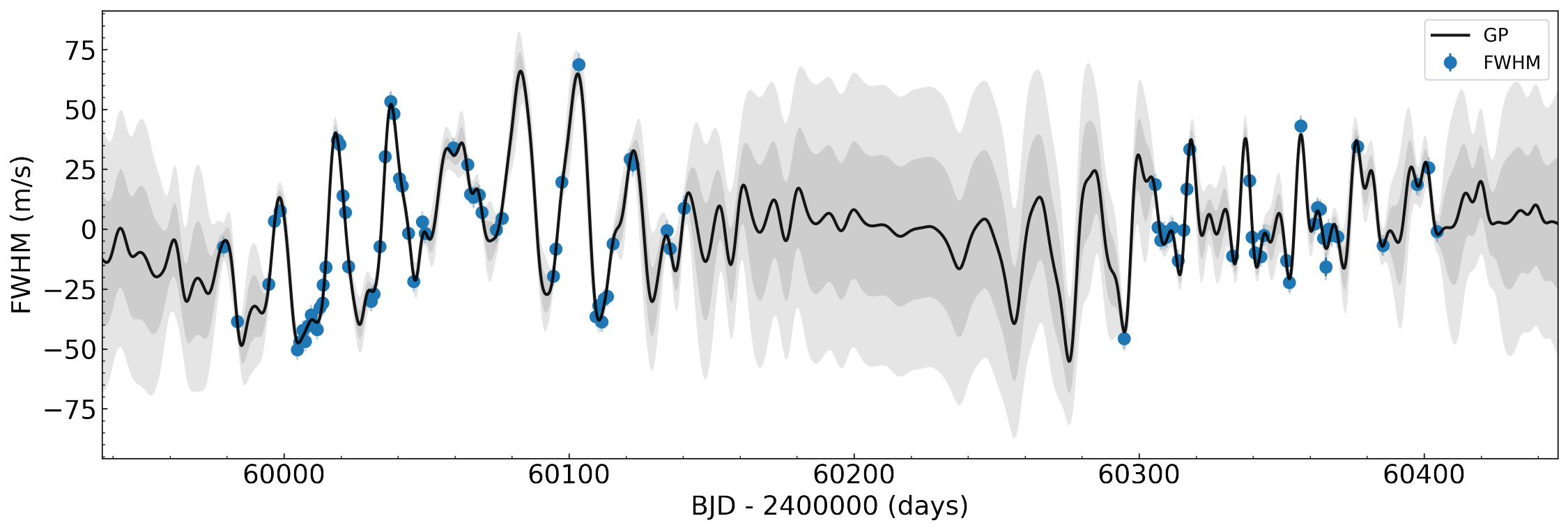}\\
    \includegraphics[width=0.33\linewidth]{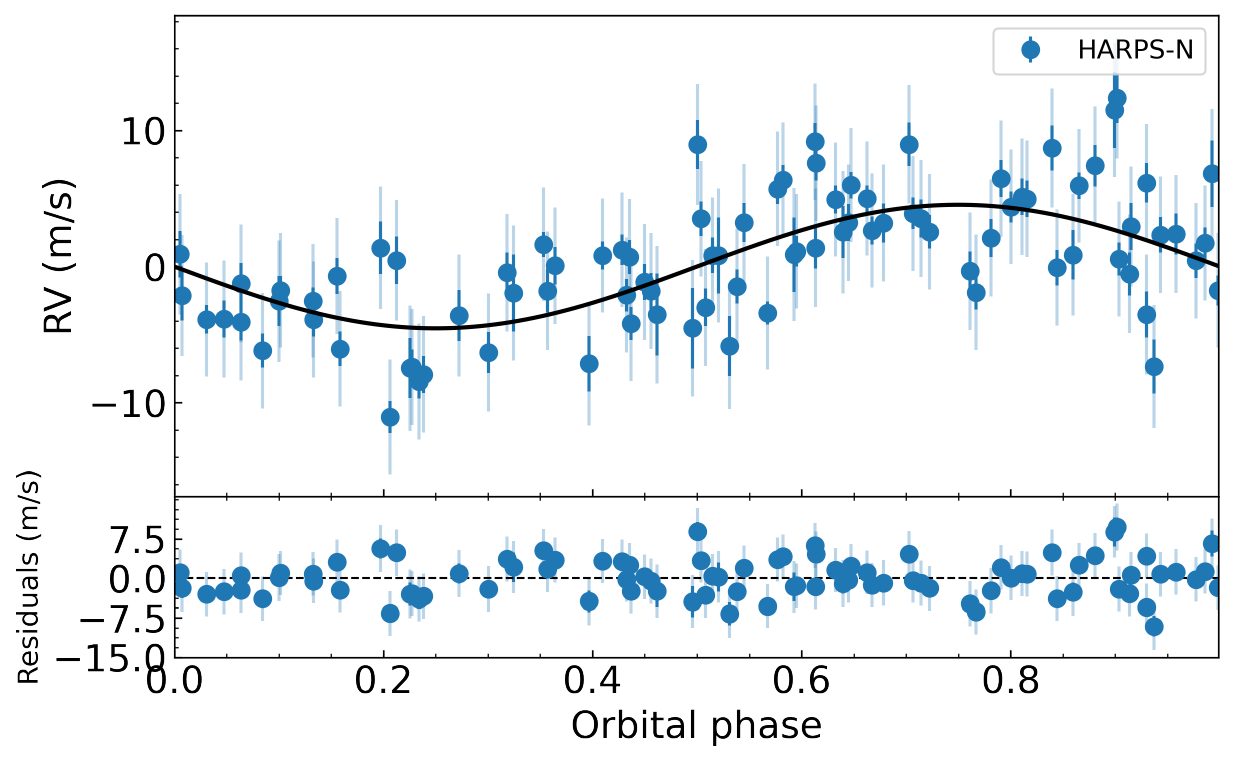}
    \includegraphics[width=0.33\linewidth]{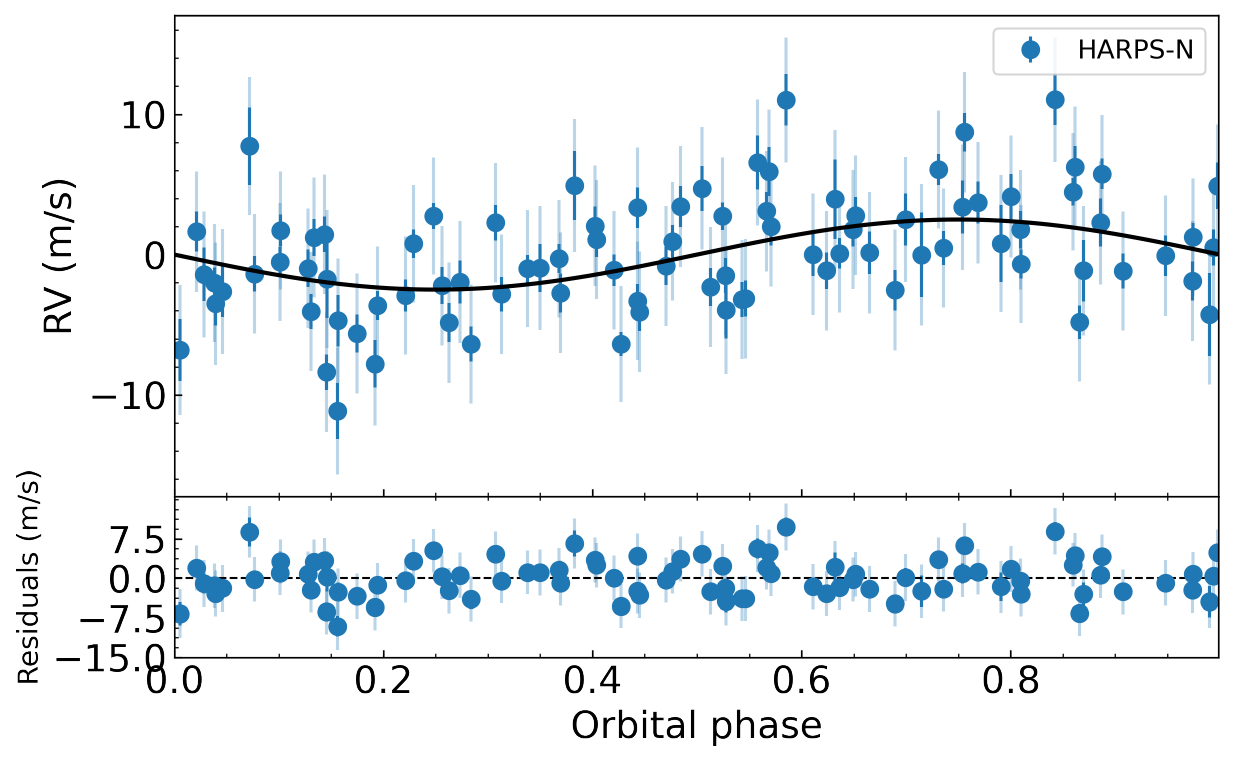}
    \includegraphics[width=0.33\linewidth]{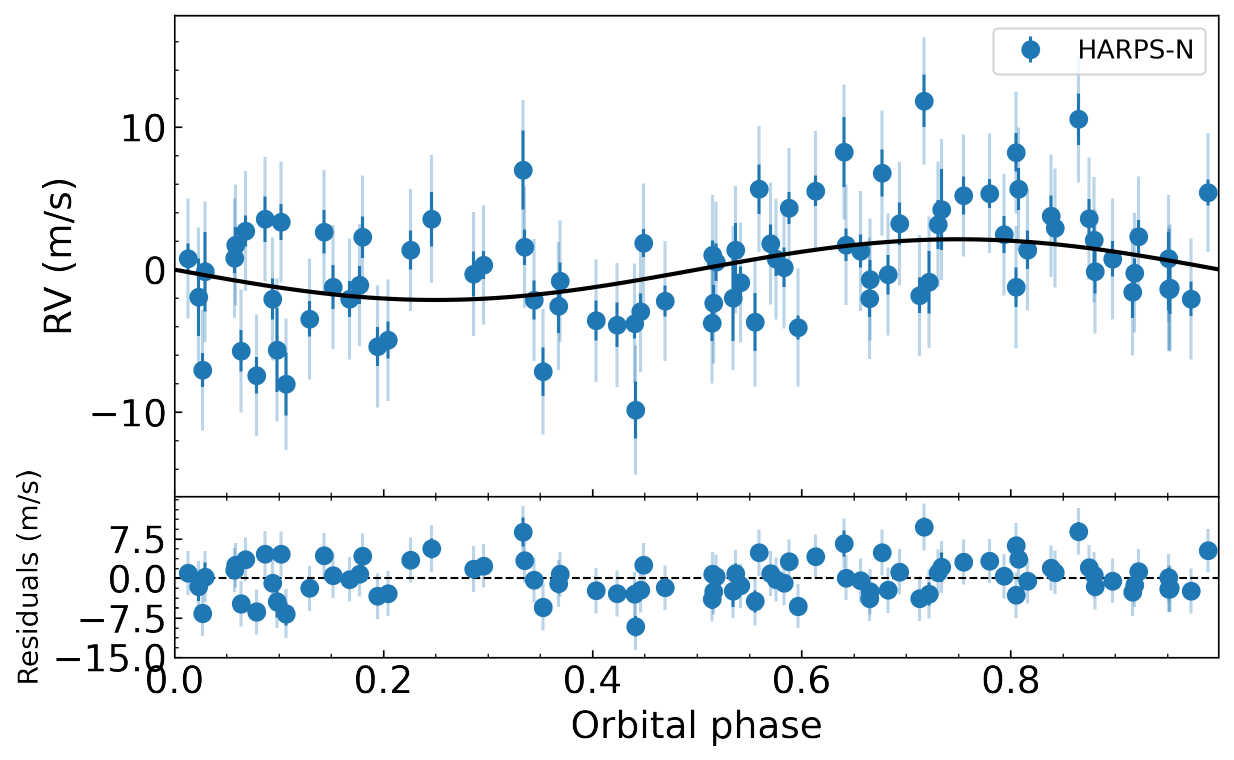}
    \includegraphics[width=0.33\linewidth]{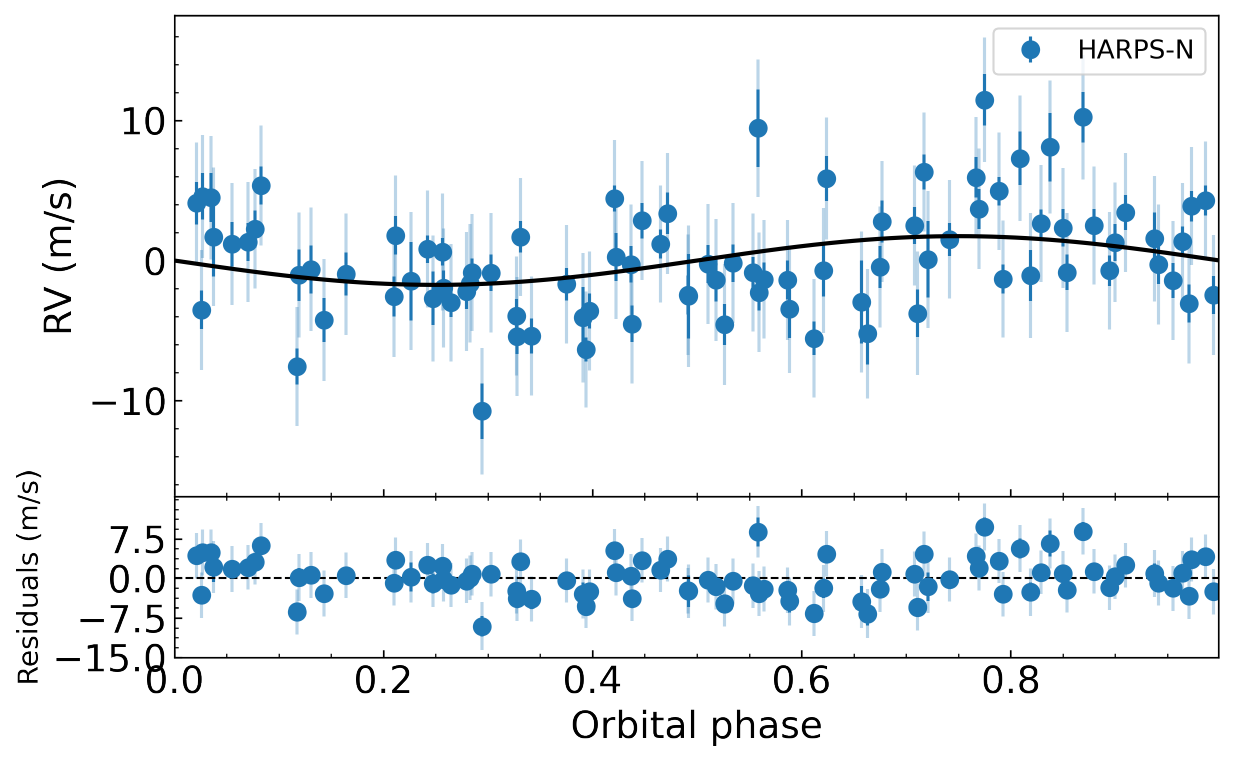}
    \includegraphics[width=0.33\linewidth]{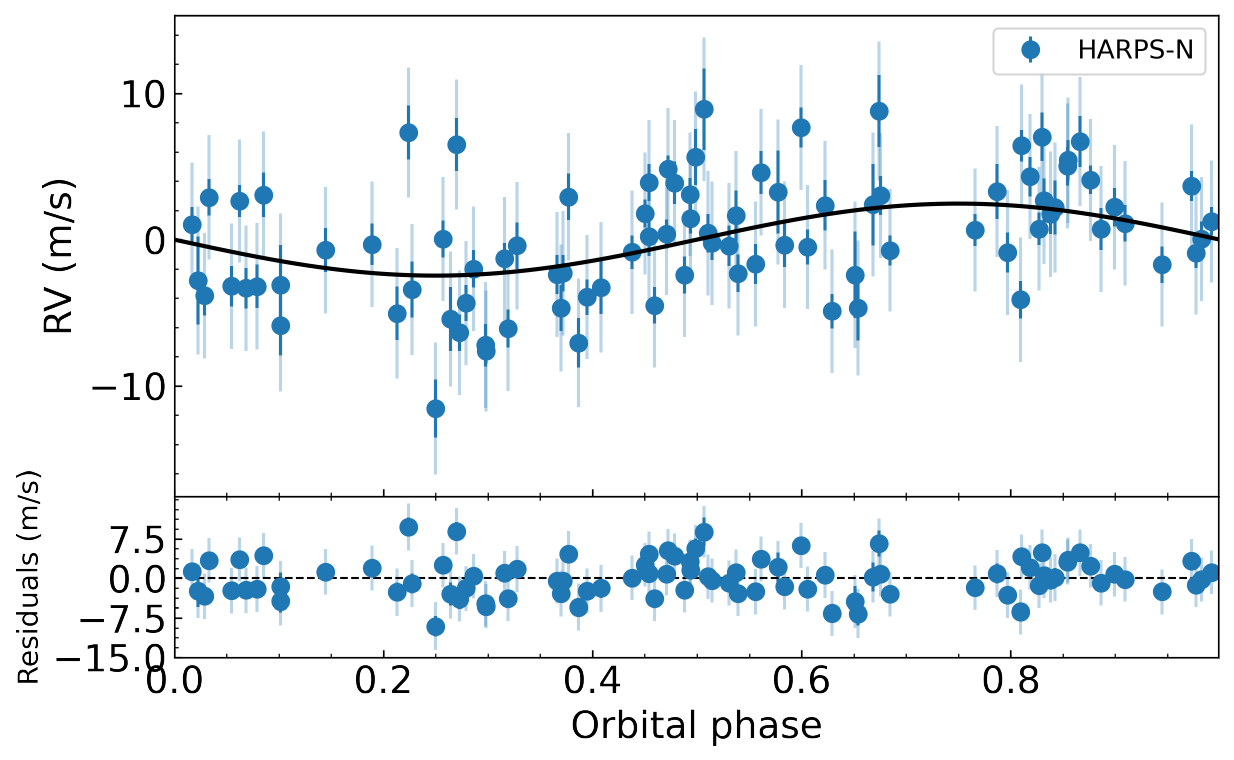}
    \caption{RV analysis performed with \texttt{pyaneti}. Top-left: Time series of the HARPS-N \texttt{TERRA} RV measurements (blue points). The inferred models for the five-planet and GP signals are shown with red and blue thick lines, respectively, whereas the combined signal is displayed in black. Top-right: Time series of the FWHM of the \texttt{DRS} cross-correlation functions (blue points). The black thick line marks the inferred GP model, whereas the grey shaded areas show its 1$\sigma$ and 2$\sigma$ credible intervals. Second and third rows, from top to bottom and left to right: Phase-folded RV curves of TOI-5624~b, c, d, e, and f, and best-fitting Keplerian models (thick black lines). In each panel, the nominal uncertainties of the data points and the jitter contribution are shown with thin and thick blue lines, respectively.}
    \label{fig:RV_curves_pyaneti}
\end{figure}

\begin{figure}
    \centering
    \begin{subfigure}[b]{0.4\textwidth}
        \includegraphics[width=\textwidth]{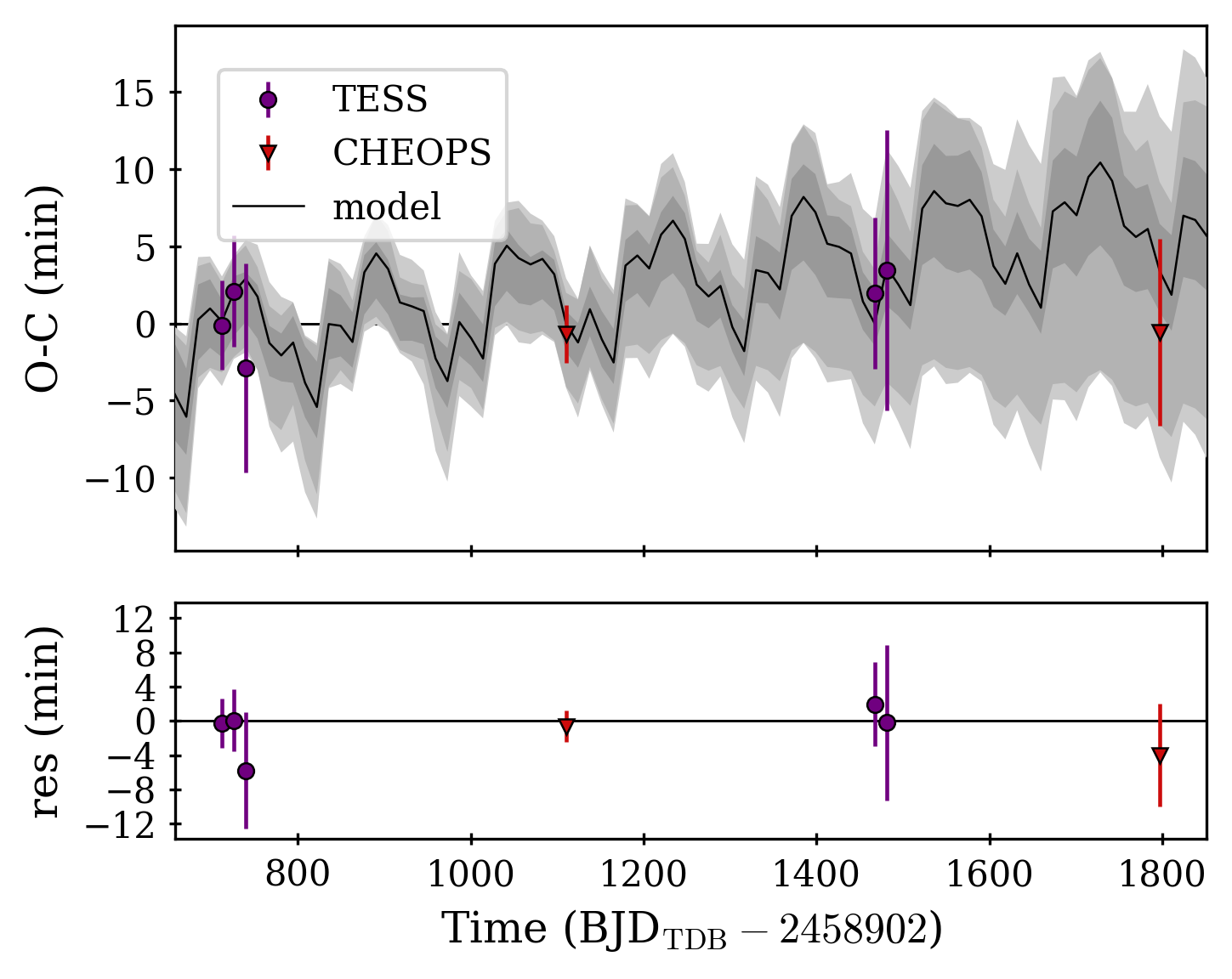}
    \end{subfigure}
    \begin{subfigure}[b]{0.4\textwidth}
        \includegraphics[width=\textwidth]{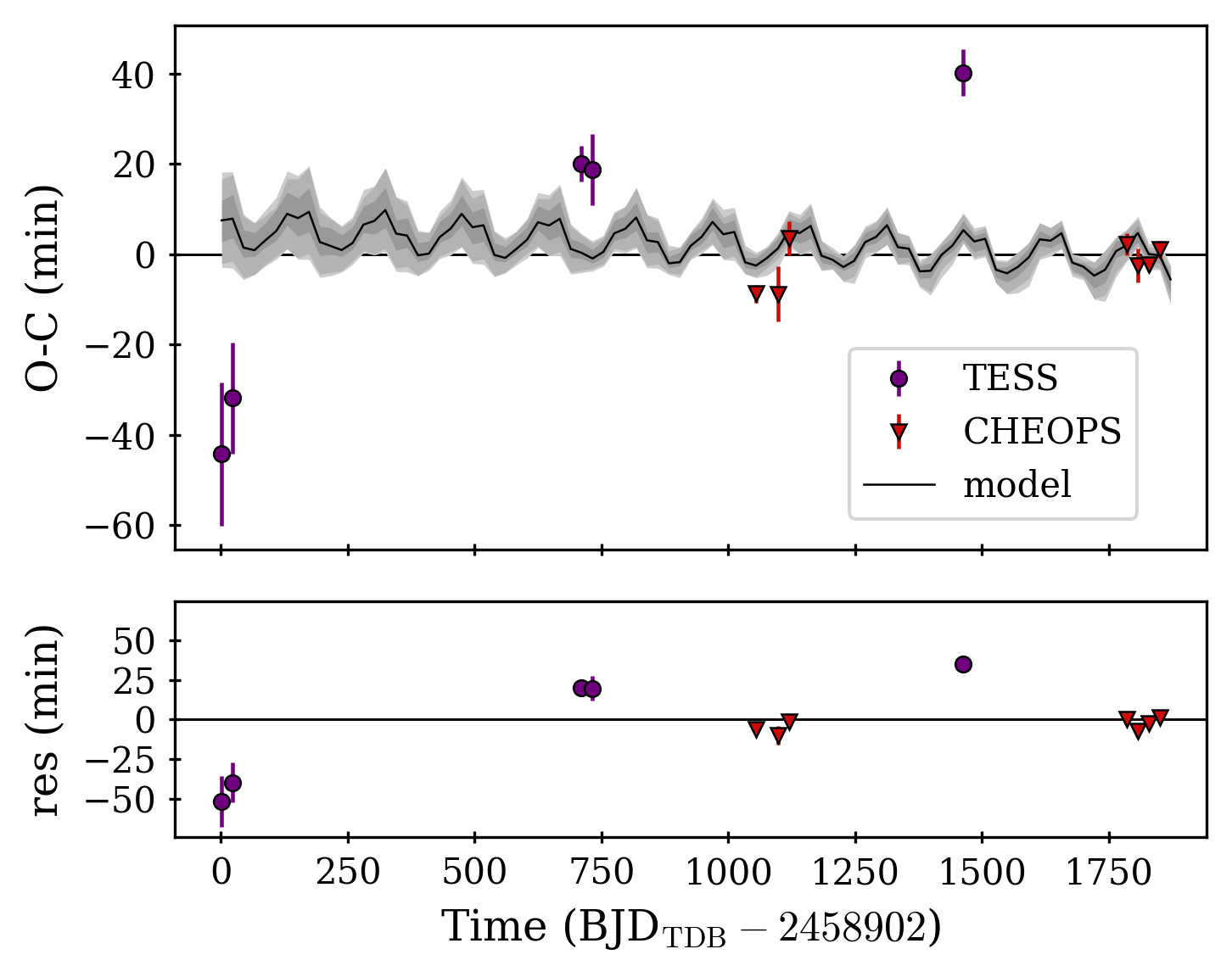}
    \end{subfigure}
    \caption{Same as Fig.~\ref{fig:oc_2}, but assuming the four-planet configuration. Left: TOI-5624\,d. Right: TOI-5624\,e.}
    \label{fig:oc_1}
\end{figure}

\begin{figure}
    \centering
    \begin{subfigure}[b]{0.4\textwidth}
        \includegraphics[width=\textwidth]{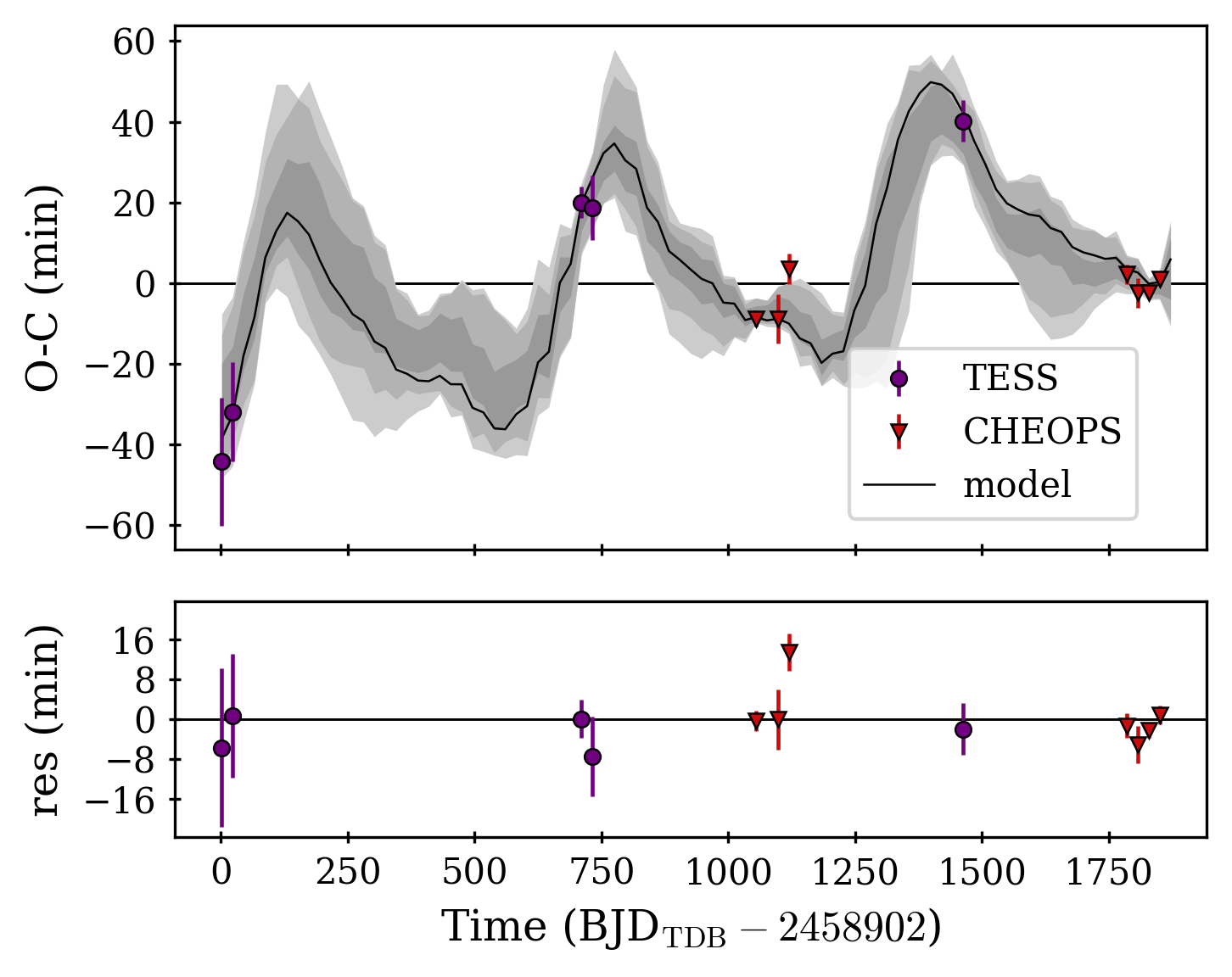}
    \end{subfigure}
    \begin{subfigure}[b]{0.4\textwidth}
        \includegraphics[width=\textwidth]{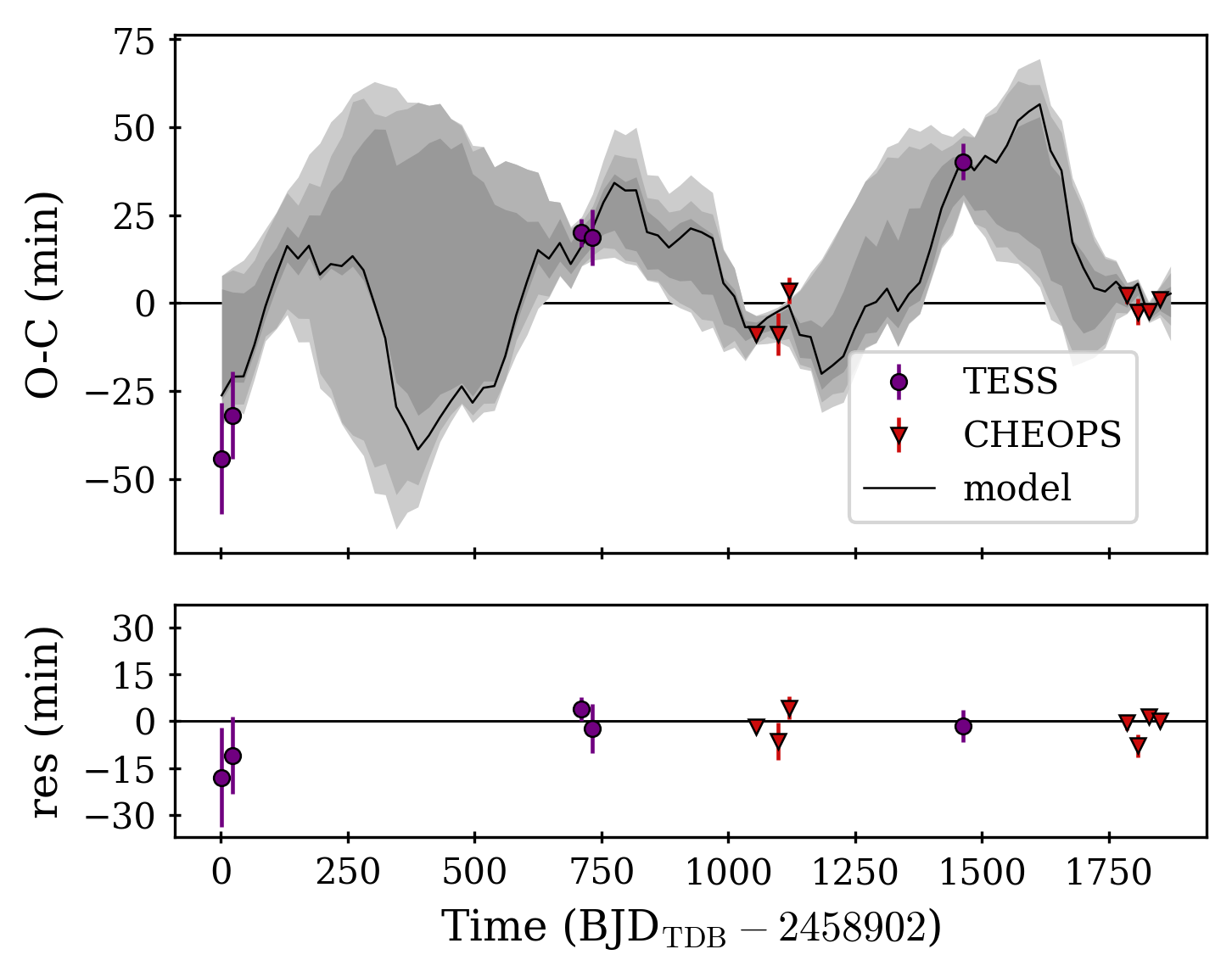}
    \end{subfigure}
    \caption{O$-$C diagrams of TOI-5624\,e for two additional dynamical analyses to test the robustness of the five-planet solution. Left: Fit obtained from the TTV+RV analysis after adopting a broader prior on the orbital period of planet~f ($P_f \in [1,50]$\,d), hence allowing configurations where the planet may lie between TOI-5624\,d and~e. Right: Fit obtained assuming again $P_f \in [1,50]$\,d, but using the transit mid-times only (i.e. excluding the RV dataset). Like the nominal analysis, both these analyses consistently conclude that only an external perturber may justify the TTVs of TOI-5624\,e.}
    \label{fig:oc_Pf_1_50}
\end{figure}

\begin{figure}
    \centering
    \includegraphics[width=0.495\linewidth]{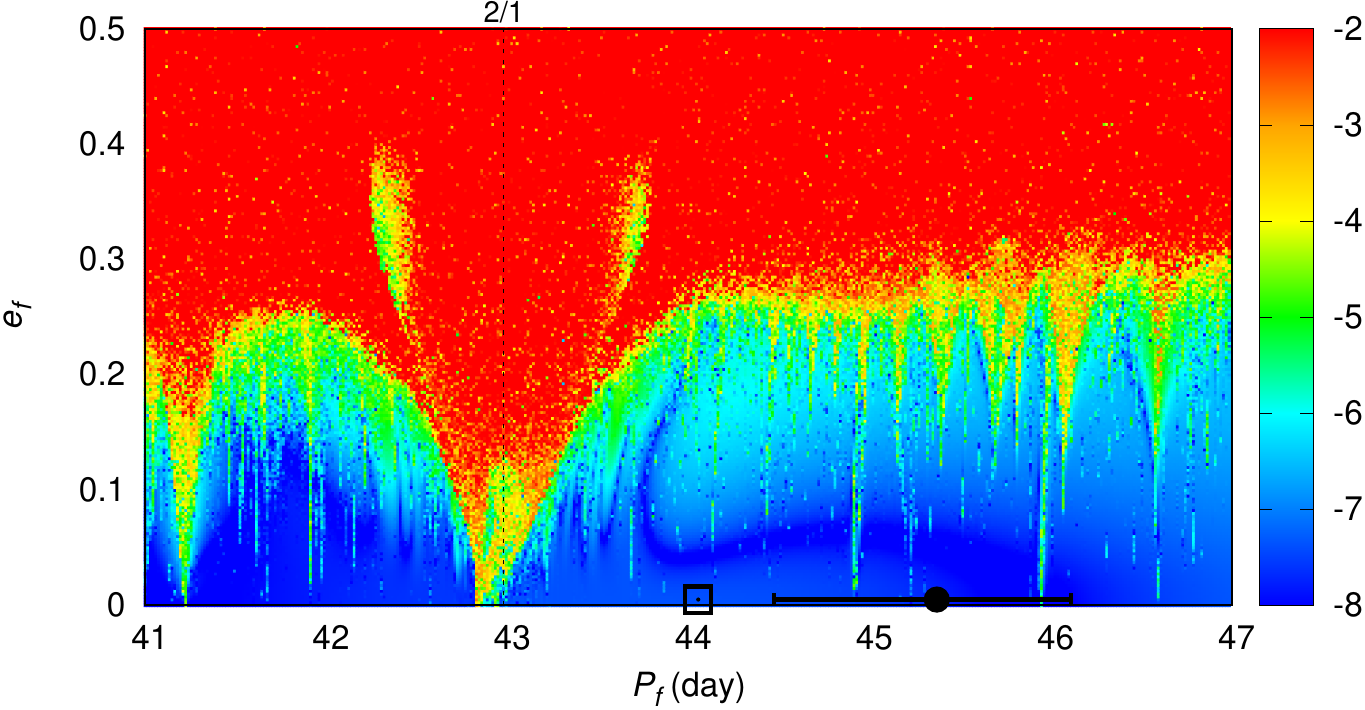}
    \includegraphics[width=0.495\linewidth]{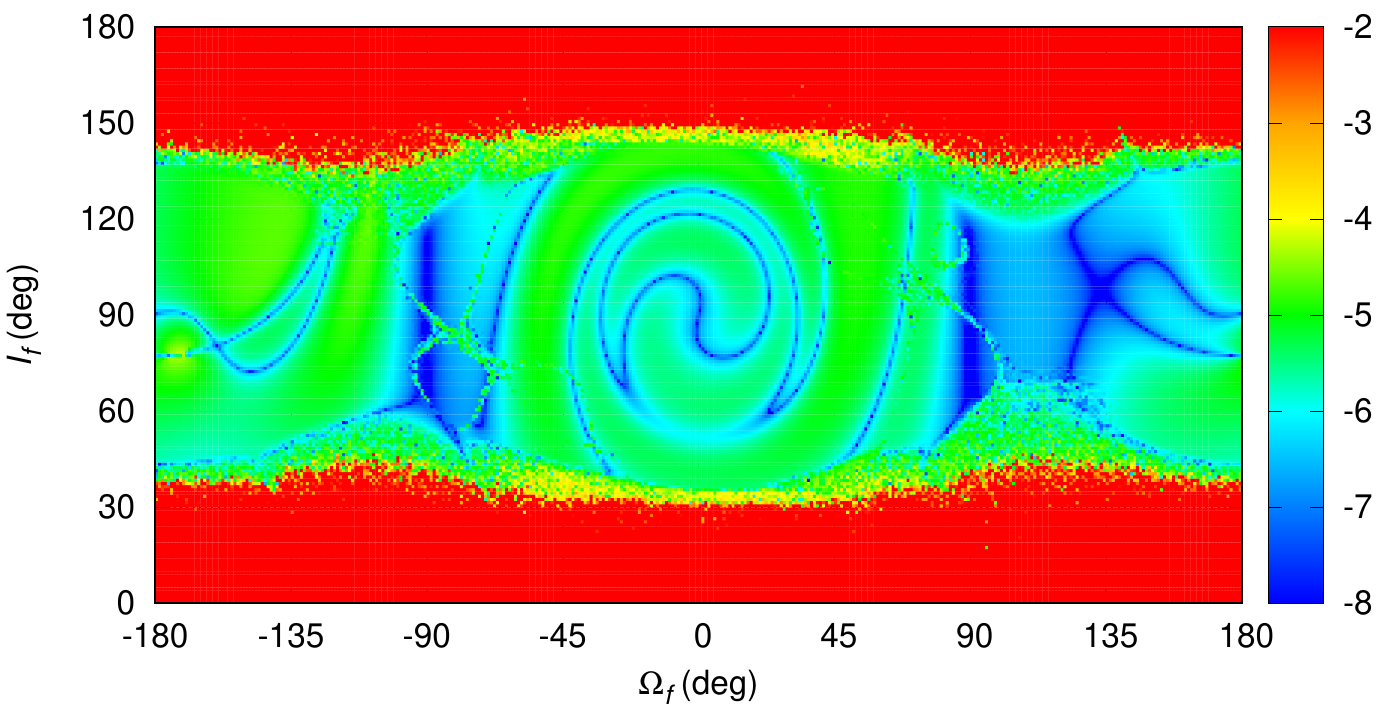}
    \caption{Stability analysis of the TOI-5624 planetary system focussing on planet f. We varied the orbital period, $P_f$ and the eccentricity $e_f$ (left) and the longitude of the node, $\Omega_f$, and the inclination, $I_f$ (right), while keeping all the remaining orbital parameters fixed at their best-fit values (Tab.~\ref{tab:planetParameters}). The step size is 0.015 day in the orbital period, 0.0025 in the eccentricity, and $1^\circ$ in the longitude of the node and inclination. Plots are colour-coded according to the stability indicator calculated from the frequency analysis of the mean longitude of TOI-5624\,f. Red points correspond to highly unstable orbits, while blue points correspond to orbits that are likely to be stable on Gyr timescales. The vertical dashed line (left) corresponds to the 2:1 MMR between planets e and f. The black dot along with its error bars highlights the RV-based best-fit solution $P_{f,\mathrm{RV}}=45.37_{-0.90}^{+0.74}$\,d from Tab.~\ref{tab:planetParameters}, while the black empty square indicates the more precise orbital period of planet f as inferred from the TTV+RV dynamical analysis ($P_{f,\mathrm{dyn}}=44.092_{-0.057}^{+0.015}$\,d, Table~\ref{Table:TRADES_results}). Both solutions (consistent at the 1.4$\sigma$ level) are compatible with a stable configuration of the system.}
    \label{fig:stabilityTOI5624f}
\end{figure}

\FloatBarrier
\section{Additional tables}
\begin{table}[!ht]
\caption{Log of CHEOPS observations.}
\label{tab:cheopsLog}
\centering
\begin{tabular}{l l c c c c l}
\hline
\hline
\noalign{\smallskip}
Counter & Filekey & Start observation & Duration & $t_{\mathrm{exp}}$ & Efficiency\tablefootmark{(a)} & Planet \\
        &        &      [UT]         &    [h]   &      [s]           &     [\%]   &    \\
\noalign{\smallskip}
\hline
\noalign{\smallskip}
CH 1 & CH\_PR110045\_TG004501\_V0300 & 2023-01-11T14:58:09.7  & 14.7 & 60 & 53 &  e    \\
CH 2 & CH\_PR110045\_TG004801\_V0300 & 2023-01-12T22:35:29.9  & 11.7 & 60 & 50 &  b    \\
CH 3 & CH\_PR110045\_TG004901\_V0300 & 2023-01-21T12:54:09.7  & 11.4 & 60 & 54 &  c    \\
CH 4 & CH\_PR110045\_TG005101\_V0300 & 2023-02-05T15:09:09.7  & 9.8 & 60 & 56 &  b    \\
CH 5 & CH\_PR110045\_TG005501\_V0300 & 2023-02-14T06:59:09.6  & 12.7 & 60 & 51 &  c    \\
CH 6 & CH\_PR110045\_TG005401\_V0300 & 2023-02-15T20:12:48.9  & 10.3 & 60 & 57 &  b    \\
CH 7 & CH\_PR110045\_TG005201\_V0300 & 2023-02-23T16:06:29.0  & 15.7 & 60 & 59 &  e    \\
CH 8 & CH\_PR110045\_TG005701\_V0300 & 2023-02-26T00:03:31.1  & 9.1 & 60 & 59 &  b    \\
CH 9 & CH\_PR110045\_TG005801\_V0300 & 2023-03-01T10:41:08.8  & 9.0 & 60 & 62 &  b    \\
CH 10 & CH\_PR110045\_TG005601\_V0300 & 2023-03-02T01:22:28.9  & 9.2 & 60 & 59 &  c    \\
CH 11 & CH\_PR110045\_TG006001\_V0300 & 2023-03-04T19:18:28.9  & 9.2 & 60 & 62 &  b    \\
CH 12 & CH\_PR110045\_TG006201\_V0300 & 2023-03-08T04:02:29.4  & 20.7 & 60 & 58 &  b d  \\
CH 13 & CH\_PR110045\_TG006202\_V0300 & 2023-03-15T00:01:30.2  & 10.0 & 60 & 57 &  b    \\
CH 14 & CH\_PR110045\_TG006401\_V0300 & 2023-03-17T06:04:30.0  & 12.0 & 60 & 54 &  e    \\
CH 15 & CH\_PR110045\_TG006301\_V0300 & 2023-03-17T18:46:29.3  & 14.2 & 60 & 56 &  c    \\
CH 16 & CH\_PR110045\_TG006203\_V0300 & 2023-03-18T09:45:08.9  & 11.7 & 60 & 59 &  b    \\
CH 17 & CH\_PR110045\_TG006302\_V0300 & 2023-03-25T16:14:29.6  & 9.2 & 60 & 61 &  c    \\
CH 18 & CH\_PR110045\_TG006204\_V0300 & 2023-03-31T21:04:09.9  & 9.8 & 60 & 59 &  b    \\
CH 19 & CH\_PR110045\_TG006303\_V0300 & 2023-04-02T13:07:30.0  & 9.8 & 60 & 57 &  c    \\
CH 20 & CH\_PR140083\_TG002101\_V0300 & 2025-01-11T10:59:57.6  & 11.7 & 60 & 55 &  e    \\
CH 21 & CH\_PR140083\_TG002701\_V0300 & 2025-01-23T00:34:29.5  & 10.6 & 60 & 53 &  d    \\
CH 22 & CH\_PR140083\_TG002501\_V0300 & 2025-01-27T19:37:09.5  & 9.0 & 60 & 59 &  b    \\
CH 23 & CH\_PR140083\_TG002801\_V0300 & 2025-02-01T22:52:09.6  & 10.6 & 60 & 57 &  e    \\
CH 24 & CH\_PR140083\_TG003201\_V0300 & 2025-02-10T09:09:37.6  & 8.9 & 60 & 58 &  b    \\
CH 25 & CH\_PR140083\_TG003101\_V0300 & 2025-02-23T07:41:57.7  & 10.7 & 60 & 60 &  e    \\
CH 26 & CH\_PR140083\_TG003401\_V0300 & 2025-02-24T10:32:30.8  & 11.4 & 60 & 57 &  c    \\
CH 27 & CH\_PR140083\_TG003901\_V0300 & 2025-03-06T03:47:30.8  & 10.2 & 60 & 62 &  b    \\
CH 28 & CH\_PR140083\_TG003902\_V0300 & 2025-03-09T12:03:30.9  & 9.7 & 60 & 60 &  b    \\
CH 29 & CH\_PR140083\_TG004201\_V0300 & 2025-03-12T04:44:30.9  & 11.4 & 60 & 61 &  c    \\
CH 30 & CH\_PR140083\_TG003903\_V0300 & 2025-03-12T21:55:29.8  & 9.2 & 60 & 65 &  b    \\
CH 31 & CH\_PR140083\_TG003904\_V0300 & 2025-03-16T06:29:17.6  & 9.1 & 60 & 66 &  b    \\
CH 32 & CH\_PR140083\_TG004001\_V0300 & 2025-03-16T20:05:50.8  & 11.4 & 60 & 62 &  e    \\
CH 33 & CH\_PR140083\_TG003905\_V0300 & 2025-03-19T15:24:10.8  & 10.2 & 60 & 64 &  b    \\
CH 34 & CH\_PR140083\_TG003906\_V0300 & 2025-03-23T01:08:29.9  & 9.2 & 60 & 67 &  b    \\
CH 35 & CH\_PR140083\_TG003907\_V0300 & 2025-03-29T21:28:29.8  & 9.2 & 60 & 68 &  b    \\
CH 36 & CH\_PR140083\_TG003908\_V0300 & 2025-04-12T09:15:57.8  & 9.2 & 60 & 67 &  b    \\
\noalign{\smallskip}
\hline
\end{tabular}
\tablefoot{\tablefoottext{a} Efficiency is the time fraction actually spent on the target, accounting for interruptions due to Earth occultations, Earth stray-light events, and crossings over the South-Atlantic anomaly (SAA).}
\end{table}

\twocolumn

\begin{table}[ht]
\caption{Polynomial detrending baseline applied to the TESS LCs within the \texttt{MCMCI} analysis.}
\label{tab:polyDetrendingTE}
\centering     
\begin{tabular}{l c c}
\hline
\hline            
\noalign{\smallskip}
LC & Planet & Detrending \\
\noalign{\smallskip}
\hline
\noalign{\smallskip}
TE 1  & c   &  $c$ \\
TE 2  & b e &  \ttt$^1$  \\
TE 3  & b   &  \ttt$^2$    \\
TE 4  & b c &  $c$   \\
TE 5  & b   &  $c$   \\
TE 6  & c   &  \ttt$^3$   \\
TE 7  & b   &  \ttt$^1$   \\
TE 8  & b   &  \ttt$^2$    \\
TE 9  & e   &  \ttt$^3$   \\
TE 10 & c   &  \ttt$^1$   \\
TE 11 & b c &  $c$ \\
TE 12 & e   &  $c$  \\
TE 13 & d   &  \ttt$^1$    \\
TE 14 & b   &  \ttdy$^1$   \\
TE 15 & b   &  \ttdy$^1$   \\
TE 16 & c   &  \ttt$^1$   \\
TE 17 & c d &  \ttt$^2$   \\
TE 18 & b   &  \ttt$^1$   \\
TE 19 & b   &  \ttt$^1$   \\
TE 20 & e   &  \ttt$^1$  \\
TE 21 & b c &  \ttdy$^3$ \\
TE 22 & d   &  \ttt$^1$   \\
TE 23 & b   &  $c$    \\
TE 24 & c   &  $c$   \\
TE 25 & b   &  $c$   \\
TE 26 & b   &  $c$   \\
TE 27 & b c &  \ttt$^1$   \\
TE 28 & b   &  $c$    \\
TE 29 & b   &  \ttt$^1$ + \ttdy$^1$   \\
TE 30 & b   &  $c$   \\
TE 31 & b   &  \ttt$^1$ \\
TE 32 & c   &  \ttdx$^1$   \\
TE 33 & b   &  \ttt$^1$    \\
TE 34 & b e &  \ttdy$^1$   \\
TE 35 & b   &  \ttdx$^1$   \\
TE 36 & c   &  \ttdx$^1$   \\
TE 37 & d   &  \ttdx$^1$   \\
TE 38 & c   & \ttxy$^1$     \\
TE 39 & b   &  \ttt$^1$   \\
TE 40 & b   &  $c$   \\
TE 41 & d   &  $c$   \\
TE 42 & c   &  \ttt$^1$ + \ttdx$^1$  \\
TE 43 & b   &  \ttdx$^1$   \\
TE 44 & c   &  $c$   \\
\noalign{\smallskip}
\hline
\end{tabular}
\tablefoot{TESS (TE) LCs are identified by a counter based on the chronological order of observation. In particular, LCs from 1 to 10, from 12 to 21, from 22 to 28, from 29 to 34, and from 35 to 44 were extracted from Sector 22, 48, 49, 75 and 76, respectively. $c$ indicates a normalisation scalar; see text for further details.}
\end{table}

\begin{table}[ht]
\caption{Polynomial detrending baseline applied to the CHEOPS LCs within the \texttt{MCMCI} analysis.}
\label{tab:polyDetrendingCH}
\centering     
\begin{tabular}{l c c}
\hline
\hline            
\noalign{\smallskip}
LC & Planet & Detrending \\
\noalign{\smallskip}
\hline
\noalign{\smallskip}
CH 1  & e   &  \GProll\ + \ttt$^1$ + \ttbg$^4$  \\
CH 2  & b   &  \GProll\ + \ttt$^2$ + \ttbg$^4$  \\
CH 3  & c   &  \GProll\ + \ttbg$^3$    \\
CH 4  & b   &  \GProll\ + \ttbg$^1$   \\
CH 5  & c   &  \GProll\ + \ttt$^2$   \\
CH 6  & b   &  \GProll\ + \ttbg$^4$   \\
CH 7  & e   &  \GProll\ + \ttt$^2$ + \ttbg$^2$ + \ttxy$^2$   \\
CH 8  & b   &  \GProll\ + \ttt$^1$ + \ttbg$^3$    \\
CH 9  & b   &  \GProll\ + \ttbg$^2$   \\
CH 10 & c   &  \GProll    \\
CH 11 & b   &  \GProll\ + \ttt$^2$ + \ttbg$^1$ \\
CH 12 & b d &  \GProll\ + \ttt$^2$ + \ttbg$^3$  \\
CH 13 & b   &  \GProll\ + \ttt$^1$ + \ttbg$^1$    \\
CH 14 & e   &  \GProll\ + \ttt$^1$ + \ttsm$^2$ + \ttbg$^4$   \\
CH 15 & c   &  \GProll\ + \ttt$^2$ + \ttbg$^2$   \\
CH 16 & b   &  \GProll\ + \ttt$^1$ + \ttbg$^1$   \\
CH 17 & c   &  \GProll\ + \ttt$^2$ + \ttbg$^1$   \\
CH 18 & b   &  \GProll\ + \ttt$^1$ + \ttbg$^2$   \\
CH 19 & c   &  \GProll\ + \ttt$^1$ + \ttbg$^1$   \\
CH 20 & e   &  \GProll\ + \ttt$^3$ + \ttbg$^1$  \\
CH 21 & d   &  \GProll\ + \ttbg$^4$ \\
CH 22 & b   &  \GProll\ + \ttt$^2$   \\
CH 23 & e   &  \GProll\ + \ttt$^2$ + \ttbg$^1$    \\
CH 24 & b   &  \GProll\ + \ttt$^2$ + \ttbg$^3$   \\
CH 25 & e   &  \GProll\ + \ttt$^2$ + \ttbg$^3$ + \ttxy$^1$   \\
CH 26 & c   &  \GProll\ + \ttbg$^4$   \\
CH 27 & b   &  \GProll\ + \ttbg$^4$ + \ttxy$^2$   \\
CH 28 & b   &  \GProll\ + \ttbg$^3$    \\
CH 29 & c   &  \GProll\ + \ttt$^1$ + \ttbg$^4$   \\
CH 30 & b   &  \GProll\ + \ttt$^2$ + \ttbg$^2$   \\
CH 31 & b   &  \GProll\ + \ttt$^1$ + \ttsm$^2$ + \ttbg$^4$ + \ttxy$^1$ \\
CH 32 & e   &  \GProll\ + \ttco$^3$ + \ttbg$^4$ + \ttxy$^2$  \\
CH 33 & b   &  \GProll\ + \ttbg$^4$ + \ttxy$^2$    \\
CH 34 & b   &  \GProll\ + \ttt$^1$ + \ttsm$^2$ + \ttco$^4$ + \ttbg$^4$   \\
CH 35 & b   &  \GProll\ + \ttt$^1$ + \ttbg$^4$ + \ttxy$^4$   \\
CH 36 & b   &  \GProll\ + \ttt$^1$ + \ttbg$^1$   \\
\noalign{\smallskip}
\hline
\end{tabular}
\tablefoot{CHEOPS (CH) LCs are identified by a counter based on the chronological order of observation as in Table~\ref{tab:cheopsLog}. \GProll\ refers to the GP modelling of the flux vs. roll angle pattern and it is the only detrending term applied before the \texttt{MCMCI} scheme; see text for further details.}
\end{table}

\begin{table*}[!ht]
\begin{center}
\begin{threeparttable}
  \caption{HARPS-N derived parameters of TOI-5624, as inferred using the multi-dimensional GP approach implemented in \texttt{pyaneti}. \label{tab:PyanetiResults}}  
  \begin{tabular}{lcc}
  \hline\hline
  \noalign{\smallskip}
  Parameter & Prior\tablefootmark{(a)}  & Derived value \\
  \noalign{\smallskip}
  \hline
    \noalign{\smallskip}
    \multicolumn{3}{l}{\emph{\bf{TOI-5624~b}}} \\
    \noalign{\smallskip}
    Orbital period, $P_{\mathrm{orb,\,b}}$ (d) &  $\mathcal{N}[3.390364,0.000011]$ & \Pb[] \\
    \noalign{\smallskip}
    Transit epoch, $T_\mathrm{0,\,b}$ (BJD$_\mathrm{TDB}\,-\,$2\,400\,000) & $\mathcal{N}[59649.1245, 0.0011]$ & \Tzerob[]  \\ 
    \noalign{\smallskip}
    Radial velocity semi-amplitude variation, $K_\mathrm{b}$ (\ms) & $\mathcal{U}[0,20]$ & \kb[] \\
    \noalign{\smallskip}
    Planet mass, $M_\mathrm{b}$ ($M_\oplus$) &  & \mpb[]  \\
    \noalign{\smallskip}
    \hline
    \noalign{\smallskip}
    \multicolumn{3}{l}{\emph{\bf{TOI-5624~c}}} \\
    \noalign{\smallskip}
    Orbital period, $P_{\mathrm{orb,\,c}}$ (d) &  $\mathcal{N}[7.885393,0.000038]$ & \Pc[] \\
    \noalign{\smallskip}
    Transit epoch, $T_\mathrm{0,\,c}$ (BJD$_\mathrm{TDB}\,-\,$2\,400\,000) & $\mathcal{N}[59650.9224,0.0010]$ & \Tzeroc[]  \\ 
    \noalign{\smallskip}
    Radial velocity semi-amplitude variation, $K_\mathrm{c}$ (\ms) & $\mathcal{U}[0,20]$ & \kc[] \\
    \noalign{\smallskip}
    Planet mass, $M_\mathrm{c}$ ($M_\oplus$) &  & \mpc[]  \\
    \noalign{\smallskip}
    \hline
    \noalign{\smallskip}
    \multicolumn{3}{l}{\emph{\bf{TOI-5624~d}}} \\
    \noalign{\smallskip}
    Orbital period, $P_{\mathrm{orb,\,d}}$ (d) &  $\mathcal{N}[13.73151,0.00014]$ & \Pd[] \\
    \noalign{\smallskip}
    Transit epoch, $T_\mathrm{0,\,d}$ (BJD$_\mathrm{TDB}\,-\,$2\,400\,000) & $\mathcal{N}[59655.2043,0.0020]$ & \Tzerod[]  \\ 
    \noalign{\smallskip}
    Radial velocity semi-amplitude variation, $K_\mathrm{d}$ (\ms) & $\mathcal{U}[0,20]$ & \kd[] \\
    \noalign{\smallskip}
    Planet mass, $M_\mathrm{d}$ ($M_\oplus$) &  & \mpd[]  \\
    \noalign{\smallskip}
    \hline
    \noalign{\smallskip}
    \multicolumn{3}{l}{\emph{\bf{TOI-5624~e}}} \\
    \noalign{\smallskip}
    Orbital period, $P_{\mathrm{orb,\,e}}$ (d) &  $\mathcal{N}[21.49031,0.00012]$ & \Pe[] \\
    \noalign{\smallskip}
    Transit epoch, $T_\mathrm{0,\,e}$ (BJD$_\mathrm{TDB}\,-\,$2\,400\,000) & $\mathcal{N}[60687.1358, 0.0019]$ & \Tzeroe[]  \\ 
    \noalign{\smallskip}
    Radial velocity semi-amplitude variation, $K_\mathrm{e}$ (\ms) & $\mathcal{U}[0,20]$ & \ke[] \\
    \noalign{\smallskip}
    Planet mass, $M_\mathrm{e}$ ($M_\oplus$) &  & \mpe[]  \\
    \noalign{\smallskip}
    \hline
    \noalign{\smallskip}
    \multicolumn{3}{l}{\emph{\bf{TOI-5624~f}}} \\
    \noalign{\smallskip}
    Orbital period, $P_{\mathrm{orb,\,f}}$ (d) &  $\mathcal{U}[38.0,50.0]$ & \Pf[] \\
    \noalign{\smallskip}
    Inferior conjunction epoch, $T_\mathrm{0,\,f}$ (BJD$_\mathrm{TDB}\,-\,$2\,400\,000) & $\mathcal{U}[60355.0,60414.0]$ & \Tzerof[]  \\ 
    \noalign{\smallskip}
    Radial velocity semi-amplitude variation, $K_\mathrm{f}$ (\ms) & $\mathcal{U}[0,20]$ & \kf[] \\
    \noalign{\smallskip}
    Planet minimum mass, $M_\mathrm{f}\sin i_\mathrm{f}$ ($M_\oplus$) &  & \mpf[]  \\
    \noalign{\smallskip}
    \hline    
    \noalign{\smallskip}
    \multicolumn{3}{l}{\emph{\bf{Additional model parameters}}} \\
    \noalign{\smallskip}
   RV offset, $A_\mathrm{RV}$ (\ms) & $\mathcal{U}[-100,100]$ & \OffsetRV[] \\
   \noalign{\smallskip}
   RV amplitude of GP, $B_\mathrm{RV}$ (\ms) & $\mathcal{U}[0,100]$ & \AmpBRV[] \\
   \noalign{\smallskip}
   RV amplitude of GP time derivative, $C_\mathrm{RV}$ (\ms\,d) & $\mathcal{U}[0,100]$ & \AmpCRV[] \\
   \noalign{\smallskip}
   RV jitter term (\ms) & $\mathcal{U}[0,100]$ & \jHARPSNN[] \\
   \noalign{\smallskip}
   FWHM offset, $A_\mathrm{FWHM}$ (\ms) & $\mathcal{U}[0,15000]$ & \OffsetFWHM[] \\
   \noalign{\smallskip}
   FWHM amplitude of GP, $B_\mathrm{FWHM}$ (\ms) & $\mathcal{U}[0,100]$ & \AmpBFWHM[] \\
   \noalign{\smallskip}   
   FWHM jitter term (\ms) & $\mathcal{U}[0,100]$ & \jFWHM[] \\
   \noalign{\smallskip}   
   Long-term evolution timescale of active regions, $\lambda_e$ (d) & $\mathcal{U}[5,100]$ & \jlambdae[] \\
   \noalign{\smallskip}
   Inverse of the harmonic complexity, $\lambda_p$ & $\mathcal{U}[0.0,2.0]$ & \jlambdap[] \\
   \noalign{\smallskip}
   GP characteristic period, $P_\mathrm{GP}$ (d) & $\mathcal{U}[15,25]$ & \jPGP[]\\
   \noalign{\smallskip}
   \hline
    \end{tabular}
    \tablefoot{\tablefoottext{a} $\mathcal{U}[a,b]$ refers to uniform priors between $a$ and $b$, whereas $\mathcal{N}[\mu,\sigma]$ refers to Gaussian priors with mean $\mu$ and standard deviation $\sigma$.}
\end{threeparttable}
\end{center}
\end{table*}

\begin{table*}
\centering
\small
\renewcommand{\arraystretch}{1.3}
\caption{Posteriors of the TOI-5624 system from the two tested dynamical scenarios.}
\begin{tabular}{l c c c r}
\hline
\hline
\noalign{\smallskip}
Configurations & & \multicolumn{1}{c}{4 Planets} & \multicolumn{1}{c}{5 Planets}  \\
\noalign{\smallskip}
\hline
\noalign{\smallskip}
Parameter & Prior & \multicolumn{2}{c}{MAP (HDI$\pm1\sigma$)} & Unit  \\
\noalign{\smallskip}
\hline
\noalign{\smallskip}
\emph{TOI-5624\,d} \rule{0pt}{12pt} & & & \\
&&& \\
\textsc{Fitted Parameters} & & &&\\
$M_{\mathrm{p}}/M_{\star}$ & \unif{0.30}{954.7} & $44_{-13}^{+4}$   & $17_{-5}^{+8}$ & $\left(\frac{M_\odot}{M_\star}\right) \times 10^{-6}$\\
Orbital Period ($P$) & \unif{13.6}{13.8} & $13.72796_{-0.00087}^{+0.00079}$ & $13.72930_{-0.00109}^{+0.00018}$  & days\\
$\sqrt{e} \cos \omega$  & \unif{-\sqrt{0.5}}{\sqrt{0.5}} & $0.117_{-0.069}^{+0.040}$  & $-0.078_{-0.047}^{+0.098}$  & \\
$\sqrt{e} \sin \omega$ & \unif{-\sqrt{0.5}}{\sqrt{0.5}} & $-0.137_{-0.028}^{+0.094}$ & $-0.004_{-0.111}^{+0.068}$ &\\
Mean Longitude ($\lambda$)& \unif{0}{360} & $315.66_{-0.43}^{+1.98}$ & $318.54_{-0.96}^{+1.04}$ & deg\\
\\
\textsc{Derived Parameters} &  & & &\\ 
Mass ($M_{\mathrm{p}}$)  & -- & $13_{-4}^{+1}$ &  $5_{-1}^{+3}$  & $M_{\oplus}$\\
Eccentricity ($e$)  & \unif{0}{0.5}  & $0.0325_{-0.0230}^{+0.0031}$  & $0.0061_{-0.0060}^{+0.0109}$ & \\
Argument of Periastron ($\omega$) & -- & $-50_{-31}^{+24}$  &  $183_{-70}^{+113}$ &deg\\
Mean Anomaly ($\mathcal{M}$) & -- & $185_{-26}^{+31}$  & $-44_{-115}^{+70}$  & deg\\
\noalign{\smallskip}
\hline
\noalign{\smallskip}
\emph{TOI-5624\,e} \rule{0pt}{12pt} & & & \\
&&& \\
\textsc{Fitted Parameters} & & &&\\
$M_{\mathrm{p}}/M_{\star}$ & \unif{0.30}{954.7} &  $60_{-12}^{+12}$ & $25_{-2}^{+13}$ & $\left(\frac{M_\odot}{M_\star}\right) \times 10^{-6}$\\
Orbital Period ($P$) & \unif{21.4}{21.6} & $21.4979_{-0.0025}^{+0.0010}$ &  $21.49253_{-0.00084}^{+0.00232}$ & days\\
$\sqrt{e} \cos \omega$  & \unif{-\sqrt{0.5}}{\sqrt{0.5}} & $0.010_{-0.094}^{+0.055}$ & $0.049_{-0.091}^{+0.019}$ & \\
$\sqrt{e} \sin \omega$ & \unif{-\sqrt{0.5}}{\sqrt{0.5}} & $-0.0905_{-0.0022}^{+0.1731}$ & $0.119_{-0.037}^{+0.046}$   &\\
Mean Longitude ($\lambda$)& \unif{0}{360} & $241.73_{-0.72}^{+1.12}$ & $241.45_{-0.40}^{+1.15}$ & deg\\
\\
\textsc{Derived Parameters} &  & & &\\ 
Mass ($M_{\mathrm{p}}$)  & -- &  $17_{-4}^{+3}$   & $7.01_{-0.66}^{+3.68}$ & $M_{\oplus}$\\
Eccentricity ($e$)  & \unif{0}{0.5}  & $0.0083_{-0.0083}^{+0.0060}$  & $0.0165_{-0.0062}^{+0.0110}$ & \\
Argument of Periastron ($\omega$) & -- & $276_{-149}^{+66}$ & $67_{-12}^{+44}$ &deg\\
Mean Anomaly ($\mathcal{M}$) & -- & $145_{-60}^{+156}$ &  $-6_{-43}^{+12}$  &deg\\
\noalign{\smallskip}
\hline
\noalign{\smallskip}
\emph{TOI-5624\,f} \rule{0pt}{12pt} & & & \\
&&& \\
\textsc{Fitted Parameters} & & &&\\
$M_{\mathrm{p}}/M_{\star}$ & \unif{0.30}{954.7} &  & $40_{-6}^{+13}$ & $\left(\frac{M_\odot}{M_\star}\right) \times 10^{-6}$\\
Orbital Period ($P$) & \unif{43}{48} &  & $44.092_{-0.057}^{+0.015}$ & days\\
$\sqrt{e} \cos \omega$  & \unif{-\sqrt{0.5}}{\sqrt{0.5}} & & $0.064_{-0.096}^{+0.143}$ & \\
$\sqrt{e} \sin \omega$ & \unif{-\sqrt{0.5}}{\sqrt{0.5}} & & $-0.143_{-0.035}^{+0.193}$  &\\
Mean Longitude ($\lambda$)& \unif{0}{360} &  & $61_{-11}^{+9}$ & deg\\
Inclination ($i$) & \unif{0}{180.0} & & $99_{-29}^{+1}$ & deg \\
\\
\textsc{Derived Parameters} &  & & &\\ 
Mass ($M_{\mathrm{p}}$)  & -- & & $11_{-2}^{+4}$   & $M_{\oplus}$ \\
Eccentricity ($e$)  & \unif{0}{0.5}  & & $0.025_{-0.024}^{+0.017}$ & \\
Argument of Periastron ($\omega$) & -- &  & $-66_{-45}^{+93}$ &deg\\
Mean Anomaly ($\mathcal{M}$) & -- & & $-53_{-96}^{+36}$ &deg\\
\noalign{\smallskip}
\hline
\end{tabular}
\tablefoot{Parameter estimates correspond to the maximum a posteriori (MAP) values, while the reported uncertainties represent the 68.3\% highest density intervals (HDIs) of the marginalised posterior distributions.}
\label{Table:TRADES_results}
\end{table*}

\end{appendix}

\end{document}